\documentclass[12pt]{article}

\usepackage{feynmf,amssymb,psfig,fullpage}
\begin{document}

\newcommand{\feynmfourp}{
        \fmfset{arrow_len}{2.5mm}
        \fmfset{wiggly_len}{1.7mm}
        \fmfset{dot_size}{0.8thick}
        \fmfleft{i1,i2} \fmfright{o1,o2}
                        }
\newcommand{\boxACAC}{
  \parbox{16mm}{
      \begin{fmfgraph}(16,11)
        \feynmfourp\fmfdotn{v}{4}
        \fmf{fermion}{i1,v1,v3,v4,v2,i2}
        \fmf{photon}{o1,v3}
        \fmf{photon}{v1,v2}
        \fmf{photon}{v4,o2}
      \end{fmfgraph}}
                        }
\newcommand{\treeACCA}{
  \parbox{16mm}{
      \begin{fmfgraph}(16,11)
        \feynmfourp\fmfdotn{v}{2}
        \fmf{fermion}{o1,v2,v1,i2}
        \fmf{photon}{i1,v1}
        \fmf{photon}{v2,o2}
      \end{fmfgraph}}
                        }
\newcommand{\treeCAAC}{
  \parbox{16mm}{
      \begin{fmfgraph}(16,11)
        \feynmfourp\fmfdotn{v}{2}
        \fmf{fermion}{i1,v1,v2,o2}
        \fmf{photon}{v1,i2}
        \fmf{photon}{o1,v2}
      \end{fmfgraph}}
                        }
\newcommand{\boxCACA}{
  \parbox{16mm}{
      \begin{fmfgraph}(16,11)
        \feynmfourp\fmfdotn{v}{4}
        \fmf{fermion}{o1,v3,v1,v2,v4,o2}
        \fmf{photon}{i1,v1}
        \fmf{photon}{v3,v4}
        \fmf{photon}{v2,i2}
      \end{fmfgraph}}
                        }
\newcommand{\cboxACAC}{
  \parbox{10mm}{
      \begin{fmfgraph}(10,8)
        \feynmfourp\fmfdotn{v}{2}
        \fmf{fermion}{i1,v1}\fmf{fermion,right,tension=0.7}{v1,v2}
        \fmf{fermion}{v2,i2}
        \fmf{photon,left,tension=0.7}{v1,v2}
        \fmf{photon}{o1,v1}
        \fmf{photon}{v2,o2}
      \end{fmfgraph}}
                        }
\newcommand{\cboxACCA}{
  \parbox{10mm}{
      \begin{fmfgraph}(10,8)
        \feynmfourp\fmfdotn{v}{2}
        \fmf{fermion}{o1,v1}\fmf{fermion,left,tension=0.7}{v1,v2}
        \fmf{fermion}{v2,i2}
        \fmf{photon,right,tension=0.7}{v1,v2}
        \fmf{photon}{i1,v1}
        \fmf{photon}{v2,o2}
      \end{fmfgraph}}
                        }
\newcommand{\ctreeACCA}{
  \parbox{10mm}{
      \begin{fmfgraph}(10,8)
        \feynmfourp\fmfdotn{v}{1}
        \fmf{photon}{i1,v1}
        \fmf{photon}{v1,o2}
        \fmf{fermion}{o1,v1}\fmf{fermion}{v1,i2}
      \end{fmfgraph}}
                        }
\newcommand{\ctreeACAC}{
  \parbox{10mm}{
      \begin{fmfgraph}(10,8)
        \feynmfourp\fmfdotn{v}{1}
        \fmf{photon}{o1,v1}
        \fmf{photon}{v1,o2}
        \fmf{fermion}{i1,v1}\fmf{fermion}{v1,i2}
      \end{fmfgraph}}
                        }
\newcommand{\ctreeCAAC}{
  \parbox{10mm}{
      \begin{fmfgraph}(10,8)
        \feynmfourp\fmfdotn{v}{1}
        \fmf{fermion}{i1,v1}\fmf{fermion}{v1,o2}
        \fmf{photon}{o1,v1}
        \fmf{photon}{v1,i2}
      \end{fmfgraph}}
                        }
\newcommand{\ctreeCACA}{
  \parbox{10mm}{
      \begin{fmfgraph}(10,8)
        \feynmfourp\fmfdotn{v}{1}
        \fmf{fermion}{o1,v1}\fmf{fermion}{v1,o2}
        \fmf{photon}{i1,v1}
        \fmf{photon}{v1,i2}
      \end{fmfgraph}}
                        }
\newcommand{\cboxCACA}{
  \parbox{10mm}{
      \begin{fmfgraph}(10,8)
        \feynmfourp\fmfdotn{v}{2}
        \fmf{fermion}{o1,v1} \fmf{fermion,left,tension=0.7}{v1,v2}
        \fmf{fermion}{v2,o2}
        \fmf{photon}{i1,v1}
        \fmf{photon}{v2,i2}
        \fmf{photon,right,tension=0.7}{v1,v2}
      \end{fmfgraph}}
                        }
\newcommand{\cboxCAAC}{
  \parbox{10mm}{
      \begin{fmfgraph}(10,8)
        \feynmfourp\fmfdotn{v}{2}
        \fmf{fermion}{i1,v1} \fmf{fermion,right,tension=0.7}{v1,v2}
        \fmf{fermion}{v2,o2}
        \fmf{photon}{o1,v1}
        \fmf{photon}{v2,i2}
        \fmf{photon,left,tension=0.7}{v1,v2}
      \end{fmfgraph}}
                        }
\newcommand{\LboxACAC}{
  \parbox{16mm}{
      \begin{fmfgraph*}(16,11)
        \feynmfourp\fmfdotn{v}{4}
        \fmf{fermion}{i1,v1,v3}
        \fmf{fermion,label=$\scriptstyle m$,label.side=right}{v3,v4}
        \fmf{fermion}{v4,v2,i2}
        \fmf{photon}{o1,v3}
        \fmf{photon,label=$\scriptstyle M$,label.side=left}{v1,v2}
        \fmf{photon}{v4,o2}
      \end{fmfgraph*}}
                        }
\newcommand{\LboxCACA}{
  \parbox{16mm}{
      \begin{fmfgraph*}(16,11)
        \feynmfourp\fmfdotn{v}{4}
        \fmf{fermion}{o1,v3,v1}
        \fmf{fermion,label=$\scriptstyle m$,label.side=left}{v1,v2}
        \fmf{fermion}{v2,v4,o2}
        \fmf{photon}{i1,v1}
        \fmf{photon,label=$\scriptstyle M$,label.side=right}{v3,v4}
        \fmf{photon}{v2,i2}
      \end{fmfgraph*}}
                        }
\newcommand{\LcboxACAC}{
  \parbox{16mm}{
      \begin{fmfgraph*}(16,11)
        \feynmfourp\fmfdotn{v}{2}
        \fmf{fermion}{i1,v1}\fmf{fermion,right,tension=0.7,
            label=$\s m$}{v1,v2}
        \fmf{fermion}{v2,i2}
        \fmf{photon,left,tension=0.7,label=$\scriptstyle M$}{v1,v2}
        \fmf{photon}{o1,v1}
        \fmf{photon}{v2,o2}
      \end{fmfgraph*}}
                        }
\newcommand{\LcboxCACA}{
  \parbox{16mm}{
      \begin{fmfgraph*}(16,11)
        \feynmfourp\fmfdotn{v}{2}
        \fmf{fermion}{o1,v1}\fmf{fermion,left,tension=0.7,
            label=$\s m$}{v1,v2}
        \fmf{fermion}{v2,o2}
        \fmf{photon,right,tension=0.7,label=$\scriptstyle M$}{v1,v2}
        \fmf{photon}{i1,v1}
        \fmf{photon}{v2,i2}
      \end{fmfgraph*}}
                        }
\newcommand{\LtreeCAAC}{
  \parbox{16mm}{
      \begin{fmfgraph*}(16,11)
        \feynmfourp\fmfdotn{v}{2}
        \fmf{fermion}{i1,v1}
        \fmf{fermion,label=$\scriptstyle m$}{v1,v2}
        \fmf{fermion}{v2,o2}
        \fmf{photon}{v1,i2}
        \fmf{photon}{o1,v2}
      \end{fmfgraph*}}
                        }
\newcommand{\LtreeACCA}{
  \parbox{16mm}{
      \begin{fmfgraph*}(16,11)
        \feynmfourp\fmfdotn{v}{2}
        \fmf{fermion}{o1,v2}
        \fmf{fermion,label=$\scriptstyle m$}{v2,v1}
        \fmf{fermion}{v1,i2}
        \fmf{photon}{v1,i1}
        \fmf{photon}{o2,v2}
      \end{fmfgraph*}}
                        }
\newcommand{\treecrACAC}{
  \parbox{16mm}{
      \begin{fmfgraph}(16,11)
                        \feynmfourp\fmfdotn{v}{2}
                        \fmf{fermion}{i1,v1}
                        \fmf{phantom}{v1,i2}
                        \fmf{fermion,tension=0.7}{v1,v2}
                        \fmf{photon}{o1,v2}
                        \fmf{phantom}{v2,o2}
        \fmffreeze
        \fmf{photon}{v1,o2}
        \fmf{fermion,rubout}{v2,i2}
      \end{fmfgraph}}
                        }
\newcommand{\boxcrCAAC}{
  \parbox{16mm}{
      \begin{fmfgraph}(16,11)
        \feynmfourp\fmfdotn{v}{4}
        \fmf{fermion}{i1,v1}
        \fmf{fermion,tension=0.7}{v1,v3}
        \fmf{fermion}{v3,v4}
        \fmf{fermion,tension=0.7}{v4,v2}\fmf{phantom}{v2,i2}
        \fmf{photon}{o1,v3}
        \fmf{photon}{v1,v2}\fmf{phantom}{v4,o2}
        \fmffreeze
        \fmf{fermion,rubout}{v2,o2}
        \fmf{photon,rubout}{v4,i2}
      \end{fmfgraph}}
                        }
\newcommand{\boxcrACCA}{
  \parbox{16mm}{
      \begin{fmfgraph}(16,11)
        \feynmfourp\fmfdotn{v}{4}
        \fmf{fermion}{o1,v3}
        \fmf{fermion,tension=0.7}{v3,v1}
        \fmf{fermion}{v1,v2}
        \fmf{fermion,tension=0.7}{v2,v4}\fmf{phantom}{v4,o2}
        \fmf{photon}{i1,v1}
        \fmf{photon}{v3,v4}\fmf{phantom}{v2,i2}
        \fmffreeze
        \fmf{photon,rubout}{v2,o2}
        \fmf{fermion,rubout}{v4,i2}
      \end{fmfgraph}}
                        }
\newcommand{\treecrCACA}{
  \parbox{16mm}{
      \begin{fmfgraph}(16,11)
                        \feynmfourp\fmfdotn{v}{2}
                        \fmf{photon}{i1,v1}
                        \fmf{phantom}{v1,i2}
                        \fmf{fermion}{o1,v2}
                        \fmf{fermion,tension=0.7}{v2,v1}
                        \fmf{phantom}{v2,o2}
        \fmffreeze
        \fmf{photon}{v2,i2}
        \fmf{fermion,rubout}{v1,o2}
      \end{fmfgraph}}
                        }
\newcommand{\treeBABA}{
  \parbox{16mm}{
      \begin{fmfgraph}(16,11)
                        \feynmfourp\fmfdotn{v}{2}
                        \fmf{fermion}{i2,v1,i1}
                        \fmf{photon}{v1,v2}
                        \fmf{fermion}{o1,v2,o2}
      \end{fmfgraph}}
                        }
\newcommand{\boxBABAh}{
  \parbox{16mm}{
      \begin{fmfgraph}(16,11)
                        \feynmfourp\fmfdotn{v}{4}
                        \fmf{fermion}{i2,v2,v1,i1}
                        \fmf{photon}{v1,v3}
                        \fmf{photon}{v2,v4}
                        \fmf{fermion}{o1,v3,v4,o2}
      \end{fmfgraph}}
                        }
\newcommand{\boxBABAv}{
  \parbox{16mm}{
      \begin{fmfgraph}(16,11)
                        \feynmfourp\fmfdotn{v}{4}
                        \fmf{fermion}{i2,v2,v4,o2}
                        \fmf{photon}{v2,v1}
                        \fmf{photon}{v4,v3}
                        \fmf{fermion}{o1,v3,v1,i1}
      \end{fmfgraph}}
                                }
\newcommand{\ctreeBABA}{
  \parbox{16mm}{
      \begin{fmfgraph}(16,11)
                        \feynmfourp\fmfdotn{v}{1}
                        \fmf{fermion}{i2,v1,o2}
                        \fmf{fermion}{o1,v1,i1}
      \end{fmfgraph}}
                        }
\newcommand{\cboxBABAh}{
  \parbox{16mm}{
      \begin{fmfgraph}(16,11)
                        \feynmfourp\fmfdotn{v}{2}
                        \fmf{fermion}{i2,v2,o2}
                        \fmf{fermion,right}{v2,v1}
                        \fmf{fermion,right}{v1,v2}
                        \fmf{fermion}{o1,v1,i1}
      \end{fmfgraph}}
                        }
\newcommand{\cboxBABAv}{
  \parbox{16mm}{
      \begin{fmfgraph}(16,11)
                        \feynmfourp\fmfdotn{v}{2}
                        \fmf{fermion}{i2,v2,o2}
                        \fmf{photon,left}{v2,v1}
                        \fmf{photon,left}{v1,v2}
                        \fmf{fermion}{o1,v1,i1}
      \end{fmfgraph}}
                        }
\newcommand{\boxBAAB}{
  \parbox{16mm}{
      \begin{fmfgraph}(16,11)
                        \feynmfourp\fmfdotn{v}{4}
                        \fmf{fermion}{i2,v2,v4,o2}
                        \fmf{photon}{v2,v1}
                        \fmf{photon}{v4,v3}
                        \fmf{fermion}{i1,v1,v3,o1}
      \end{fmfgraph}}
                        }
\newcommand{\cboxBAAB}{
  \parbox{16mm}{
      \begin{fmfgraph}(16,11)
                        \feynmfourp\fmfdotn{v}{2}
                        \fmf{fermion}{i2,v2,o2}
                        \fmf{photon,left}{v2,v1}
                        \fmf{photon,left}{v1,v2}
                        \fmf{fermion}{i1,v1,o1}
      \end{fmfgraph}}
                        }
\newcommand{\treeBACC}{
  \parbox{16mm}{
      \begin{fmfgraph}(16,11)
                        \feynmfourp\fmfdotn{v}{2}
                        \fmf{fermion}{i2,v1,v2,o2}
                        \fmf{photon}{i1,v1}
                        \fmf{photon}{o1,v2}
      \end{fmfgraph}}
                        }
\newcommand{\boxBACC}{
  \parbox{16mm}{
      \begin{fmfgraph}(16,11)
                        \feynmfourp\fmfdotn{v}{4}
                        \fmf{fermion}{i2,v2,v1,v3,v4,o2}
                        \fmf{photon}{v2,v4}
                        \fmf{photon}{i1,v1}
                        \fmf{photon}{o1,v3}
      \end{fmfgraph}}
                        }
\newcommand{\ctreeBACC}{
  \parbox{16mm}{
      \begin{fmfgraph}(16,11)
                        \feynmfourp\fmfdotn{v}{1}
                        \fmf{fermion}{i2,v1,o2}
                        \fmf{photon}{i1,v1,o1}
      \end{fmfgraph}}
                        }
\newcommand{\cboxBACC}{
  \parbox{16mm}{
      \begin{fmfgraph}(16,11)
                        \feynmfourp\fmfdotn{v}{2}
                        \fmf{fermion}{i2,v2,o2}
                        \fmf{fermion,right}{v2,v1}
                        \fmf{fermion,right}{v1,v2}
                        \fmf{photon}{i1,v1,o1}
      \end{fmfgraph}}
                        }
\newcommand{\boxABBA}{
  \parbox{16mm}{
      \begin{fmfgraph}(16,11)
                        \feynmfourp\fmfdotn{v}{4}
                        \fmf{fermion}{o2,v4,v2,i2}
                        \fmf{fermion}{o1,v3,v1,i1}
                        \fmf{photon}{v1,v2}
                        \fmf{photon}{v3,v4}
      \end{fmfgraph}}
                        }
\newcommand{\cboxABBA}{
  \parbox{16mm}{
      \begin{fmfgraph}(16,11)
                        \feynmfourp\fmfdotn{v}{2}
                        \fmf{fermion}{o2,v2,i2}
                        \fmf{fermion}{o1,v1,i1}
                        \fmf{photon,left}{v1,v2}
                        \fmf{photon,left}{v2,v1}
      \end{fmfgraph}}
                        }
\newcommand{\treeABAB}{
  \parbox{16mm}{
      \begin{fmfgraph}(16,11)
                        \feynmfourp\fmfdotn{v}{2}
                        \fmf{fermion}{i1,v1,i2}
                        \fmf{fermion}{o2,v2,o1}
                        \fmf{photon}{v1,v2}
      \end{fmfgraph}}
                        }
\newcommand{\boxABABh}{
  \parbox{16mm}{
      \begin{fmfgraph}(16,11)
                        \feynmfourp\fmfdotn{v}{4}
                        \fmf{fermion}{i1,v1,v2,i2}
                        \fmf{fermion}{o2,v4,v3,o1}
                        \fmf{photon}{v1,v3}
                        \fmf{photon}{v2,v4}
      \end{fmfgraph}}
                        }
\newcommand{\boxABABv}{
  \parbox{16mm}{
      \begin{fmfgraph}(16,11)
                        \feynmfourp\fmfdotn{v}{4}
                        \fmf{fermion}{o2,v4,v2,i2}
                        \fmf{fermion}{i1,v1,v3,o1}
                        \fmf{photon}{v1,v2}
                        \fmf{photon}{v3,v4}
      \end{fmfgraph}}
                        }
\newcommand{\ctreeABAB}{
  \parbox{16mm}{
      \begin{fmfgraph}(16,11)
                        \feynmfourp\fmfdotn{v}{1}
                        \fmf{fermion}{i1,v1,i2}
                        \fmf{fermion}{o2,v1,o1}
      \end{fmfgraph}}
                        }
\newcommand{\cboxABABh}{
  \parbox{16mm}{
      \begin{fmfgraph}(16,11)
                        \feynmfourp\fmfdotn{v}{2}
                        \fmf{fermion}{i1,v1,o1}
                        \fmf{fermion}{o2,v2,i2}
                        \fmf{fermion,left}{v1,v2}
                        \fmf{fermion,left}{v2,v1}
      \end{fmfgraph}}
                        }
\newcommand{\cboxABABv}{
  \parbox{16mm}{
      \begin{fmfgraph}(16,11)
                        \feynmfourp\fmfdotn{v}{2}
                        \fmf{fermion}{i1,v1,o1}
                        \fmf{fermion}{o2,v2,i2}
                        \fmf{photon,left}{v1,v2}
                        \fmf{photon,left}{v2,v1}
      \end{fmfgraph}}
                        }
\newcommand{\treeABCC}{
  \parbox{16mm}{
      \begin{fmfgraph}(16,11)
                        \feynmfourp\fmfdotn{v}{2}
                        \fmf{fermion}{o2,v2,v1,i2}
                        \fmf{photon}{i1,v1}
                        \fmf{photon}{o1,v2}
      \end{fmfgraph}}
                        }
\newcommand{\boxABCC}{
  \parbox{16mm}{
      \begin{fmfgraph}(16,11)
                        \feynmfourp\fmfdotn{v}{4}
                        \fmf{fermion}{o2,v4,v3,v1,v2,i2}
                        \fmf{photon}{v2,v4}
                        \fmf{photon}{i1,v1}
                        \fmf{photon}{o1,v3}
      \end{fmfgraph}}
                        }
\newcommand{\ctreeABCC}{
  \parbox{16mm}{
      \begin{fmfgraph}(16,11)
                        \feynmfourp\fmfdotn{v}{1}
                        \fmf{fermion}{o2,v1,i2}
                        \fmf{photon}{i1,v1}
                        \fmf{photon}{o1,v1}
      \end{fmfgraph}}
                        }
\newcommand{\cboxABCC}{
  \parbox{16mm}{
      \begin{fmfgraph}(16,11)
                        \feynmfourp\fmfdotn{v}{2}
                        \fmf{fermion}{o2,v2,i2}
                        \fmf{fermion,left}{v1,v2}
                        \fmf{fermion,left}{v2,v1}
                        \fmf{photon}{i1,v1}
                        \fmf{photon}{o1,v1}
      \end{fmfgraph}}
                        }
\newcommand{\treeCCBA}{
  \parbox{16mm}{
      \begin{fmfgraph}(16,11)
                        \feynmfourp\fmfdotn{v}{2}
                        \fmf{fermion}{o1,v2,v1,i1}
                        \fmf{photon}{i2,v1}
                        \fmf{photon}{o2,v2}
      \end{fmfgraph}}
                        }
\newcommand{\boxCCBA}{
  \parbox{16mm}{
      \begin{fmfgraph}(16,11)
                        \feynmfourp\fmfdotn{v}{4}
                        \fmf{fermion}{o1,v3,v4,v2,v1,i1}
                        \fmf{photon}{v1,v3}
                        \fmf{photon}{i2,v2}
                        \fmf{photon}{o2,v4}
      \end{fmfgraph}}
                        }
\newcommand{\ctreeCCBA}{
  \parbox{16mm}{
      \begin{fmfgraph}(16,11)
                        \feynmfourp\fmfdotn{v}{1}
                        \fmf{fermion}{o1,v1,i1}
                        \fmf{photon}{i2,v1}
                        \fmf{photon}{o2,v1}
      \end{fmfgraph}}
                        }
\newcommand{\cboxCCBA}{
  \parbox{16mm}{
      \begin{fmfgraph}(16,11)
                        \feynmfourp\fmfdotn{v}{2}
                        \fmf{fermion}{o1,v1,i1}
                        \fmf{photon}{o2,v2,i2}
                        \fmf{fermion,right}{v1,v2}
                        \fmf{fermion,right}{v2,v1}
      \end{fmfgraph}}
                        }
\newcommand{\treeCCAB}{
  \parbox{16mm}{
      \begin{fmfgraph}(16,11)
                        \feynmfourp\fmfdotn{v}{2}
                        \fmf{fermion}{i1,v1,v2,o1}
                        \fmf{photon}{i2,v1}
                        \fmf{photon}{o2,v2}
      \end{fmfgraph}}
                        }
\newcommand{\boxCCAB}{
  \parbox{16mm}{
      \begin{fmfgraph}(16,11)
                        \feynmfourp\fmfdotn{v}{4}
                        \fmf{fermion}{i1,v1,v2,v4,v3,o1}
                        \fmf{photon}{v1,v3}
                        \fmf{photon}{i2,v2}
                        \fmf{photon}{o2,v4}
      \end{fmfgraph}}
                        }
\newcommand{\ctreeCCAB}{
  \parbox{16mm}{
      \begin{fmfgraph}(16,11)
                        \feynmfourp\fmfdotn{v}{1}
                        \fmf{fermion}{i1,v1,o1}
                        \fmf{photon}{i2,v1}
                        \fmf{photon}{o2,v1}
      \end{fmfgraph}}
                        }
\newcommand{\cboxCCAB}{
  \parbox{16mm}{
      \begin{fmfgraph}(16,11)
                        \feynmfourp\fmfdotn{v}{2}
                        \fmf{fermion}{i1,v1,o1}
                        \fmf{photon}{o2,v2,i2}
                        \fmf{fermion,left}{v2,v1}
                        \fmf{fermion,left}{v1,v2}
      \end{fmfgraph}}
                        }
\newcommand{\boxCCCCl}{
  \parbox{16mm}{
      \begin{fmfgraph}(16,11)
                        \feynmfourp\fmfdotn{v}{4}
                        \fmf{fermion}{v1,v2,v4,v3,v1}
                        \fmf{photon}{i2,v2}\fmf{photon}{i1,v1}
                        \fmf{photon}{o2,v4}\fmf{photon}{o1,v3}
      \end{fmfgraph}}
                        }
\newcommand{\boxCCCCr}{
  \parbox{16mm}{
      \begin{fmfgraph}(16,11)
                        \feynmfourp\fmfdotn{v}{4}
                        \fmf{fermion}{v1,v3,v4,v2,v1}
                        \fmf{photon}{i2,v2}\fmf{photon}{i1,v1}
                        \fmf{photon}{o2,v4}\fmf{photon}{o1,v3}
      \end{fmfgraph}}
                        }
\newcommand{\cboxCCCCl}{
  \parbox{16mm}{
      \begin{fmfgraph}(16,11)
                        \feynmfourp\fmfdotn{v}{2}
                        \fmf{photon}{i2,v2,o2}
                        \fmf{photon}{i1,v1,o1}
                        \fmf{fermion,left}{v1,v2}
                        \fmf{fermion,left}{v2,v1}
      \end{fmfgraph}}
                        }
\newcommand{\cboxCCCCr}{
  \parbox{16mm}{
      \begin{fmfgraph}(16,11)
                        \feynmfourp\fmfdotn{v}{2}
                        \fmf{photon}{i2,v2,o2}
                        \fmf{photon}{i1,v1,o1}
                        \fmf{fermion,right}{v2,v1}
                        \fmf{fermion,right}{v1,v2}
      \end{fmfgraph}}
                        }
\newcommand{\boxcrBABA}{
  \parbox{16mm}{
      \begin{fmfgraph}(16,11)
                        \feynmfourp\fmfdotn{v}{4}
                        \fmf{fermion}{i2,v2}
                        \fmf{fermion,tension=0}{v2,v4}
                        \fmf{fermion}{v4,o2}
                        \fmf{fermion}{o1,v3}
                        \fmf{fermion,tension=0}{v3,v1}
                        \fmf{fermion}{v1,i1}
                        \fmf{photon,tension=0.5}{v1,v4}
                        \fmf{photon,tension=0.5,rubout}{v2,v3}
      \end{fmfgraph}}
                        }
\newcommand{\treecrABBA}{
  \parbox{16mm}{
      \begin{fmfgraph}(16,11)
                        \feynmfourp\fmfdotn{v}{2}
                        \fmf{phantom}{i1,v1,i2}
                        \fmf{photon,tension=0.7}{v1,v2}
                        \fmf{phantom}{o1,v2,o2}
                        \fmffreeze
                        \fmf{fermion}{o2,v1}\fmf{fermion}{v1,i1}
                        \fmf{fermion,rubout}{v2,i2}
                        \fmf{fermion,rubout}{o1,v2}
      \end{fmfgraph}}
                        }
\newcommand{\boxcrABBAh}{
  \parbox{16mm}{
      \begin{fmfgraph}(16,11)
                        \feynmfourp\fmfdotn{v}{4}
                        \fmf{fermion}{v1,i1}\fmf{fermion}{v2,i2}
                        \fmf{fermion}{o2,v4}\fmf{fermion}{o1,v3}
                        \fmf{phantom,tension=0.5}{v1,v2}
                        \fmf{phantom,tension=0.5}{v3,v4}
                        \fmf{photon,tension=0.5}{v1,v3}
                        \fmf{photon,tension=0.5}{v2,v4}
                        \fmffreeze
                        \fmf{plain}{v3,v}
                        \fmf{fermion,rubout}{v,v2}
                        \fmf{fermion,rubout}{v4,v1}
      \end{fmfgraph}}
                        }
\newcommand{\boxcrABBAv}{
  \parbox{16mm}{
      \begin{fmfgraph}(16,11)
                        \feynmfourp\fmfdotn{v}{4}
                        \fmf{fermion}{v1,i1}\fmf{fermion}{v2,i2}
                        \fmf{fermion}{o2,v4}\fmf{fermion}{o1,v3}
                        \fmf{phantom,tension=0.5}{v1,v2}
                        \fmf{phantom,tension=0.5}{v3,v4}
                        \fmf{fermion,tension=0.5}{v3,v1}
                        \fmf{fermion,tension=0.5}{v4,v2}
                        \fmffreeze
                        \fmf{photon}{v4,v1}
                        \fmf{photon,rubout}{v3,v2}
      \end{fmfgraph}}
                        }
\newcommand{\ctreeABBA}{
  \parbox{16mm}{
      \begin{fmfgraph}(16,11)
                        \feynmfourp\fmfdotn{v}{1}
                        \fmf{fermion}{o1,v1,i2}
                        \fmf{fermion}{o2,v1,i1}
      \end{fmfgraph}}
                        }
\newcommand{\treecrCCBA}{
  \parbox{16mm}{
      \begin{fmfgraph}(16,11)
                        \feynmfourp\fmfdotn{v}{2}
                        \fmf{phantom}{i1,v1,i2}
                        \fmf{fermion,tension=0.7}{v2,v1}
                        \fmf{phantom}{o1,v2,o2}
                        \fmffreeze
                        \fmf{fermion}{o1,v2}\fmf{fermion}{v1,i1}
                        \fmf{phantom}{v1,v2,v3,v4,v1}
                        \fmf{photon}{v2,i2}
                        \fmf{photon,rubout}{o2,v1}
      \end{fmfgraph}}
                        }
\newcommand{\boxcrCCBA}{
  \parbox{16mm}{
      \begin{fmfgraph}(16,11)
                        \feynmfourp\fmfdotn{v}{4}
                        \fmf{fermion}{v1,i1}
                        \fmf{fermion}{o1,v3}
                        \fmf{phantom}{v2,i2}
                        \fmf{phantom}{v4,o2}
                        \fmf{photon,tension=0.7}{v1,v3}
                        \fmf{fermion}{v3,v4}
                        \fmf{fermion,tension=0.7}{v4,v2}
                        \fmf{fermion}{v2,v1}
                        \fmffreeze
                        \fmf{photon}{v2,o2}
                        \fmf{photon,rubout}{v4,i2}
      \end{fmfgraph}}
                        }
\newcommand{\treecrBAAB}{
  \parbox{16mm}{
      \begin{fmfgraph}(16,11)
                        \feynmfourp\fmfdotn{v}{2}
                        \fmf{phantom}{i1,v1,i2}
                        \fmf{photon,tension=0.7}{v1,v2}
                        \fmf{phantom}{o1,v2,o2}
                        \fmffreeze
                        \fmf{fermion}{i1,v1}\fmf{fermion}{v1,o2}
                        \fmf{fermion,rubout}{i2,v2}
                        \fmf{fermion,rubout}{v2,o1}
      \end{fmfgraph}}
                        }
\newcommand{\boxcrBAABh}{
  \parbox{16mm}{
      \begin{fmfgraph}(16,11)
                        \feynmfourp\fmfdotn{v}{4}
                        \fmf{fermion}{i1,v1}\fmf{fermion}{i2,v2}
                        \fmf{fermion}{v4,o2}\fmf{fermion}{v3,o1}
                        \fmf{phantom,tension=0.5}{v1,v2}
                        \fmf{phantom,tension=0.5}{v3,v4}
                        \fmf{photon,tension=0.5}{v1,v3}
                        \fmf{photon,tension=0.5}{v2,v4}
                        \fmffreeze
                        \fmf{plain,rubout}{v1,v}
                        \fmf{fermion,rubout}{v,v4}
                        \fmf{fermion,rubout}{v2,v3}
      \end{fmfgraph}}
                        }
\newcommand{\boxcrBAABv}{
  \parbox{16mm}{
      \begin{fmfgraph}(16,11)
                        \feynmfourp\fmfdotn{v}{4}
                        \fmf{fermion}{i1,v1}\fmf{fermion}{i2,v2}
                        \fmf{fermion}{v4,o2}\fmf{fermion}{v3,o1}
                        \fmf{phantom,tension=0.5}{v1,v2}
                        \fmf{phantom,tension=0.5}{v3,v4}
                        \fmf{fermion,tension=0.5}{v1,v3}
                        \fmf{fermion,tension=0.5}{v2,v4}
                        \fmffreeze
                        \fmf{photon}{v4,v1}
                        \fmf{photon,rubout}{v3,v2}
      \end{fmfgraph}}
                        }
\newcommand{\ctreeBAAB}{
  \parbox{16mm}{
      \begin{fmfgraph}(16,11)
                        \feynmfourp\fmfdotn{v}{1}
                        \fmf{fermion}{i2,v1,o2}
                        \fmf{fermion}{i1,v1,o1}
      \end{fmfgraph}}
                        }
\newcommand{\cboxBAABh}{
  \parbox{16mm}{
      \begin{fmfgraph}(16,11)
                        \feynmfourp\fmfdotn{v}{2}
                        \fmf{fermion}{i2,v2,o2}
                        \fmf{fermion,left}{v2,v1}
                        \fmf{fermion,left}{v1,v2}
                        \fmf{fermion}{i1,v1,o1}
      \end{fmfgraph}}
                        }
\newcommand{\cboxBAABv}{
  \parbox{16mm}{
      \begin{fmfgraph}(16,11)
                        \feynmfourp\fmfdotn{v}{2}
                        \fmf{fermion}{i2,v2,o2}
                        \fmf{photon,left}{v2,v1}
                        \fmf{photon,left}{v1,v2}
                        \fmf{fermion}{i1,v1,o1}
      \end{fmfgraph}}
                        }
\newcommand{\boxcrABAB}{
  \parbox{16mm}{
      \begin{fmfgraph}(16,11)
                        \feynmfourp\fmfdotn{v}{4}
                        \feynmfourp\fmfdotn{v}{4}
                        \fmf{fermion}{i1,v1}\fmf{fermion}{v2,i2}
                        \fmf{fermion}{o2,v4}\fmf{fermion}{v3,o1}
                        \fmf{phantom,tension=0.5}{v1,v2}
                        \fmf{phantom,tension=0.5}{v3,v4}
                        \fmf{fermion,tension=0.5}{v1,v3}
                        \fmf{fermion,tension=0.5}{v4,v2}
                        \fmffreeze
                        \fmf{photon}{v4,v1}
                        \fmf{photon,rubout}{v3,v2}
      \end{fmfgraph}}
                        }
\newcommand{\treecrCCAB}{
  \parbox{16mm}{
      \begin{fmfgraph}(16,11)
                        \feynmfourp\fmfdotn{v}{2}
                        \fmf{phantom}{i1,v1,i2}
                        \fmf{fermion,tension=0.7}{v1,v2}
                        \fmf{phantom}{o1,v2,o2}
                        \fmffreeze
                        \fmf{fermion}{v2,o1}\fmf{fermion}{i1,v1}
                        \fmf{phantom}{v1,v2,v3,v4,v1}
                        \fmf{photon}{v2,i2}
                        \fmf{photon,rubout}{o2,v1}
      \end{fmfgraph}}
                        }
\newcommand{\boxcrCCAB}{
  \parbox{16mm}{
      \begin{fmfgraph}(16,11)
                        \feynmfourp\fmfdotn{v}{4}
                        \fmf{fermion}{i1,v1}
                        \fmf{fermion}{v3,o1}
                        \fmf{phantom}{v2,i2}
                        \fmf{phantom}{v4,o2}
                        \fmf{photon,tension=0.7}{v1,v3}
                        \fmf{fermion}{v4,v3}
                        \fmf{fermion,tension=0.7}{v2,v4}
                        \fmf{fermion}{v1,v2}
                        \fmffreeze
                        \fmf{photon}{v2,o2}
                        \fmf{photon,rubout}{v4,i2}
      \end{fmfgraph}}
                        }
\newcommand{\treecrBACC}{
  \parbox{16mm}{
      \begin{fmfgraph}(16,11)
                        \feynmfourp\fmfdotn{v}{2}
                        \fmf{phantom}{i1,v1,i2}
                        \fmf{fermion,tension=0.7}{v2,v1}
                        \fmf{phantom}{o1,v2,o2}
                        \fmffreeze
                        \fmf{photon}{i1,v1}\fmf{fermion}{v1,o2}
                        \fmf{fermion,rubout}{i2,v2}
                        \fmf{photon}{v2,o1}
      \end{fmfgraph}}
                        }
\newcommand{\boxcrBACC}{
  \parbox{16mm}{
      \begin{fmfgraph}(16,11)
                        \feynmfourp\fmfdotn{v}{4}
                        \fmf{photon}{i1,v1}
                        \fmf{photon}{v3,o1}
                        \fmf{phantom,tension=0.5}{v1,v2}
                        \fmf{phantom,tension=0.5}{v3,v4}
                        \fmf{fermion}{i2,v2}
                        \fmf{fermion}{v4,o2}
                        \fmf{fermion,tension=0.5}{v3,v1}
                        \fmf{photon,tension=0.5}{v4,v2}
                        \fmffreeze
                        \fmf{plain}{v1,v}
                        \fmf{fermion,rubout}{v,v4}
                        \fmf{fermion,rubout}{v2,v3}
      \end{fmfgraph}}
                        }
\newcommand{\treecrABCC}{
  \parbox{16mm}{
      \begin{fmfgraph}(16,11)
                        \feynmfourp\fmfdotn{v}{2}
                        \fmf{phantom}{i1,v1,i2}
                        \fmf{fermion,tension=0.7}{v1,v2}
                        \fmf{phantom}{o1,v2,o2}
                        \fmffreeze
                        \fmf{photon}{i1,v1}\fmf{fermion}{o2,v1}
                        \fmf{fermion,rubout}{v2,i2}
                        \fmf{photon}{v2,o1}
      \end{fmfgraph}}
                        }
\newcommand{\boxcrABCC}{
  \parbox{16mm}{
      \begin{fmfgraph}(16,11)
                        \feynmfourp\fmfdotn{v}{4}
                        \fmf{photon}{i1,v1}
                        \fmf{photon}{v3,o1}
                        \fmf{phantom,tension=0.5}{v1,v2}
                        \fmf{phantom,tension=0.5}{v3,v4}
                        \fmf{fermion}{v2,i2}
                        \fmf{fermion}{o2,v4}
                        \fmf{fermion,tension=0.5}{v1,v3}
                        \fmf{photon,tension=0.5}{v4,v2}
                        \fmffreeze
                        \fmf{plain}{v4,v}
                        \fmf{fermion,rubout}{v,v1}
                        \fmf{fermion,rubout}{v3,v2}
      \end{fmfgraph}}
                        }
\newcommand{\boxcrCCCCl}{
  \parbox{16mm}{
      \begin{fmfgraph}(16,11)
                        \feynmfourp\fmfdotn{v}{4}
                        \fmf{photon}{i2,v2}\fmf{photon}{i1,v1}
                        \fmf{photon}{o2,v4}\fmf{photon}{o1,v3}
                        \fmf{fermion,tension=0.5}{v1,v3}
                        \fmf{fermion,tension=0.5}{v2,v4}
                        \fmf{phantom,tension=0.5}{v1,v2}
                        \fmf{phantom,tension=0.5}{v3,v4}
                        \fmffreeze
                        \fmf{plain}{v4,v}
                        \fmf{fermion,rubout}{v,v1}
                        \fmf{fermion,rubout}{v3,v2}
      \end{fmfgraph}}
                        }
\newcommand{\boxcrCCCCr}{
  \parbox{16mm}{
      \begin{fmfgraph}(16,11)
                        \feynmfourp\fmfdotn{v}{4}
                        \fmf{photon}{i2,v2}\fmf{photon}{i1,v1}
                        \fmf{photon}{o2,v4}\fmf{photon}{o1,v3}
                        \fmf{fermion,tension=0.5}{v3,v1}
                        \fmf{fermion,tension=0.5}{v4,v2}
                        \fmf{phantom,tension=0.5}{v1,v2}
                        \fmf{phantom,tension=0.5}{v3,v4}
                        \fmffreeze
                        \fmf{fermion,rubout}{v,v4}
                        \fmf{plain,rubout}{v1,v}
                        \fmf{fermion,rubout}{v2,v3}
      \end{fmfgraph}}
                        }
\newcommand{\treeAAAA}{
  \parbox{16mm}{
      \begin{fmfgraph}(16,11)
                        \feynmfourp\fmfdotn{v}{2}
                        \fmf{fermion}{i1,v1,i2}
                        \fmf{photon}{v1,v2}
                        \fmf{fermion}{o1,v2,o2}
      \end{fmfgraph}}
                        }
\newcommand{\treecrAAAA}{
  \parbox{16mm}{
      \begin{fmfgraph}(16,11)
                        \feynmfourp\fmfdotn{v}{2}
                        \fmf{fermion}{i1,v1}\fmf{phantom}{v1,i2}
                        \fmf{photon}{v1,v2}
                        \fmf{fermion}{o1,v2}\fmf{phantom}{v2,o2}
                        \fmffreeze
                        \fmf{fermion,rubout}{v1,o2}
                        \fmf{fermion,rubout}{v2,i2}
      \end{fmfgraph}}
                        }
\newcommand{\boxAAAA}{
  \parbox{16mm}{
      \begin{fmfgraph}(16,11)
                        \feynmfourp\fmfdotn{v}{4}
                        \fmf{fermion}{i1,v1,v2,i2}
                        \fmf{fermion}{o1,v3,v4,o2}
                        \fmf{photon}{v1,v3}
                        \fmf{photon}{v2,v4}
      \end{fmfgraph}}
                        }
\newcommand{\boxcrAAAA}{
  \parbox{16mm}{
      \begin{fmfgraph}(16,11)
                        \feynmfourp\fmfdotn{v}{4}
                        \fmf{fermion}{i1,v1}\fmf{fermion}{v2,i2}
                        \fmf{fermion}{v4,o2}\fmf{fermion}{o1,v3}
                        \fmf{phantom,tension=0.5}{v1,v2}
                        \fmf{phantom,tension=0.5}{v3,v4}
                        \fmf{photon,tension=0.5}{v1,v3}
                        \fmf{photon,tension=0.5}{v2,v4}
                        \fmffreeze
                        \fmf{plain,rubout}{v1,v}
                        \fmf{fermion,rubout}{v,v4}
                        \fmf{fermion,rubout}{v3,v2}
      \end{fmfgraph}}
                        }
\newcommand{\ctreeAAAA}{
  \parbox{16mm}{
      \begin{fmfgraph}(16,11)
                        \feynmfourp\fmfdotn{v}{1}
                        \fmf{fermion}{i1,v1,i2}
                        \fmf{fermion}{o1,v1,o2}
      \end{fmfgraph}}
                        }
\newcommand{\cboxAAAA}{
  \parbox{16mm}{
      \begin{fmfgraph}(16,11)
                        \feynmfourp\fmfdotn{v}{2}
                        \fmf{fermion}{i1,v1}
                        \fmf{fermion,left}{v1,v2}
                        \fmf{fermion}{v2,i2}
                        \fmf{fermion}{o1,v1}
                        \fmf{fermion,right}{v1,v2}
                        \fmf{fermion}{v2,o2}
      \end{fmfgraph}}
                        }
\newcommand{\LboxAAAA}{
  \parbox{16mm}{
      \begin{fmfgraph*}(16,11)
                        \feynmfourp\fmfdotn{v}{4}
                        \fmf{fermion}{i1,v1}
                        \fmf{fermion,label=$\s m$,label.side=left}{v1,v2}
                        \fmf{fermion}{v2,i2}
                        \fmf{fermion}{o1,v3}
                        \fmf{fermion,label=$\s m$,label.side=right}{v3,v4}
                        \fmf{fermion}{v4,o2}
                        \fmf{photon,label=${}_M$,l.side=right}{v1,v3}
                        \fmf{photon,label=${}_M$,l.side=left}{v2,v4}
      \end{fmfgraph*}}
                        }
\newcommand{\LcboxAAAA}{
  \parbox{16mm}{
      \begin{fmfgraph*}(16,11)
                        \feynmfourp\fmfdotn{v}{2}
                        \fmf{fermion}{i1,v1}
                        \fmf{fermion,left,label=$\s m$}{v1,v2}
                        \fmf{fermion}{v2,i2}
                        \fmf{fermion}{o1,v1}
                        \fmf{fermion,right,label=$\s m$}{v1,v2}
                        \fmf{fermion}{v2,o2}
      \end{fmfgraph*}}
                        }
\newcommand{\fffgtwotree}{
  \parbox{16mm}{
      \begin{fmfgraph}(16,11)
                \fmfset{dot_size}{0.8thick}
                        \fmfleft{i1}
                        \fmfright{o1}
                        \fmf{plain}{i1,o1}
      \end{fmfgraph}}
                        }
\newcommand{\fffgtwooneloop}{
  \parbox{16mm}{
      \begin{fmfgraph}(16,11)
                \fmfset{dot_size}{0.8thick}
                        \fmfleft{i1}
                        \fmfright{o1}
                        \fmfdotn{v}{2}
                        \fmf{plain}{i1,v1}
                        \fmf{plain,left,tension=0.5}{v1,v2,v1}
                        \fmf{plain}{v2,o1}
      \end{fmfgraph}}
                        }
\newcommand{\fffgthreetreeL}{
  \parbox{16mm}{
      \begin{fmfgraph}(16,11)
                \fmfset{dot_size}{0.8thick}
                        \fmfleft{i1,i2}
                        \fmfright{o1}
                        \fmfdotn{v}{1}
                        \fmf{plain}{i1,v1,i2}
                        \fmf{plain}{v1,o1}
      \end{fmfgraph}}
                        }
\newcommand{\fffgthreetreeR}{
  \parbox{16mm}{
      \begin{fmfgraph}(16,11)
                \fmfset{dot_size}{0.8thick}
                        \fmfright{o1,o2}
                        \fmfleft{i1}
                        \fmfdotn{v}{1}
                        \fmf{plain}{o1,v1,o2}
                        \fmf{plain}{v1,i1}
      \end{fmfgraph}}
                        }
\newcommand{\fffgthreeoneloopL}{
  \parbox{16mm}{
      \begin{fmfgraph}(16,11)
                \fmfset{dot_size}{0.8thick}
                        \fmfleft{i1,i2}
                        \fmfright{o1}
                        \fmfdotn{v}{3}
                        \fmf{plain}{i1,v1,v3,v2,i2}
                        \fmf{plain}{v3,o1}
                        \fmf{plain,tension=0.5}{v1,v2}
      \end{fmfgraph}}
                        }
\newcommand{\fffgthreeoneloopR}{
  \parbox{16mm}{
      \begin{fmfgraph}(16,11)
                \fmfset{dot_size}{0.8thick}
                        \fmfright{o1,o2}
                        \fmfleft{i1}
                        \fmfdotn{v}{3}
                        \fmf{plain}{o1,v1,v3,v2,o2}
                        \fmf{plain}{v3,i1}
                        \fmf{plain,tension=0.5}{v1,v2}
      \end{fmfgraph}}
                        }
\newcommand{\fffgfourtree}{
  \parbox{16mm}{
      \begin{fmfgraph}(16,11)
                \fmfset{dot_size}{0.8thick}
                        \fmfleft{i1,i2}
                        \fmfright{o1,o2}
                        \fmfdotn{v}{2}
                        \fmf{plain}{i1,v1,i2}
                        \fmf{plain}{v1,v2}
                        \fmf{plain}{o1,v2,o2}
      \end{fmfgraph}}
                        }
\newcommand{\fffgfourtreeS}{
  \parbox{16mm}{
      \begin{fmfgraph}(16,11)
                \fmfset{dot_size}{0.8thick}
                        \fmfleft{i1,i2}
                        \fmfright{o1,o2}
                        \fmfdotn{v}{2}
                        \fmf{plain}{i1,v1,v2,i2}
                        \fmf{plain}{o1,v1,v2,o2}
      \end{fmfgraph}}
                        }
\newcommand{\fffgfourtreepc}{
  \parbox{16mm}{
      \begin{fmfgraph}(16,11)
                \fmfset{dot_size}{0.8thick}
                        \fmfleft{i1,i2}
                        \fmfright{o1,o2}
                        \fmfdotn{v}{4}
                        \fmf{plain}{i1,v1}
                        \fmf{plain}{i2,v1,v2}
			\fmf{plain,left,tension=0.5}{v2,v3}
			\fmf{plain,left,tension=0.5}{v3,v2}
			\fmf{plain}{v3,v4,o2}
                        \fmf{plain}{o1,v4}
      \end{fmfgraph}}
                        }
\newcommand{\fffgfourtreeSpc}{
  \parbox{16mm}{
      \begin{fmfgraph}(16,11)
                \fmfset{dot_size}{0.8thick}
                        \fmfleft{i1,i2}
                        \fmfright{o1,o2}
                        \fmfdotn{v}{4}
                        \fmf{plain}{i1,v1,v2}
                        \fmf{plain}{o1,v1,v2}
			\fmf{plain,left,tension=0.5}{v2,v3}
			\fmf{plain,left,tension=0.5}{v3,v2}
                        \fmf{plain}{v3,v4,o2}
			\fmf{plain}{v3,v4,i2}
      \end{fmfgraph}}
                        }
\newcommand{\fffgfourtreelc}{
  \parbox{16mm}{
      \begin{fmfgraph}(16,11)
                \fmfset{dot_size}{0.8thick}
                        \fmfleft{i1,i2}
                        \fmfright{o1,o2}
                        \fmfdotn{v}{4}
                        \fmf{plain}{i1,v1}\fmf{phantom}{v1,i2}
			\fmf{plain}{v1,v4}
			\fmf{plain}{o1,v4,o2}
			\fmffreeze
                        \fmf{plain,tension=1.5}{v1,v2}
			\fmf{plain,left}{v2,v3}
			\fmf{plain,left}{v3,v2}
			\fmf{plain,tension=1.5}{v3,i2}

      \end{fmfgraph}}
                        }
\newcommand{\fffgfourtreeSlc}{
  \parbox{16mm}{
      \begin{fmfgraph}(16,11)
                \fmfset{dot_size}{0.8thick}
                        \fmfleft{i1,i2}
                        \fmfright{o1,o2}
                        \fmfdotn{v}{4}
                        \fmf{plain}{i1,v1,v2,o2}\fmf{phantom}{v2,i2}
			\fmf{plain}{o1,v1}
			\fmffreeze
                        \fmf{plain,tension=2.5}{v2,v3}
			\fmf{phantom,tension=2.5}{v3,v4}
			\fmf{plain,tension=2.5}{v4,i2}
			\fmffreeze
			\fmf{plain,left,tension=0.8}{v3,v4}
			\fmf{plain,left,tension=0.8}{v4,v3}
      \end{fmfgraph}}
                        }
\newcommand{\fffgfourtreeL}{
  \parbox{16mm}{
      \begin{fmfgraph}(16,11)
                \fmfset{dot_size}{0.8thick}
                        \fmfleft{i1,i2}
                        \fmfright{o1,o2}
                        \fmfdotn{v}{4}
			\fmf{phantom}{i1,v2,i2}
                        \fmf{plain}{o1,v4,o2}
                        \fmf{plain}{v2,v4}
                        \fmffreeze
                        \fmf{plain}{i1,v1,v2,v3,i2}
                        \fmf{plain,tension=0.5,left}{v1,v3}
      \end{fmfgraph}}
                        }
\newcommand{\fffgfourtreeST}{
  \parbox{16mm}{
      \begin{fmfgraph}(16,11)
                \fmfset{dot_size}{0.8thick}
                        \fmfleft{i1,i2}
                        \fmfright{o1,o2}
                        \fmfdotn{v}{4}
			\fmf{plain}{i1,v1,v2}\fmf{phantom}{v2,i2}
			\fmf{plain}{o1,v1,v2}\fmf{phantom}{v2,o2}
                        \fmf{plain}{v2,v3,i2}
                        \fmf{plain}{v2,v4,o2}
			\fmffreeze
                        \fmf{plain,left}{v3,v4}
      \end{fmfgraph}}
                        }
\newcommand{\fffgfourtreeSB}{
  \parbox{16mm}{
      \begin{fmfgraph}(16,11)
                \fmfset{dot_size}{0.8thick}
                        \fmfleft{i1,i2}
                        \fmfright{o1,o2}
                        \fmfdotn{v}{4}
			\phantom{i1,v2}
			\phantom{o1,v4}
			\fmf{plain}{i1,v1,v2}
			\fmf{plain,tension=0.8}{v2,v4}\fmf{plain}{v4,i2}
			\fmf{plain}{o1,v3,v2}
			\fmf{plain}{v4,o2}
			\fmffreeze
                        \fmf{plain,right}{v1,v3}
      \end{fmfgraph}}
                        }
\newcommand{\fffgfourtreeR}{
  \parbox{16mm}{
      \begin{fmfgraph}(16,11)
                \fmfset{dot_size}{0.8thick}
                        \fmfleft{i1,i2}
                        \fmfright{o1,o2}
                        \fmfdotn{v}{4}
                        \fmf{plain}{i1,v1,i2}
                        \fmf{plain}{v1,v2}
			\fmf{phantom}{o1,v2,o2}
			\fmffreeze
                        \fmf{plain}{o1,v3,v2,v4,o2}
                        \fmf{plain,tension=0.5,right}{v3,v4}
      \end{fmfgraph}}
                        }
\newcommand{\fffgfourtreecr}{
  \parbox{16mm}{
      \begin{fmfgraph}(16,11)
                \fmfset{dot_size}{0.8thick}
                        \fmfleft{i1,i2}
                        \fmfright{o1,o2}
                        \fmfdotn{v}{2}
                        \fmf{plain}{i1,v1}
                        \fmf{phantom}{v1,i2}
                        \fmf{plain}{v1,v2}
                        \fmf{plain}{o1,v2}
                        \fmf{phantom}{v2,o2}
                        \fmffreeze
                        \fmf{plain,rubout}{v1,o2}
                        \fmf{plain,rubout}{v2,i2}
      \end{fmfgraph}}
                        }
\newcommand{\fffgfouroneloop}{
  \parbox{16mm}{
      \begin{fmfgraph}(16,11)
                \fmfset{dot_size}{0.8thick}
                        \fmfleft{i1,i2}
                        \fmfright{o1,o2}
                        \fmfdotn{v}{4}
                        \fmf{plain}{i1,v1,v2,i2}
                        \fmf{plain}{o1,v3,v4,o2}
                        \fmf{plain}{v1,v3}
                        \fmf{plain}{v2,v4}
      \end{fmfgraph}}
                        }
\newcommand{\fffgfouroneloopcrh}{
  \parbox{16mm}{
      \begin{fmfgraph}(16,11)
                \fmfset{dot_size}{0.8thick}
                        \fmfleft{i1,i2}
                        \fmfright{o1,o2}
                        \fmfdotn{v}{4}
                        \fmf{plain}{i1,v1}
                        \fmf{plain,tension=0.7}{v1,v3}
                        \fmf{plain}{v3,o1}
                        \fmf{plain}{i2,v2}
                        \fmf{plain,tension=0.7}{v2,v4}
                        \fmf{plain}{v4,o2}
                        \fmf{phantom}{v1,v2}
                        \fmf{phantom}{v3,v4}
                        \fmffreeze
                        \fmf{plain,rubout}{v1,v4}
                        \fmf{plain,rubout}{v2,v3}
      \end{fmfgraph}}
                        }
\newcommand{\fffgfouroneloopcrv}{
  \parbox{16mm}{
      \begin{fmfgraph}(16,11)
                \fmfset{dot_size}{0.8thick}
                        \fmfleft{i1,i2}
                        \fmfright{o1,o2}
                        \fmfdotn{v}{4}
                        \fmf{plain}{i1,v1}
                        \fmf{plain,tension=0.7}{v1,v2}
                        \fmf{plain}{v2,i2}
                        \fmf{plain}{o1,v3}
                        \fmf{plain,tension=0.7}{v3,v4}
                        \fmf{plain}{v4,o2}
                        \fmf{phantom}{v1,v3}
                        \fmf{phantom}{v2,v4}
                        \fmffreeze
                        \fmf{plain,rubout}{v1,v4}
                        \fmf{plain,rubout}{v2,v3}
      \end{fmfgraph}}
                        }
\newcommand{\FFFgfourtree}{
  \parbox{16mm}{
      \begin{fmfgraph}(16,11)
                \fmfset{dot_size}{0.8thick}
                \fmfset{dash_len}{2mm}
                        \fmfleft{i1,i2}
                        \fmfright{o1,o2}
                        \fmfdotn{v}{2}
                        \fmf{dashes}{i1,v1,i2}
                        \fmf{plain}{v1,v2}
                        \fmf{dashes}{o1,v2,o2}
      \end{fmfgraph}}
                        }
\newcommand{\FFFgfourtreecr}{
  \parbox{16mm}{
      \begin{fmfgraph}(16,11)
                \fmfset{dot_size}{0.8thick}
                \fmfset{dash_len}{2mm}
                        \fmfleft{i1,i2}
                        \fmfright{o1,o2}
                        \fmfdotn{v}{2}
                        \fmf{dashes}{i1,v1}
                        \fmf{phantom}{v1,i2}
                        \fmf{plain}{v1,v2}
                        \fmf{dashes}{o1,v2}
                        \fmf{phantom}{v2,o2}
                        \fmffreeze
                        \fmf{dashes,rubout}{v1,o2}
                        \fmf{dashes,rubout}{v2,i2}
      \end{fmfgraph}}
                        }
\newcommand{\FFFgfouroneloop}{
  \parbox{16mm}{
      \begin{fmfgraph}(16,11)
                \fmfset{dot_size}{0.8thick}
                \fmfset{dash_len}{2mm}
                        \fmfleft{i1,i2}
                        \fmfright{o1,o2}
                        \fmfdotn{v}{4}
                        \fmf{dashes}{i1,v1}
                        \fmf{plain}{v1,v2}
                        \fmf{dashes}{v2,i2}
                        \fmf{dashes}{o1,v3}
                        \fmf{plain}{v3,v4}
                        \fmf{dashes}{v4,o2}
                        \fmf{plain}{v1,v3}
                        \fmf{plain}{v2,v4}
      \end{fmfgraph}}
                        }
\newcommand{\FFFgfouroneloopcrh}{
  \parbox{16mm}{
      \begin{fmfgraph}(16,11)
                \fmfset{dot_size}{0.8thick}
                \fmfset{dash_len}{2mm}
                        \fmfleft{i1,i2}
                        \fmfright{o1,o2}
                        \fmfdotn{v}{4}
                        \fmf{dashes}{i1,v1}
                        \fmf{plain,tension=0.7}{v1,v3}
                        \fmf{dashes}{v3,o1}
                        \fmf{dashes}{i2,v2}
                        \fmf{plain,tension=0.7}{v2,v4}
                        \fmf{dashes}{v4,o2}
                        \fmf{phantom}{v1,v2}
                        \fmf{phantom}{v3,v4}
                        \fmffreeze
                        \fmf{plain,rubout}{v1,v4}
                        \fmf{plain,rubout}{v2,v3}
      \end{fmfgraph}}
                        }
\newcommand{\FFFgfouroneloopcrv}{
  \parbox{16mm}{
      \begin{fmfgraph}(16,11)
                \fmfset{dot_size}{0.8thick}
                \fmfset{dash_len}{2mm}
                        \fmfleft{i1,i2}
                        \fmfright{o1,o2}
                        \fmfdotn{v}{4}
                        \fmf{dashes}{i1,v1}
                        \fmf{plain,tension=0.7}{v1,v2}
                        \fmf{dashes}{v2,i2}
                        \fmf{dashes}{o1,v3}
                        \fmf{plain,tension=0.7}{v3,v4}
                        \fmf{dashes}{v4,o2}
                        \fmf{phantom}{v1,v3}
                        \fmf{phantom}{v2,v4}
                        \fmffreeze
                        \fmf{plain,rubout}{v1,v4}
                        \fmf{plain,rubout}{v2,v3}
      \end{fmfgraph}}
                        }
\newcommand{\Cfffgfourtree}{
  \parbox{16mm}{
      \begin{fmfgraph}(16,11)
                \fmfset{dot_size}{0.8thick}
                        \fmfleft{i1,i2}
                        \fmfright{o1,o2}
                        \fmfdotn{v}{1}
                        \fmf{plain}{i1,v1,i2}
                        \fmf{plain}{o1,v1,o2}
      \end{fmfgraph}}
                        }
\newcommand{\Cfffgfouroneloop}{
  \parbox{16mm}{
      \begin{fmfgraph}(16,11)
                \fmfset{dot_size}{0.8thick}
                        \fmfleft{i1,i2}
                        \fmfright{o1,o2}
                        \fmfdotn{v}{2}
                        \fmf{plain}{i1,v1}
                        \fmf{plain,left,tension=0.7}{v1,v2}
                        \fmf{plain}{v2,i2}
                        \fmf{plain}{o1,v1}
                        \fmf{plain,right,tension=0.7}{v1,v2}
                        \fmf{plain}{v2,o2}
      \end{fmfgraph}}
                        }
\newcommand{\CFFFgfourtree}{
  \parbox{16mm}{
      \begin{fmfgraph}(16,11)
                \fmfset{dot_size}{1.2thick}
                \fmfset{dash_len}{2mm}
                        \fmfleft{i1,i2}
                        \fmfright{o1,o2}
                        \fmfdotn{v}{1}
                        \fmf{dashes}{i1,v1,i2}
                        \fmf{dashes}{o1,v1,o2}
      \end{fmfgraph}}
                        }
\newcommand{\CFFFgfouroneloop}{
  \parbox{16mm}{
      \begin{fmfgraph}(16,11)
                \fmfset{dot_size}{1.2thick}
                \fmfset{dash_len}{2mm}
                        \fmfleft{i1,i2}
                        \fmfright{o1,o2}
                        \fmfdotn{v}{2}
                        \fmf{dashes}{i1,v1}
                        \fmf{plain,left,tension=0.7}{v1,v2}
                        \fmf{dashes}{v2,i2}
                        \fmf{dashes}{o1,v1}
                        \fmf{plain,right,tension=0.7}{v1,v2}
                        \fmf{dashes}{v2,o2}
      \end{fmfgraph}}
                        }
\newcommand{\genBABA}{
  \parbox{16mm}{
      \begin{fmfgraph}(16,11)
                        \feynmfourp
      \fmfv{decor.shape=circle,decor.filled=empty,decor.size=.20w}{v1}
                        \fmf{fermion}{i2,v1,o2}
                        \fmf{fermion}{o1,v1,i1}
      \end{fmfgraph}}
                        }
\newcommand{\genBAAB}{
  \parbox{16mm}{
      \begin{fmfgraph}(16,11)
                        \feynmfourp
      \fmfv{decor.shape=circle,decor.filled=empty,decor.size=.20w}{v1}
                        \fmf{fermion}{i2,v1,o2}
                        \fmf{fermion}{i1,v1,o1}
      \end{fmfgraph}}
                        }
\newcommand{\genBACC}{
  \parbox{16mm}{
      \begin{fmfgraph}(16,11)
                        \feynmfourp
      \fmfv{decor.shape=circle,decor.filled=empty,decor.size=.20w}{v1}
                        \fmf{fermion}{i2,v1,o2}
                        \fmf{photon}{i1,v1,o1}
      \end{fmfgraph}}
                        }
\newcommand{\genABBA}{
  \parbox{16mm}{
      \begin{fmfgraph}(16,11)
                        \feynmfourp
      \fmfv{decor.shape=circle,decor.filled=empty,decor.size=.20w}{v1}
                        \fmf{fermion}{o2,v1,i2}
                        \fmf{fermion}{o1,v1,i1}
      \end{fmfgraph}}
                        }
\newcommand{\genABAB}{
  \parbox{16mm}{
      \begin{fmfgraph}(16,11)
                        \feynmfourp
      \fmfv{decor.shape=circle,decor.filled=empty,decor.size=.20w}{v1}
                        \fmf{fermion}{o2,v1,i2}
                        \fmf{fermion}{i1,v1,o1}
      \end{fmfgraph}}
                        }
\newcommand{\genABCC}{
  \parbox{16mm}{
      \begin{fmfgraph}(16,11)
                        \feynmfourp
      \fmfv{decor.shape=circle,decor.filled=empty,decor.size=.20w}{v1}
                        \fmf{fermion}{o2,v1,i2}
                        \fmf{photon}{i1,v1,o1}
      \end{fmfgraph}}
                        }
\newcommand{\genCCBA}{
  \parbox{16mm}{
      \begin{fmfgraph}(16,11)
                        \feynmfourp
      \fmfv{decor.shape=circle,decor.filled=empty,decor.size=.20w}{v1}
                        \fmf{photon}{o2,v1,i2}
                        \fmf{fermion}{o1,v1,i1}
      \end{fmfgraph}}
                        }
\newcommand{\genCCAB}{
  \parbox{16mm}{
      \begin{fmfgraph}(16,11)
                        \feynmfourp
      \fmfv{decor.shape=circle,decor.filled=empty,decor.size=.20w}{v1}
                        \fmf{photon}{o2,v1,i2}
                        \fmf{fermion}{i1,v1,o1}
      \end{fmfgraph}}
                        }
\newcommand{\genCCCC}{
  \parbox{16mm}{
      \begin{fmfgraph}(16,11)
                        \feynmfourp
      \fmfv{decor.shape=circle,decor.filled=empty,decor.size=.20w}{v1}
                        \fmf{photon}{o2,v1,i2}
                        \fmf{photon}{i1,v1,o1}
      \end{fmfgraph}}
                        }
\newcommand{\BAKm}{
  \parbox{16mm}{
      \begin{fmfgraph}(16,11)
                \fmfset{dot_size}{1.2thick}
                \fmfset{arrow_len}{2.5mm}
                \fmfset{wiggly_len}{1.7mm}
                        \fmfleft{i1,i2}
                        \fmfright{o1,o2}
                        \fmfdotn{v}{2}
                        \fmf{phantom}{i1,v1,o1}
                        \fmf{fermion}{i2,v2,o2}
                        \fmf{fermion,right}{v1,v2}
                        \fmf{fermion,right}{v2,v1}
      \end{fmfgraph}}
                        }
\newcommand{\BAKM}{
  \parbox{16mm}{
      \begin{fmfgraph}(16,11)
                \fmfset{dot_size}{1.2thick}
                \fmfset{arrow_len}{2.5mm}
                \fmfset{wiggly_len}{1.7mm}
                        \fmfleft{i1,i2}
                        \fmfright{o1,o2}
                        \fmfdotn{v}{2}
                        \fmf{phantom}{i1,v1,o1}
                        \fmf{fermion}{i2,v2,o2}
                        \fmf{photon,right}{v1,v2}
                        \fmf{photon,right}{v2,v1}
      \end{fmfgraph}}
                        }
\newcommand{\ABKm}{
  \parbox{16mm}{
      \begin{fmfgraph}(16,11)
                \fmfset{dot_size}{1.2thick}
                \fmfset{arrow_len}{2.5mm}
                \fmfset{wiggly_len}{1.7mm}
                        \fmfleft{i1,i2}
                        \fmfright{o1,o2}
                        \fmfdotn{v}{2}
                        \fmf{phantom}{i1,v1,o1}
                        \fmf{fermion}{o2,v2,i2}
                        \fmf{fermion,left}{v1,v2}
                        \fmf{fermion,left}{v2,v1}
      \end{fmfgraph}}
                        }
\newcommand{\ABKM}{
  \parbox{16mm}{
      \begin{fmfgraph}(16,11)
                \fmfset{dot_size}{1.2thick}
                \fmfset{arrow_len}{2.5mm}
                \fmfset{wiggly_len}{1.7mm}
                        \fmfleft{i1,i2}
                        \fmfright{o1,o2}
                        \fmfdotn{v}{2}
                        \fmf{phantom}{i1,v1,o1}
                        \fmf{fermion}{o2,v2,i2}
                        \fmf{photon,right}{v1,v2}
                        \fmf{photon,right}{v2,v1}
      \end{fmfgraph}}
                        }
\newcommand{\CCKm}{
  \parbox{16mm}{
      \begin{fmfgraph}(16,11)
                \fmfset{dot_size}{1.2thick}
                \fmfset{arrow_len}{2.5mm}
                \fmfset{wiggly_len}{1.7mm}
                        \fmfleft{i1,i2}
                        \fmfright{o1,o2}
                        \fmfdotn{v}{2}
                        \fmf{phantom}{i1,v1,o1}
                        \fmf{photon}{o2,v2,i2}
                        \fmf{fermion,right}{v1,v2}
                        \fmf{fermion,right}{v2,v1}
      \end{fmfgraph}}
                        }
\newcommand{\CCKml}{
  \parbox{16mm}{
      \begin{fmfgraph}(16,11)
                \fmfset{dot_size}{1.2thick}
                \fmfset{arrow_len}{2.5mm}
                \fmfset{wiggly_len}{1.7mm}
                        \fmfleft{i1,i2}
                        \fmfright{o1,o2}
                        \fmfdotn{v}{2}
                        \fmf{phantom}{i1,v1,o1}
                        \fmf{photon}{o2,v2,i2}
                        \fmf{fermion,left}{v1,v2}
                        \fmf{fermion,left}{v2,v1}
      \end{fmfgraph}}
                        }
\newcommand{\streeACAC}{
  \parbox{16mm}{
      \begin{fmfgraph}(16,11)
        \feynmfourp\fmfdotn{v}{2}
        \fmf{fermion}{i1,v1,v2,i2}
        \fmf{photon}{o1,v1}
        \fmf{photon}{v2,o2}
      \end{fmfgraph}}
                        }
\newcommand{\streeACCA}{
  \parbox{16mm}{
      \begin{fmfgraph}(16,11)
        \feynmfourp\fmfdotn{v}{2}
        \fmf{fermion}{o1,v1,v2,i2}
        \fmf{photon}{i1,v1}
        \fmf{photon}{v2,o2}
      \end{fmfgraph}}
                        }
\newcommand{\streeCAAC}{
  \parbox{16mm}{
      \begin{fmfgraph}(16,11)
        \feynmfourp\fmfdotn{v}{2}
        \fmf{fermion}{i1,v1,v2,o2}
        \fmf{photon}{o1,v1}
        \fmf{photon}{v2,i2}
      \end{fmfgraph}}
                        }
\newcommand{\streeCACA}{
  \parbox{16mm}{
      \begin{fmfgraph}(16,11)
        \feynmfourp\fmfdotn{v}{2}
        \fmf{fermion}{o1,v1,v2,o2}
        \fmf{photon}{i1,v1}
        \fmf{photon}{v2,i2}
      \end{fmfgraph}}
                        }
\newcommand{\streeBABA}{
  \parbox{16mm}{
      \begin{fmfgraph}(16,11)
        \feynmfourp\fmfdotn{v}{2}
        \fmf{fermion}{o1,v1,i1}
        \fmf{photon}{v1,v2}
        \fmf{fermion}{i2,v2,o2}
      \end{fmfgraph}}
                        }
\newcommand{\streeBAAB}{
  \parbox{16mm}{
      \begin{fmfgraph}(16,11)
        \feynmfourp\fmfdotn{v}{2}
        \fmf{fermion}{i1,v1,o1}
        \fmf{photon}{v1,v2}
        \fmf{fermion}{i2,v2,o2}
      \end{fmfgraph}}
                        }
\newcommand{\streeABBA}{
  \parbox{16mm}{
      \begin{fmfgraph}(16,11)
        \feynmfourp\fmfdotn{v}{2}
        \fmf{fermion}{o1,v1,i1}
        \fmf{photon}{v1,v2}
        \fmf{fermion}{o2,v2,i2}
      \end{fmfgraph}}
                        }
\newcommand{\streeABAB}{
  \parbox{16mm}{
      \begin{fmfgraph}(16,11)
        \feynmfourp\fmfdotn{v}{2}
        \fmf{fermion}{i1,v1,o1}
        \fmf{photon}{v1,v2}
        \fmf{fermion}{o2,v2,i2}
      \end{fmfgraph}}
                        }
\newcommand{\LGamtil}{
  \parbox{20mm}{
      \begin{fmfgraph*}(20,12)
        \fmfleft{i1,i2} \fmfright{o1,o2}
        \fmf{plain}{i1,v1} \fmflabel{$k$}{i1}
        \fmf{plain}{v1,i2} \fmflabel{$i$}{i2}
        \fmf{plain}{o1,v1} \fmflabel{$l$}{o1}
        \fmf{plain}{v1,o2} \fmflabel{$j$}{o2}
        \fmfv{decor.shape=circle,decor.filled=empty,decor.size=8thick}{v1}
      \end{fmfgraph*}}
                         }
\newcommand{\Gamtil}{
  \parbox{20mm}{
      \begin{fmfgraph}(20,15)
        \fmfleft{i1,i2} \fmfright{o1,o2}
        \fmf{plain}{i1,v1,i2}
        \fmfblob{.20w}{v1}
        \fmf{plain}{o1,v1,o2}
      \end{fmfgraph}}
                         }
\newcommand{\Gamtils}{
  \parbox{20mm}{
      \begin{fmfgraph}(20,15)
        \fmfleft{i1,i2} \fmfright{o1,o2}
        \fmf{plain}{i1,v1,v2,v3,i2}
        \fmfblob{.20w}{v1,v3}
        \fmf{plain,tension=0.1}{v1,v2}
        \fmf{plain,tension=0.1}{v2,v3}
        \fmfv{decor.shape=circle,decor.filled=empty,decor.size=2thick}{v2}
        \fmf{plain}{o1,v1,v2,v3,o2}
      \end{fmfgraph}}
                         }
\newcommand{\Gamtilt}{
  \parbox{20mm}{
      \begin{fmfgraph}(20,15)
        \fmfleft{i1,i2} \fmfright{o1,o2}
        \fmf{plain}{i1,v1,i2}
        \fmfblob{.20w}{v1,v3}
        \fmf{plain}{v1,v2,v3}
        \fmfv{decor.shape=circle,decor.filled=empty,decor.size=2thick}{v2}
        \fmf{plain}{o1,v3,o2}
      \end{fmfgraph}}
                         }
\newcommand{\Gamtilu}{
  \parbox{20mm}{
      \begin{fmfgraph}(20,15)
        \fmfleft{i1,i2} \fmfright{o1,o2}
        \fmf{plain}{i1,v1,o2}         \fmf{phantom}{i1,v1,i2}
        \fmf{plain,rubout}{o1,v3,i2}  \fmf{phantom}{o1,v3,o2}
        \fmf{plain,tension=0.2}{v1,v2}\fmf{plain,tension=0.2}{v2,v3}
        \fmfblob{.20w}{v1,v3}
        \fmfv{decor.shape=circle,decor.filled=empty,decor.size=2thick}{v2}
      \end{fmfgraph}}
                         }
\newcommand{\LGamtilt}{
  \parbox{20mm}{
      \begin{fmfgraph*}(20,15)
        \fmfleft{i1,i2} \fmfright{o1,o2}
        \fmf{plain}{i1,v1,i2}

\fmfv{decor.shape=circle,decor.filled=empty,decor.size=.30w,label=$t$,
label.dist=-.05w}{v1}
        \fmf{plain}{o1,v1,o2}
      \end{fmfgraph*}}
                         }
\newcommand{\LGamtilu}{
  \parbox{20mm}{
      \begin{fmfgraph*}(20,15)
        \fmfleft{i1,i2} \fmfright{o1,o2}
        \fmf{plain}{i1,v1,i2}

\fmfv{decor.shape=circle,decor.filled=empty,decor.size=.30w,label=$u$,
label.dist=-.07w}{v1}
        \fmf{plain}{o1,v1,o2}
      \end{fmfgraph*}}
                         }
\newcommand{\genCAAC}{
  \parbox{16mm}{
      \begin{fmfgraph}(16,11)
                        \feynmfourp
      \fmfv{decor.shape=circle,decor.filled=empty,decor.size=.20w}{v1}
                        \fmf{fermion}{i1,v1,o2}
                        \fmf{photon}{o1,v1,i2}
      \end{fmfgraph}}
                        }
\newcommand{\genBCCB}{
  \parbox{16mm}{
      \begin{fmfgraph}(16,11)
                        \feynmfourp
      \fmfv{decor.shape=circle,decor.filled=empty,decor.size=.20w}{v1}
                        \fmf{fermion}{i2,v1,o1}
                        \fmf{photon}{i1,v1,o2}
      \end{fmfgraph}}
                        }
\newcommand{\genACCA}{
  \parbox{16mm}{
      \begin{fmfgraph}(16,11)
                        \feynmfourp
      \fmfv{decor.shape=circle,decor.filled=empty,decor.size=.20w}{v1}
                        \fmf{fermion}{o1,v1,i2}
                        \fmf{photon}{i1,v1,o2}
      \end{fmfgraph}}
                        }
\newcommand{\genCBBC}{
  \parbox{16mm}{
      \begin{fmfgraph}(16,11)
                        \feynmfourp
      \fmfv{decor.shape=circle,decor.filled=empty,decor.size=.20w}{v1}
                        \fmf{fermion}{o2,v1,i1}
                        \fmf{photon}{o1,v1,i2}
      \end{fmfgraph}}
                        }
\newcommand{\cboxSBAAB}{
  \parbox{16mm}{
      \begin{fmfgraph}(16,11)
                        \feynmfourp\fmfdotn{v}{2}
                        \fmf{fermion}{i1,v1}
			\fmf{fermion}{i2,v1}
			\fmf{fermion}{v2,o1}
			\fmf{fermion}{v2,o2}
                        \fmf{fermion,left}{v1,v2}
                        \fmf{fermion,right}{v1,v2}
      \end{fmfgraph}}
                        }
\newcommand{\streecrBAAB}{
  \parbox{16mm}{
      \begin{fmfgraph}(16,11)
        \feynmfourp\fmfdotn{v}{2}
        \fmf{fermion}{i1,v1}\fmf{phantom}{v1,o1}
        \fmf{fermion}{i2,v2}\fmf{phantom}{v2,o2}
	\fmf{photon}{v1,v2}
	\fmffreeze
        \fmf{fermion,rubout}{v1,o2}
        \fmf{fermion,rubout}{v2,o1}
      \end{fmfgraph}}
                        }
\newcommand{\sboxcrBAAB}{
  \parbox{16mm}{
      \begin{fmfgraph}(16,11)
                        \feynmfourp\fmfdotn{v}{4}
                        \feynmfourp\fmfdotn{v}{4}
                        \fmf{fermion}{i1,v1}\fmf{fermion}{i2,v2}
                        \fmf{fermion}{v4,o2}\fmf{fermion}{v3,o1}
			\fmf{photon,tension=0.7}{v1,v2}
			\fmf{photon,tension=0.7}{v3,v4}
                        \fmf{phantom,tension=0.8}{v1,v3}
                        \fmf{phantom,tension=0.8}{v2,v4}
			\fmffreeze
                        \fmf{fermion,rubout}{v1,v4}
                        \fmf{plain,rubout}{v2,v}\fmf{fermion}{v,v3}
      \end{fmfgraph}}
                        }
\newcommand{\cboxSABBA}{
  \parbox{16mm}{
      \begin{fmfgraph}(16,11)
                        \feynmfourp\fmfdotn{v}{2}
                        \fmf{fermion}{v1,i1}
			\fmf{fermion}{v1,i2}
			\fmf{fermion}{o1,v2}
			\fmf{fermion}{o2,v2}
                        \fmf{fermion,left}{v2,v1}
                        \fmf{fermion,right}{v2,v1}
      \end{fmfgraph}}
                        }
\newcommand{\streecrABBA}{
  \parbox{16mm}{
      \begin{fmfgraph}(16,11)
        \feynmfourp\fmfdotn{v}{2}
        \fmf{fermion}{v1,i1}\fmf{phantom}{o1,v1}
        \fmf{fermion}{v2,i2}\fmf{phantom}{o2,v2}
	\fmf{photon}{v1,v2}
	\fmffreeze
        \fmf{fermion,rubout}{o2,v1}
        \fmf{fermion,rubout}{o1,v2}
      \end{fmfgraph}}
                        }
\newcommand{\sboxcrABBA}{
  \parbox{16mm}{
      \begin{fmfgraph}(16,11)
                        \feynmfourp\fmfdotn{v}{4}
                        \feynmfourp\fmfdotn{v}{4}
                        \fmf{fermion}{v1,i1}\fmf{fermion}{v2,i2}
                        \fmf{fermion}{o2,v4}\fmf{fermion}{o1,v3}
			\fmf{photon,tension=0.7}{v1,v2}
			\fmf{photon,tension=0.7}{v3,v4}
                        \fmf{phantom,tension=0.8}{v1,v3}
                        \fmf{phantom,tension=0.8}{v2,v4}
			\fmffreeze
                        \fmf{fermion,rubout}{v4,v1}
                        \fmf{plain}{v3,v}\fmf{fermion,rubout}{v,v2}
      \end{fmfgraph}}
                        }


\begin{fmffile}{bbs}

\newcommand{\be}{\begin{eqnarray}}
\newcommand{\ee}{\end{eqnarray}}
\newcommand{\ba}{\begin{array}}
\newcommand{\ea}{\end{array}}
\newcommand{\nn}{\nonumber}
\newcommand{\ra}{\rightarrow}

\def\s{\scriptstyle}
\def\ss{\scriptscriptstyle}
\def\cl{\centerline}
\def\ds{\displaystyle} \def\d{\dagger} \def\m{\mu} \def\G{\Gamma}
\def\gt{\tilde\Gamma} \def\p{\partial} \def\parm{\partial_{\m}}
\def\f{\phi} \def\p{\psi} \def\k{\kappa} \def\fd{\f^{\d}}
\def\o{\over} \def\i{\infty} \def\a{\alpha} \def\L{\Lambda}
\def\zf{Z_{\s\phi}} \def\tg{\tilde\Gamma}
\def\zp{Z_{\s\psi}}
\def\zfft{Z_{b,t}^{\ss \f\f\f\f}}
\def\zfpt{Z_{b,t}^{\ss \f\f\p\p}}
\def\zpft{Z_{b,t}^{\ss \p\p\f\f}}
\def\zppt{Z_{b,t}^{\ss \p\p\p\p}}
\def\zffu{Z_{b,u}^{\ss \f\f\f\f}}
\def\zfpu{Z_{b,u}^{\ss \f\f\p\p}}
\def\zpfu{Z_{b,u}^{\ss \p\p\f\f}}
\def\zppu{Z_{b,u}^{\ss \p\p\p\p}}
\def\bffb{B_B^{\ss \f\f\f\f}}
\def\bfpb{B_B^{\ss \f\f\p\p}}
\def\bpfb{B_B^{\ss \p\p\f\f}}
\def\bppb{B_B^{\ss \p\p\p\p}}
\def\bff{B^{\ss \f\f\f\f}}
\def\bfp{B^{\ss \f\f\p\p}}
\def\bpf{B^{\ss \p\p\f\f}}
\def\bpp{B^{\ss \p\p\p\p}}
\def\affb{A_B^{\ss \f\f\f\f}}
\def\afpb{A_B^{\ss \f\f\p\p}}
\def\apfb{A_B^{\ss \p\p\f\f}}
\def\appb{A_B^{\ss \p\p\p\p}}
\def\aff{A^{\ss \f\f\f\f}}
\def\afp{A^{\ss \f\f\p\p}}
\def\apf{A^{\ss \p\p\f\f}}
\def\app{A^{\ss \p\p\p\p}}

\setcounter{page}{0}
\thispagestyle{empty}
\begin{flushright}
ICN-UNAM-98-01\\
February 4, 1998
\end{flushright}
\vskip 0.5 truein
\begin{center}
{\huge Quantum Field Theory in the Limit 
$x\ll 1\,${\LARGE\footnote{This work was
supported by Conacyt grant 3298P--E9608.}}}
\vskip 0.4truein
{\large
\bf  C.R.\ Stephens\footnote{e--mail: stephens@nuclecu.unam.mx},
A. Weber\footnote{Supported during part of the work
by fellowships of the DAAD and the Mexican
Government; e--mail: axel@nuclecu.unam.mx},
J.C. L\'opez Vieyra\footnote{e--mail: vieyra@nuclecu.unam.mx}
and P.O. Hess\footnote{e--mail: hess@nuclecu.unam.mx}}\\
\vskip 0.25truein
Instituto de Ciencias Nucleares, UNAM, \\
Circuito Exterior C.U., A. Postal 70-543, \\
04510 M\'exico D.F., Mexico, \\
Tel.\ (52)-5-622 46 90, Fax (52)-5-622 46 93
\end{center}
\vskip 1truein
{\bf Abstract:}\ The asymptotic high momentum behaviour of 
quantum field theories with cubic interactions is 
investigated using renormalization group
techniques in the asymmetric limit $x\ll1$.
Particular emphasis is paid to theories with interactions
involving more than one field where it is found that a matrix renormalization
is necessary. Asymptotic scaling forms, in agreement with Regge theory, are
derived for the elastic two-particle scattering amplitude and verified in
one-loop renormalization group improved perturbation theory,
corresponding to the summation of leading logs to all orders. We give
explicit forms for the Regge trajectories of different scalar theories
in this approximation and
determine the signatures.
\vskip 0.35truein \noindent
PACS numbers: 11.10.Gh, 11.10.Jj, 11.10.St, 12.40.Nn \\
Keywords: renormalization group, Regge limit, high-energy behaviour
\vfill\eject

\section{Introduction}

Asymptotic behaviour at large energy scales, or alternatively small
distance scales, has been, and will continue to be, one of the prime
foci for research in quantum field theory, and therefore by implication
in particle physics. One obvious example of where short distance
physics plays a crucial role is in the perturbative treatment of
continuum quantum field theory wherein the existence of quantum
fluctuations leads to the apparent nonsense of ultraviolet divergences.
The renormalization programme, in general, has dealt with this
problem very successfully in a physically important class of field
theories.

One of the most important members of this class of theories is QCD
which possesses the important property of being asymptotically free in
the ultraviolet. The renormalization group (RG) has proved to be
an extremely powerful tool in accessing this limit via its non-perturbative
resummation of classes of Feynman diagrams. In particular
RG methods, often combined with other non-perturbative techniques
such as the operator product expansion, have been very
successful in describing the
logarithmic corrections seen experimentally in deep inelastic scattering
(see ref.\ \cite{deepinel} for a review). Asymptotically free theories
have a very important
advantage relative to others  --- that a description of their true high energy
behaviour does not depend on lower energy ``infrared scales''. This
means that a very simple renormalization procedure such as minimal
subtraction is sufficient to access all the relevant physics associated
with the fixed point and corrections to it.

In terms of asymptotic behaviour as a function of
momentum use of the RG, as based on minimal subtraction for example,
has essentially been restricted to symmetric
limits where all momenta are scaled equally by some factor.
Such a symmetric limit is associated with isotropic scale changes and
has been sufficient to describe the deep-inelastic limit.
Very often however systems are anisotropic and/or inhomogeneous or
are probed in an asymmetric way. This is precisely the case when
describing semi-hard processes in QCD in
the limit $Q^2\gg\Lambda^2_{\s QCD}$, $x\ll 1$ where
$Q^2$ and $x$ are the Bjorken scaling variables. This corresponds to the
diffractive, or Regge limit, which is of great experimental
relevance (see for instance ref.\ \cite{desy} for recent results from HERA),
and where standard RG
improved perturbation theory breaks down. Various methods have been
proposed and used to tackle this problem with varying degrees of success.
One of the most common is the summation of leading logs \cite{lipatov} via
an inspection of the perturbation expansion, a technique which has a long
history (see for instance ref.\ \cite{eden} and references therein).
Besides an intrinsic degree of arbitrariness, summing sets of logs can
be quite difficult combinatorially. Other methods include: multiperipheral
models \cite{multi}, absorption models \cite{absorp} and eikonal models
\cite{eikonal}.

There is also a strong secondary motivation for studying the Regge limit that
is associated with the remarkable property of Regge theory that relates high
energy behaviour in one channel with infrared behaviour in the crossed channel,
and in particular the physics of bound states, the solutions of the equation
$\alpha(s=m^2)=l$ yielding the masses of the $s$-channel bound states, $l$
being the angular momentum of the state.

In this paper we wish to present an RG methodology that is quite general
and can describe both symmetric (hard) and asymmetric (semi-hard)
asymptotic limits
of the high momentum behaviour in quantum field theory (some preliminary
results appeared in ref.\ \cite{plbus}). In particular we will treat quite
generally ``cubic'' field theories, wherein the asymmetric Regge limit is a
particularly interesting one.
We will illustrate the principal concepts in the context of scalar theories
to avoid complications due to spin and gauge invariance. We emphasize however
that the methodology is equally applicable to these cases too.
RG methods have been used in the context of Reggeonic field theory
\cite{migdal}, however,
the latter is an effective field theory wherein one makes an ansatz for the
theory that describes the Reggeonic effective degrees of freedom. For instance,
in
this approach the ``bare'' Regge slope and intercept are not calculated but are
input data.

One of the reasons the RG does not seem to have been utilized to investigate
asymmetric asymptotic behaviour is its association with the problem of
ultraviolet divergences. In the Regge limit the breakdown of perturbation
theory
has nothing to do with the latter. However, terms such as $\ln s$ or
$\ln t$, in the limits $s$, $t\ra\i$, do lead to divergences. These
divergences in contradistinction to symmetric short distance behaviour are very
asymmetric. The reason they appear can be traced to the nature of the
effective degrees of freedom in the problem. For small $s$ and $t$
they are four-dimensional, whereas in the Regge limit, as is well
known, there is a ``kinematic'' dimensional reduction to two
dimensions owing to the extreme anisotropy between the longitudinal
and transverse sectors. This type of dimensional reduction has much in
common with ``geometric'' dimensional reductions, such as occur
in finite temperature field theory in the vicinity of a second order
phase transition \cite{mplame}. Such crossovers between effective degrees of
freedom of one type and another, qualitatively completely different,
are ubiquitous in physics. Indeed the crossover between asymptotic
freedom and confinement offers a perfect paradigm.
Given the close relationship, via Regge theory, between high energy behaviour
in one channel and low energy behaviour in the crossed channel we thus
expect the results herein to be of use in understanding the crossover between
bound states and unbound states. Yet a third motivation, additional to
studying high momentum behaviour and the physics of bound states is that
of studying the crossover between string-like behaviour and particle-like
behaviour. Study of the Regge limit and the subsequent introduction of the
Veneziano
model led directly to a string formulation. Now the emphasis is more on
deriving point-particle field theories as effective field theories in the low
energy
limit of string theory rather than vice-versa. It remains a challenge however
to understand fully to what extent a point particle field theory may manifest
stringy behaviour in some effective field theory limit.

To describe systematically such crossovers using RG methods one
requires a RG that can interpolate between different effective degrees
of freedom as a function of ``scale'', where scale could mean
temperature, momentum, size etc.  Such an RG, that can be applied to a
myriad of other crossover situations, has been developed under the
name of ``environmentally friendly'' renormalization \cite{envfri} in
recognition of the fact that a crossover very often can be thought of
as taking place due to the effect of some ``environmental'' parameter,
such as temperature. Using these methods it has been possible, for
instance, to access the dimensional crossover in finite temperature
field theory \cite{fintemp} between an effective four dimensional
theory at low temperatures to an effectively three dimensional theory
near a second or weakly first order phase transition. In this paper
we will apply the philosophy and techniques of environmentally friendly
renormalization to an investigation of the asymmetric, high momentum
limit of ``cubic'' quantum field theories.

The format of the paper will be as follows: in section 2 we will give a
brief discussion of relativistic Regge theory and derive the non-perturbative
asymptotic scaling form for the two-particle on-shell scattering amplitude
in the large-$t$ limit. In section 3 we will introduce the RG techniques
we will use to access the Regge limit, while section 4 deals with the case 
of interactions involving different fields leading to mixing between the
effective vertices in the asymptotic regime. In secion 5 we will treat fully 
to one loop the example of a scalar, non-gauge version of QED with interaction
$\fd\phi\psi$. Finally in section 6 we will discuss the results found and
draw some conclusions. There are also three appendices: one showing
that our techniques also work to two loops, another proving that
indeed to one loop our renormalization sums up all leading logs whilst the
third gathers together some technical results needed in the main text.

\section{Relativistic Regge Theory}

In this section we will briefly review the fundamentals of Regge theory for 
relativistic particle scattering. A more detailed account can be found, for 
instance, in ref.\ \cite{russian}. One advantage of Regge theory is that it
yields totally non-perturbative information, albeit at the expense of
several assumptions, about the asymptotic forms of scattering amplitudes.
Additionally it will yield directly the connection between bound state
information in one channel with high energy behaviour in the crossed channel.

Let $A(s,t)$ be the amplitude for the scattering of two spinless bosons, 
where the two
outgoing bosons may be different from the incoming ones. $A(s,t)$ is
given by the amputated connected four-point function $\tilde{\Gamma}^{\s ijkl}$
with the external momenta on mass shell, and $s$ an $t$ are the Mandelstam
variables. For the sake of simplicity, in this section we will write down the 
formulas for the case where the masses $m$ of all external particles are equal.

The central assumption of Regge theory (and of S-matrix theory in general)
is that $A(s,t)$ is an analytic function of $s$ and
$t$ and satisfies a dispersion relation of the following form:
\be
A(s,t) = \frac{1}{\pi} \int_{t_0}^\infty \mbox{d}t'\, 
\frac{A_t(s,t')}{t' - t} \: + \: \frac{1}{\pi} 
\int_{u_0}^{\infty} 
\mbox{d}u'\, \frac{A_u(s,u')}{u' - u} \:, \label{spectral}
\ee
where $A_t$ and $A_u$ are the absorptive parts of $A$, i.e.\ the 
discontinuties across the cuts in the $t$ and $u$ channels, and are
assumed to be analytic themselves with the respective cuts. The third
Mandelstam variable $u$ is given in terms of the other two by the
on-shell relation
\be
s + t + u = 4m^2 \:. \label{mandelstam}
\ee
The positions of the branch points in (\ref{spectral}) are 
$t_0 = u_0 = 4m^2$, provided that there are no lighter
particles in the theory, hence the cuts lie outside the physical region
$t,u < 0$ for $s$-channel scattering.

The exchange of particles leads to additional pole terms in (\ref{spectral}),
which are usually omitted in the dispersion relation. They play no role
in the following discussion. Similarly, subtractions may be necessary in
(\ref{spectral}) due to the behaviour of $A(s,t)$ for asymptotic values of
$t$, but they will not affect the following reasoning.

For $s > 4 m^2$, we decompose $A$ into partial waves,
\be
A(s,t) = A(s,z) = \sum_{l=0}^\infty (2l + 1) a_l(s) P_l(z) \:,
\ee
where $z$ is the cosine of the scattering angle in the center-of-mass system
for $s$-channel scattering, 
\be
z = 1 + \frac{2t}{s - 4m^2} = -1 - \frac{2u}{s - 4m^2} \:. \label{angle}
\ee

We can rewrite the dispersion relation (\ref{spectral}) in terms of the
scattering angle as
\be
A(s,z) = \frac{1}{\pi} \int_{1 + 2t_0/(s-4m^2)}^\infty \mbox{d}z'\, 
\frac{A_t(s,z')}{z' - z} \: + \: \frac{1}{\pi} 
\int_{-1 - 2u_0/(s-4m^2)}^{-\infty} 
\mbox{d}z'\, \frac{A_u(s,-z')}{z' - z} \:, \label{dispersion}
\ee
from where we extract the partial wave amplitudes
\be
a_l(s) = \frac{1}{\pi} \int_{1 + 2t_0/(s-4m^2)}^\infty \mbox{d}z'\, 
A_t(s,z') Q_l(z') \: - \: \frac{1}{\pi} \int_{1 + 2u_0/(s-4m^2)}^{\infty} 
\mbox{d}z'\, A_u(s,z') Q_l(-z')
\ee
with the Legendre functions of the second kind $Q_l$. 

The partial wave amplitudes $a_l$ do not allow directly for a 
Sommerfeld-Watson transform due to the behaviour of $Q_l(z)$ for large $|l|$, 
and we have to introduce two different
amplitudes $a^{\pm}(l,s)$ with ``signatures'' $+$ and $-$, respectively. 
They are given by the Froissart-Gribov formula \cite{froissart}
\be
a^{\pm}(l,s) = \frac{1}{\pi} \int_{1 + 2t_0/(s-4m^2)}^\infty \mbox{d}z'\, 
A_t(s,z') Q_l(z') \: \pm \: \frac{1}{\pi} \int_{1 + 2u_0/(s-4m^2)}^{\infty} 
\mbox{d}z'\, A_u(s,z') Q_l(z') \:. \label{amplitudes}
\ee
Both amplitudes can be continued uniquely to complex values of $l$ by
virtue of Carlson's theorem (see, e.g., ref.\ \cite{carlson}). 
As we will see later, $a^+$ is associated with 
physical states of even angular momentum and $a^-$ with odd angular momentum 
ones. 

The corresponding scattering amplitudes
\be
A^{\pm}(s,z) = \sum_{l=0}^\infty (2l + 1) a^{\pm}(l,s) P_l(z)
\ee
then allow separately for a Sommerfeld-Watson transform \cite{watson}. 
Assuming that
the partial wave amplitudes $a^{\pm}$, considered as analytic funtions in
the complex $l$-plane, possess simple poles at 
\be
l = \alpha^{\pm}_i(s) \:, \quad 
\mbox{Re}(\alpha^{\pm}_i(s)) > -\frac{1}{2}\:,
\ee
we get the result
\begin{eqnarray}
A^{\pm}(s,z) &=& \frac{i}{2} \int_{-1/2 - i \infty}^{-1/2 + i \infty}
\mbox{d}l \, \frac{(2 l + 1) a^{\pm}(s,l) P_l(-z)}{\sin(\pi l)} 
\nonumber \\
& & {}-\: \pi \sum_i \frac{(2 \alpha^{\pm}_i(s) + 1) \beta^{\pm}_i(s) 
P_{\alpha^{\pm}_i(s)}(-z)}{\sin(\pi \alpha^{\pm}_i(s))} \:,
\label{sowat}
\end{eqnarray}
where $\beta^{\pm}_i(s)$ is the residue of $a^{\pm}$ at 
$l = \alpha^{\pm}_i(s)$. 

In the case that subtractions are necessary
in (\ref{spectral}), we have to assume in addition the absence of
``elementary poles'' for (\ref{sowat}) to be valid. Furthermore, since
in fact the functions $a^{\pm}$ possess besides simple poles also branch 
cuts in the $l$-plane giving rise to additional 
contributions in (\ref{sowat}), we will assume that in the asymptotic 
limit to be discussed in the following, the dominant contributions 
originate from the poles only. The poles $\alpha^{\pm}_i(s)$ of $a^{\pm}$
are known as Regge poles \cite{regge}, while the functional dependence of a 
Regge pole on $s$ is referred to as its Regge trajectory.

Let us now consider the limit $z \to \infty$, which for $s > 4 m^2$ 
corresponds to $t \to \infty$.
Using the asymptotic behaviour of the Legendre functions (for this and other
properties of the Legendre functions see, for instance, ref.\ 
\cite{bateman}),
\be
P_{\alpha}(z) \approx \frac{\Gamma(2\alpha + 1)}{2^{\alpha} (\Gamma(\alpha
+ 1))^2} \, z^{\alpha} \:, \quad |z| \to \infty \:,
\ee
we have from (\ref{sowat}) that asymptotically
\be
A^{\pm}(s,z) = \sum_i \frac{{\beta_i'}^{\pm}(s)}
{\sin(\pi \alpha^{\pm}_i(s))} \, z^{\alpha^{\pm}_i(s)} \:, \label{asympt}
\ee
where
\be
{\beta_i'}^{\pm}(s) = - \frac{\pi \Gamma(2\alpha^{\pm}_i(s) + 2) 
\beta^{\pm}_i(s) e^{-i \pi \alpha^{\pm}_i(s)}}{2^{\alpha^{\pm}_i(s)} 
(\Gamma(\alpha^{\pm}_i(s) + 1))^2} \:. \label{residues}
\ee

We can then determine the asymptotic behaviour of the full scattering
amplitude,
\be
A(s,z) = \frac{1}{2} \left( A^+(s,z) + A^+(s,-z) + A^-(s,z) - A^-(s,-z)
\right) \:.
\ee
Plugging in (\ref{asympt}), we obtain
\be
A(s,z) = \sum_i \frac{{\beta_i'}^+(s) (1 + e^{i \pi \alpha^+_i(s)})}
{2 \sin(\pi \alpha^+_i(s))} \, z^{\alpha^+_i(s)} + 
\sum_i \frac{{\beta_i'}^-(s) (1 - e^{i \pi \alpha^-_i(s)})}
{2 \sin(\pi \alpha^-_i(s))} \, z^{\alpha^-_i(s)}
\ee
for $z \to \infty$. Replacing $z$ via (\ref{angle}), we have asymptotically
\be
A(s,t) = \sum_i \frac{\tilde{\beta}^+_i(s) (1 + e^{i \pi \alpha^+_i(s)})}
{2 \sin(\pi \alpha^+_i(s))} \, t^{\alpha^+_i(s)} + 
\sum_i \frac{\tilde{\beta}^-_i(s) (1 - e^{i \pi \alpha^-_i(s)})}
{2 \sin(\pi \alpha^-_i(s))} \, t^{\alpha^-_i(s)} \label{regge}
\ee
as $t \to \infty$, where
\be
\tilde{\beta}^{\pm}_i(s) = \left( \frac{2}{s - 4 m^2} 
\right)^{\alpha^{\pm}_i(s)} {\beta_i'}^{\pm}(s) \:.
\ee

Up to now, we have shown the validity of (\ref{regge}) for $s > 4 m^2$ only. 
Due to the
analyticity of $A_t$ and $A_u$, however, we can analytically continue
in $s$ to the region $s < 4 m^2$, starting with (\ref{amplitudes}) in the
form
\begin{eqnarray}
a^{\pm}(l,s) &=& \frac{1}{\pi} \int_{t_0}^\infty \frac{2\,\mbox{d}t'}
{s - 4m^2} \, A_t(s,t') \, Q_l \left( 1 + \frac{2 t'}{s - 4m^2} \right)
\nonumber \\ 
&& {} \pm \: \frac{1}{\pi} \int_{u_0}^{\infty} \frac{2\,\mbox{d}u'}
{s - 4m^2} \, A_u(s,u') \, Q_l \left( 1 + \frac{2 u'}{s - 4m^2} \right) \:.
\end{eqnarray}
Then (\ref{regge}) remains unchanged in the limit $t \to \infty$, with
\be
\tilde{\beta}^{\pm}_i(s) = - \frac{\pi \Gamma(2\alpha^{\pm}_i(s) + 2) 
\beta^{\pm}_i(s) e^{-2 i \pi \alpha^{\pm}_i(s)}}
{(\Gamma(\alpha^{\pm}_i(s) + 1))^2 (4 m^2 - s)^{\alpha^{\pm}_i(s)}} 
\label{tresidues} \:
\ee
for $s < 4 m^2$. Note that the phase factors in 
(\ref{residues}--\ref{tresidues}) follow from taking $s$ on the
upper edge of the right-hand cut in the $s$-plane, i.e.\
$s \to s + i\epsilon$ for $s > 4m^2$, and likewise for $t$.

For comparison with our results to be described in sections 3, 4 and 5
we determine the $t$- and $u$-contributions in (\ref{regge}). The latter are
identified with the contributions to $A(s,t)$ from the corresponding cuts in 
(\ref{spectral}), and can be associated with certain classes of
diagrams in the limit $t \to \infty$. Following the derivation of
(\ref{regge}), we can easily identify the terms
\be
\sum_i \frac{\tilde{\beta}^+_i(s)}
{2 \sin(\pi \alpha^+_i(s))} \, t^{\alpha^+_i(s)} + 
\sum_i \frac{\tilde{\beta}^-_i(s)}
{2 \sin(\pi \alpha^-_i(s))} \, t^{\alpha^-_i(s)} \label{tcontribut}
\ee
with the $t$-contributions, while
\be
\sum_i \frac{\tilde{\beta}^+_i(s) e^{i \pi \alpha^+_i(s)}}
{2 \sin(\pi \alpha^+_i(s))} \, t^{\alpha^+_i(s)} - 
\sum_i \frac{\tilde{\beta}^-_i(s) e^{i \pi \alpha^-_i(s)}}
{2 \sin(\pi \alpha^-_i(s))} \, t^{\alpha^-_i(s)} \label{ucontribut}
\ee
arise from the $u$-contributions.

Formula (\ref{regge}) is the main result of Regge theory. It states that
the asymptotic behaviour is determined by the Regge trajectories, and that
the dominant contribution comes from the right-most Regge pole
$\alpha^{\pm}_0(s)$ in the complex $l$-plane. The Regge limit $t \to \infty$
corresponds for $s < 0$ to the physical high-energy limit for scattering in 
the $t$-channel. For $s > 0$, the scattering amplitude is not
experimentally accessible, but has to be obtained by analytic continuation
from the physical region.

From (\ref{regge}) we see that $A$ has a pole
in $s$ whenever $\alpha^+_i(s)$ ($\alpha^-_i(s)$) equals an even (odd)
non-negative real integer. The signature factor
\be
1 \pm e^{i \pi \alpha^{\pm}_i(s)}
\ee
leads to the cancellation of the poles for odd $\alpha^+_i(s)$
and even $\alpha^-_i(s)$. We can read off from (\ref{sowat}) that the 
existence and position of these poles is independent of the value of $z$
(or $t$). In particular, they occur inside the region $-1 < z < 1$
and therefore correspond to bound states (for $s < 4 m^2$) or resonances 
(Re$(s) > 4 m^2$) in the $s$-channel with the respective angular momenta. 
In this way, Regge theory provides a link between the high-energy behaviour 
in one channel and bound state formation in the crossed channel.

We will conclude this section with a qualitative description 
of the Regge trajectories in the complex $l$-plane. The leading trajectory
$\alpha^{\pm}_0(s)$ starts out at $l = -1$ for $s \to -\infty$, and
with increasing $s$
moves along the real axis towards larger values of $l$. If we assume the
absence of vacuum condensation, it should cross $l = 0$ (the first
physical state) with a positive value of $s$. It then produces a series of
excited bound states with increasing angular momenta, until at $s = 4m^2$ 
we find a branch point in $s$, and the trajectory moves into the upper half 
plane. It may stay close to the real axis for a while, producing resonances 
with angular momenta corresponding to the real parts of $l$, whenever the
real part passes through an integer value. Finally, the Regge trajectory
will return through the upper half plane to $l = -1$ for $s \to \infty$.
Fig.\ 1 shows a ``typical'' trajectory in the complex $l$-plane,
corresponding to an interaction through exchange of massive particles.
\begin{figure}
\begin{center}
\unitlength1cm
\begin{picture}(14,9)
\put(12.3,0.6){$\mbox{Re} \, l$}
\put(4.6,8.5){$\mbox{Im} \, l$}
\put(11.5,7.6){$\alpha_0(s)$}
\put(0,-1){\psfig{figure=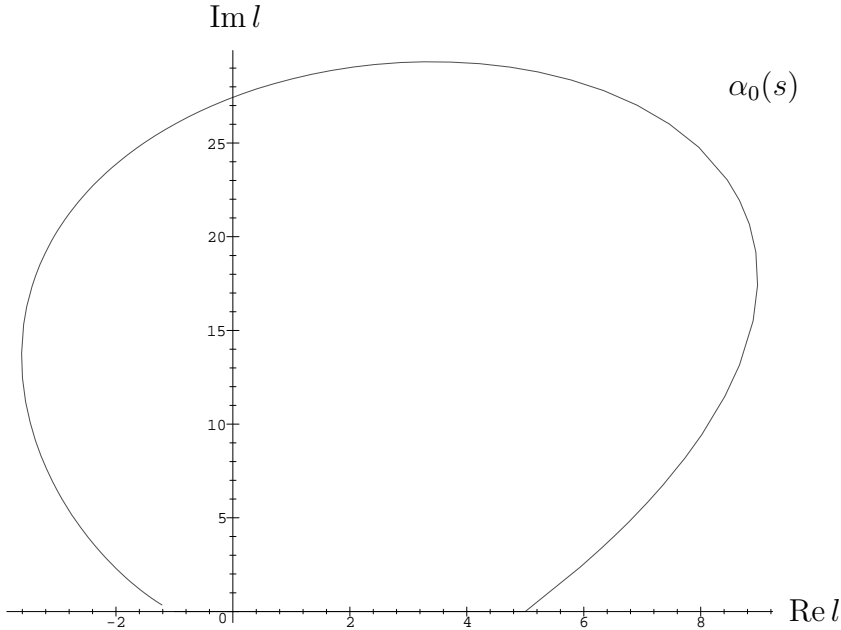,width=14cm,angle=-90}}
\end{picture}
\parbox{13cm}{\caption{A representative leading Regge trajectory $\alpha_0$ 
in the complex $l$-plane, starting and ending at $l = -1$.}}
\end{center}
\end{figure}

How far the trajectory penetrates the right half plane depends on the
dynamics of the theory. It may produce just one bound state at $l = 0$,
or only one resonance, or no physical state at all. There are also
Regge trajectories in the case of repulsive interactions. In this case
the leading trajectory, starting at $l = -1$, would move to the left in 
the complex plane, eventually returning to $l = -1$ through the lower
half plane. It will never produce bound states nor resonances. Finally,
there are in general subleading Regge trajectories which behave similarly
to the leading trajectory, but start out at lower values
$l = -2, -3, -4, \ldots$ for $s \to -\infty$.

\section{Scaling and the Renormalization Group}

In this section we will present the basic ideas behind the
renormalization we will use. In particular we will introduce the
concepts of environmentally friendly renormalization within the
simplified context of a cubic scalar field theory with Lagrangian
\be
{\cal L}={1\o2}\parm\f\parm\f+{1\o2}m_B^2\f^2+{g_B\o 6}\f^3 \:.
\ee
The index $B$ will be used to denote bare quantities
throughout. 

We will illustrate the formalism within the context of ``cubic''
scalar theories
for reasons of sheer simplicity, and because cubic couplings also appear
in gauge theories. Although these bosonic theories suffer from vacuum
instability, perturbation theory around the ``perturbative'' vacuum is
well-defined order by order. Furthermore, they have been used extensively
as model theories to investigate Regge behaviour in perturbation theory
(see, for instance, ref.\ \cite{eden}). In principle, one may always think
of adding a stabilizing $\phi^4$ term but with an arbitrarily small
coupling such that it plays no role in what we are about to discuss.

The general
notation we will use will be the following: an $n$-point connected
(bare) Greens function will be denoted, $G_B^{i_1...i_n}(p_1,...,p_n)$ where
we use the superscript notation $i_1...i_n$ to denote the external
legs, i.e.\ $i_1,...,i_n$ can take different values depending on the
field content of the theory. For instance, in $\f^3$ theory,
$i_k=\phi$ for all $k$.  For a theory with interaction term $\fd\f\p$,
$i_k$ can represent $\phi$, $\fd$ or $\psi$. Thus
$G_B^{\ss \phi\phi}$ represents propagation of the charged $\phi$-particle
and $G_B^{\ss\fd\fd}$ propagation of the corresponding antiparticle,
whilst $G_B^{\ss\psi\psi}$ represents
propagation  of the neutral particle $\psi$.

Rather than deal with the
connected Greens functions we will find it useful to consider the
quantities  $\tilde\Gamma_B^{i_1...i_n}(p_1,...,p_n)$ which are
closely related to the $S$-matrix elements and are obtained from the
connected Greens functions by removing the external legs,
\begin{equation}
{\tg}^{i_1...i_n}_B(p_1,...,p_n) =
\prod_{k=1}^{n}\Gamma^{i_k i_k}_B(p_k)
G^{i_1...i_n}_B(p_1,...,p_n) \:.
\end{equation}
The relation to fully one-particle irreducible Greens functions is found using
the standard ``tree theorem''. For instance, for the four-point function 
introduced in the last section one has the decomposition
\begin{eqnarray}
{\tilde \Gamma}^{ijkl}_B(s,t,u) &=&
\Gamma^{ijkl}_B(s,t,u) +
\Gamma^{ijm}_B(s) G^{mm}_B(s)\Gamma^{mkl}_B(s) \nonumber\\
&& {}+ \Gamma^{ikm}_B(t) G^{mm}_B(t)\Gamma^{mjl}_B(t) +
\Gamma^{ilm}_B(u) G^{mm}_B(u)\Gamma^{mjk}_B(u) \:,
\end{eqnarray}
or diagrammatically
\vspace{2mm}
\begin{equation}
{\tilde \Gamma}^{ijkl}_B = \hspace{4mm}
\LGamtil \hspace{4mm} =\Gamtil+\Gamtils+\Gamtilt+\Gamtilu
\end{equation} 
\vspace{1mm}

\noindent As indicated above,
on mass shell $\tilde{\Gamma}^{ijkl}_B$ depends on the momenta only
through the Mandelstam variables $s,t,u$, which are related by
(\ref{mandelstam}).

We will first consider the large-$t$ limit of the four-point function,
and afterwards the ``dual'' limit $s \to \infty$. It is then
convenient to consider the particles $k$ and $l$ as incoming and $i$ and
$j$ as outgoing, i.e.\ the $s$-channel process is $k l \to i j$.
Consequently we define
\be
s = - (p_1 + p_2)^2 \:, \quad t = - (p_1 - p_3)^2 \:, \quad
u = - (p_1 - p_4)^2 \:,
\label{mandeldef}
\ee
working in Euclidean space throughout.\footnote{Note that the present
definitions of the Mandelstam variables differ by a sign from the ones
used in ref.\ \cite{plbus}.} The minus signs in (\ref{mandeldef})
are introduced for convenience, in order that the Mandelstam variables take
their usual Minkowskian values.

Let us first discuss the philosophy behind the renormalization.
To illustrate this we will consider the lowest order functions,
$\tg_B^{\ss\phi\phi}$, $\tg_B^{\ss\phi\phi\phi}$ and
$\tg_B^{\ss\phi\phi\phi\phi}$
for $\f^3$. Explicitly, to one loop
\begin{eqnarray}
\tg^{\ss\phi\phi}_B(p_1,p_2,m_B,g_B,\L)&=&\fffgtwotree+\fffgtwooneloop\\
\tg^{\ss\phi\phi\phi}_B(p_1,p_2,p_3,m_B,g_B,\L)&=&
\fffgthreetreeL+\fffgthreeoneloopL\\
\tg^{\ss\phi\phi\phi\phi}_B(p_1,p_2,p_3,p_4,m_B,g_B,\L)
&=&\fffgfourtree+\fffgfourtreepc+(s\ {\rm  and}\ u \ {\rm perms})+\nonumber\\
&&\fffgfourtreeL+\fffgfourtreeR+(s\ {\rm  and}\ u \ {\rm perms})+\nonumber\\
&&\fffgfouroneloop+\fffgfouroneloopcrh+\fffgfouroneloopcrv \label{phi3one}
\end{eqnarray}
where the $s$ and $u$ permutations indicate the same diagrams but in the
crossed channels.
We have introduced an ultraviolet cutoff in order to regularize any UV
divergences
present. If we restrict attention to four dimensions for simplicity then the
only
UV divergence occurs in $\ss\fffgtwooneloop$ which diverges logarithmically.
This UV divergence can be removed in the standard fashion via a normalization
condition, for instance that the renormalized two-point vertex function have a
zero
on the mass shell, i.e.
\be
\tg^{\ss\phi\phi}(p^2=-m^2,m,g)=0 \:. \label{massren}
\ee
After this mass renormalization, then as far as UV divergences are concerned
the theory is totally finite. However, we will now consider the large momentum
behaviour of the theory. In particular we will consider
two-particle scattering in the asymmetric limit $t\ra\i$ for fixed $s$.

In this limit and considering the on-shell amplitude we can classify
the
one-loop diagrams according to whether they contain a factor of $\ln t$, a
factor
of $\ln(-t)$, or are ``finite''. This separation can in principle be carried
out to all
orders leading to the following decomposition for $\Gamma_B^{ijkl}$:
\begin{equation}
\Gamma^{ijkl}_B(s,t,u)=
A^{ijkl}_{B,t}(s,t) + A^{ijkl}_{B,u}(s,u) + A^{ijkl}_{B,s}(t,u) \:,
\end{equation}
where $A^{ijkl}_{B,t}$ includes the one-particle irreducible diagrams 
with powers of $\ln(-t)$ (``$t$-con\-tri\-bu\-tions''),
$A^{ijkl}_{B,u}$ includes the diagrams with  powers of $\ln t$
(``$u$-contributions''),
and $A^{ijkl}_{B,s}$ includes the remaining diagrams (``$s$-contributions'').
There is of course an analogous decomposition
associated with the limits $s\ra\i$ and $u\ra\i$.

For the $\f^3$ theory at one loop $A^{ijkl}_{B,t}$ is given by the planar
box diagram in (\ref{phi3one}), $A^{ijkl}_{B,u}$ by the first and
$A^{ijkl}_{B,s}$ by the second ``crossed'' box.
The relation of the
present definition of $t$- and $u$-contributions to the one given in the
last section (above formula (\ref{tcontribut})) is easy to see: The
planar box diagram has a threshold cut in $t$, and the first crossed box
diagram a corresponding cut in $u$. Consequently they determine to one loop
the contributions from the cuts in (\ref{spectral}). The second crossed
box diagram possesses threshold cuts in $t$ and $u$, but its contributions
are suppressed in the limit $t \to \infty$.

It is worth mentioning here
one subtlety associated with the above: at some order in the loop expansion
there will appear terms with logarithms that are associated with
singularities in the complex angular momentum plane
other than Regge poles. For instance, in the present theory at five loops
the non-planar ladder type diagram with three iterated crosses will yield
a term of the form $\sim g_B^{12}t^{-1}\ln t$ which arises from a pinch type
singularity rather than an end-point singularity in the integration over
Feynman parameters and should be interpreted as a contribution to
an essential singularity related to the Gribov-Pomeranchuk phenomenon
\cite{eden}. There will also be terms that correspond to
Regge cuts, once again these will be associated with non-planar diagrams.
One can take different attitudes towards this problem. One is simply
to include them as contributions to the functions $A^{ijkl}_{B,t}$
even though they
really correspond to singularities other than Regge poles. This is in
some sense analogous to the ``poles only'' assumption in Regge theory.
A more satisfactory resolution is to find appropriate renormalizations
that resum the diagrammatic series so as to reproduce the correct
type of singularity. Neither way of doing things is ``wrong''. The RG
viewed as a method of reorganization and resummation of perturbative
diagrammatic series via reparametrizations cannot give incorrect answers,
however, it can give bad approximations. One is seeking a reorganization
and resummation that best captures the physics of interest. In the case
at hand as these matters are all associated with much higher
loop orders we will not consider them further here.

At the level of the functions
$\tg_B^{ijkl}$ the natural combinations for investigating the large-$t$
behaviour are the functions $B_{B,t}^{ijkl}$ for the $t$-contributions,
and $B_{B,u}^{ijkl}$ for the $u$-contributions, defined as
\begin{eqnarray}
B^{ijkl}_{B,t}(s,t) &=& A^{ijkl}_{B,t}(s,t) + \Gamma^{ikm}_{B}(t)G^{mn}_B(t)
\Gamma^{mjl}_B(t) \nonumber\\
&=&
 \LGamtilt
 +
\Gamtilt
\end{eqnarray}
\begin{eqnarray}
B^{ijkl}_{B,u}(s,u) &=& A^{ijkl}_{B,u}(s,u) + \Gamma^{ilm}_{B}(u)G^{mn}_B(u)
\Gamma^{mjk}_B(u) \nonumber\\
&=&
\LGamtilu+\Gamtilu
\end{eqnarray}
with an analogous definition for $B_{B,s}^{ijkl}$ for the $s$-contributions.
We hence have the decomposition
\be
\tg^{ijkl}_B(s,t,u)=
B^{ijkl}_{B,t}(s,t) + B^{ijkl}_{B,u}(s,u) + B^{ijkl}_{B,s}(t,u) \:.
\ee

The function $B^{ijkl}_{B,s}$ contains no large logarithms and will not
play an important role in the asymptotic $t$ behaviour. However, in the
large-$t$ limit both $B^{ijkl}_{B,t}$ and $B^{ijkl}_{B,u}$ become
perturbatively uncontrollable
due to large logarithmic corrections. One way to ameliorate this problem is to
try to sum up the large logarithmic terms thus generating a non-perturbative
result. This has had a great deal of success but it is heuristically based.

We will here implement an RG approach: Consider the function
$B^{ijkl}_{B,t}(s,t)$ in the large-$t$ limit. In the case of $\f^3$ theory,
we have explicitly to one loop (omitting the trivial upper indices on
$B_{B,t}$)
\be
B_{B,t}(s,t)={g_B^2\o -t}+{g_B^4\o -t}K(s)\ln(-t)+
O \left( \frac{g_B^4}{t} \right)
\label{phi3onellarget} \:,
\ee
where the first term on the r.h.s.\ stems from the $t$-channel tree diagram
and the second from the planar box diagram, while the contributions
of the other one-particle reducible diagrams are relatively suppressed 
in the large-$t$ limit. The function
\be
K(s)={1\o16\pi^2}\int_0^1{\mbox{d} \beta \o m^2 - \beta(1-\beta)s}
\label{Kfeynman}
\ee
corresponds
to a two-dimensional loop integration owing to the well known dimensional
reduction associated with the Regge limit (see section 5 for a short
derivation of this result). It is clear that in the $t\ra\i$
limit the above perturbative expansion is ill defined. Two-loop diagrams
lead to terms $\sim g_B^6 t^{-1}K^2(s)(\ln t)^2$ and $\sim g_B^6 t^{-1}K'(s)
\ln t$
where $K'$ is $t$-independent. Similarly, the function $B_{B,u}(s,u)$
exhibits the behaviour
\be
B_{B,u}(s,u(t,s))={g_B^2\o t}+{g_B^4\o t}K(s)\ln t+
O \left( \frac{g_B^4}{t} \right)
\label{phi3onellargeu} \:.
\ee

It is clear that in both cases some form of resummation is needed. We will
achieve that here by utilizing the RG. It is
not difficult to see, due to the fact that there are two sets of
large logarithms, one associated with $B_{B,t}$ and another set with $B_{B,u}$,
that an overall multiplicative renormalization of
the connected four-point function will not be sufficient
to make the perturbation series well defined. This is an important point
well worth emphasizing here. Normally  we associate renormalization
with a renormalization of a ``coupling constant'', i.e. a parameter in the
original Lagrangian, or an overall renormalization of a Greens function or
vertex function. One may think of this as being associated with the
fact that in a certain asymptotic limit there is only one singularity to
exponentiate. For instance, in critical phenomena in the
massless theory terms of the form $\epsilon^n(\ln k)^n$ exponentiate
to give $k^{-\eta}$. In the present case however there is more than one
singularity as we shall see. Hence, to exponentiate these
singularities one needs to get inside the Greens functions and
identify the parts that will renormalize ``naturally''. In one sense
our renormalizations can be seen as just reorganizations of
perturbative series. No reorganization is wrong, in that it cannot
affect the exact answer, however, to a given order in
perturbation theory one type of reorganization may capture much
more of the actual behaviour of the entire function than another one.

To achieve an adequate renormalization we will define multiplicative
renormalizations of $B_{B,t}$ and $B_{B,u}$ separately. Explicitly for
$B_{B,t}$
\be
B_t(s,t,g(\k),m(\k),\k)=Z_t \, B_{B,t}(s,t,g_B,m_B,\L) \:,
\ee
where the precise form of $Z_t$ of course depends on the specific
normalization condition chosen.
Hence, $B_t$ satisfies an RG equation that follows naturally from the
$\k$-independence of the bare theory,
\be
\k{d\o
d\k}B_t(s,t,g(\k),m(\k),\k)=\gamma_t \, B_t(s,t,g(\k),m(\k),\k) \:,
\label{rge}
\ee
where $\gamma_{t}=d\ln Z_t/d\ln\k$ is the anomalous dimension of $B_t$.

There is of course the question of what are we going to use as our RG scale?
The natural answer to this follows from consideration of the physics of the
problem. In the large-$t$ regime there is a natural decoupling between the
transverse and longitudinal sectors whose consequence is an effective
dimensional reduction. In other words the effective degrees of freedom of
the system are four dimensional for small $t$ and two dimensional for large
$t$. As the physics changes as a function of $t$, it is natural to choose an
arbitrary, fiducial value of $t$ as the RG parameter. This is also consistent
with the philosophy of environmentally friendly renormalization, i.e.\ that
one should use a renormalization that depends on the ``environmental''
variable (in this case asymmetric momentum) that induces the crossover.
We have assumed in the above renormalizations of the coupling constant
and mass. This may or may not be neccessary. In the current scalar theory
in four dimensions it is not essential as there are no large $t$-logarithms
in the vertex or mass that require exponentiation. For instance, the
one-loop vertex correction is $\sim g_B^3(\ln t)^2/t$, whilst the one-loop
mass correction, after removing the ultraviolet divergence by minimal
subtraction, is $\sim g_B^2\ln t$.
The former tells us that asymptotically $g_B$ is a very good approximation to
the vertex function. The latter when inserted into the four-point function
gives
a term of $O(t^{-2})$ which is negligible compared to $O(t^{-1})$ terms. In
six dimensions however, where the theory is renormalizable, a renormalization
of the coupling is essential. In this respect it is more akin to the case of
QCD.

With the above choice of the RG scale we have $\gamma_t = \gamma_t(s, g(\k),
m(\k), \k)$ which we will abbreviate in the following as $\gamma_t(s, \k)$.
Integrating equation (\ref{rge}) yields
\be
B_t(s,t,g(\k),m(\k),\k)={\rm e}^{-\int^{\rho\k}_{\k}
\gamma_{t}(s,x){dx\o x}}B_t(s,t,g(\rho\k),m(\rho\k),\rho\k)
\ee
with an analogous equation for $B_u$. As the renormalization scale $\rho$ is
arbitrary we choose $\rho\k=t$, hence
\be
B_t(s,t,g(\k),m(\k),\k)={\rm e}^{-\int^{t}_{\k}
\gamma_{t}(s,x){dx\o x}}B_t(s,t,g(t),m(t),t) \:.
\ee
Now, in the limit of large $t$ and $\k$, as we will see, there exists a 
``fixed point'' wherein
$g(t)\ra g(\i)$ and $\gamma_t$ is purely a function of $s$, i.e.
the total dependence of $\gamma_t$ on $\k$ through $g(\k)$, $m(\k)$ and
$\k$ disappears. In that the anomalous dimension, $\gamma_t$, depends 
continuously on $s$ it might be better to speak of a ``line'' of fixed points
as in the well known case of the two-dimensional XY-model. The fact that 
the anomalous scaling dimension
of $B_t$ for $\k\ra\i$ is independent of $\k$ is a direct result of the
term by term factorization in perturbation theory in this limit.
Note that $\gamma_t$
will depend on $g(\i)$, which is a non-universal parameter. In this sense the
fixed point has more in common with the ``infinite mass'', or mean field
fixed point in critical phenomena. The fact that fluctuation corrections
to $g(t)$ and $m(t)$ also vanish in the large-$t$ limit lends further support
to this analogy.
 For $\k$ and $t$ sufficiently large we have that
\be
B_t(s,t,g(\k),m(\k),\k)={g^2(t)\o -\k}\left({t\o\k}\right)^{\alpha(s)}
F_t\left({s\o t},
{g(t)\o t^{1/2}},{m(t)\o t^{1/2}}\right) \:,
\ee
where $\alpha(s)=-1-\gamma_{t}(s)$ and $F_t$ is a three-variable scaling
function. Note that we have removed a factor $g^2(t)/t$ from $B_t$.
As a consequence of the irrelevance of the operators 
associated with $g(t)$ and $m(t)$ with respect to 
the $t\ra\i$ ``fixed point'', the $t$-dependence of the
scaling funcion $F_t$ disappears to leave an amplitude
$G_t((s/g(\i),s/m(\i))$, i.e.\ in terms of momentum variables the amplitude
is only a function of $s$.\footnote{It is instructive to verify the
statements made in this general discussion explicitly for the full
crossover function in $\f^3$ theory presented in ref.\ \cite{plbus}.}
Just as the ``critical'' exponent, $\alpha$, depends
on a continuous parameter, $s$, so too does the ``critical amplitude'' $G_t$.
Thus using the RG we can recover the generic form expected via Regge theory,
cf.\ (\ref{tcontribut}). 

To obtain the correct ratio of $t$- and $u$-contributions, we have to
consider the role played by the $Z$ factors. Let us first recall that
it is the bare function $B_{B,t}$ that forms part of the scattering amplitude. 
According to the above, it is given by
\be
B_{B,t}(s,t,g_B,m_B,\Lambda) = \frac{g^2(t)}{-\k} \, \frac{G_t(s)}
{Z_t(s,g(\k),m(\k),\k)} \left( \frac{t}{\k} \right)^{\alpha(s)} \:,
\ee
where we have reexpressed $Z_t$ in terms of the renormalized parameters.
The factor $Z_t^{-1}$ is obviously necessary to cancel the unphysical
$\k$-dependence. Mathematically,
$Z_t$ is represented by a formal power series. To be
able to extract the signature given by the ratio of $u$- and $t$-contributions,
we have to make sure that $Z_u = Z_t$, at least in a formal sense. As we
will see below and in the following section in more detail, this implies
considering the large-$u$ limit for the $u$-contributions.
Proceeding in exactly the same way as for the $t$-contributions above 
(just replacing $t$ by $u$ everywhere) we find for the function $B_u$
\be
B_u(s,u,g(\k),m(\k),\k)={\rm e}^{-\int^{u}_{\k}
\gamma_{u}(s,x){dx\o x}}B_u(s,u,g(u),m(u),u) \:.
\ee
Once again in the limit of large $u$ and $\k$ for fixed $s$, 
$\gamma_u$ is independent of
$\k$ and the corresponding scaling function $F_u\ra G_u(s)$.
To one loop in the large-$t$ limit, as we will see,
$\gamma_t=\gamma_u$ and $G_t=G_u$. It is apparent that this holds to all 
orders in the case of the simple theory under consideration, since for
every diagram that contributes to $\gamma_t$ and $G_t$ there is a crossed
diagram with the incoming or outgoing legs interchanged that contributes 
identically in the large-$u$ limit to $\gamma_u$ and $G_u$.

We can now replace $u$ by $-t$, to be interpreted as $e^{i\pi} t$ (see below),
and add the $t$- and $u$-contributions to find (with $Z \equiv Z_t = Z_u$
and similarly for $G$)
\be
\tilde\Gamma^{\ss\f\f\f\f}_B(s,t)={g^2(t)\o -\k} \, \frac{G(s)
(1+e^{i\pi\alpha(s)})}{Z(s,g(\k),m(\k),\k)}
\left({t\o\k}\right)^{\alpha(s)}+ \: B_s(t,u) \:,
\ee
where $B_s(t,u)$ contains the
finite (in the large-$t$ limit) $s$-contributions.
Rather than thinking of calculating $Z$ in the denominator we should
use an experimental result on the two-point scattering amplitude at
some value of $t$ for fixed $s$ to determine it. 
We shall see this more explicitly below.
Thus we see it is possible using the RG to produce signatured amplitudes
as in (\ref{regge}). In the present case as we are dealing with identical
bosons only positive signature states enter. Note that in the above
we are only considering the leading Regge trajectory associated with
the tree level value ($g_B\ra 0$) $\alpha=-1$. There are, as mentioned,
subleading
trajectories associated with $\alpha=-n$, $n>1$. One can in fact regard
them as corrections to the dominant scaling given by the leading Regge
trajectory. In principle the RG
techniques we develop here are capable of accessing these ``correction
to scaling'' trajectories
as well. The problem of achieving it is the rather technical one of
being able to project out from the Feynman diagrams the contributions
of $O(t^{-n})$ and their associated logarithms. This can be done by
analysing the Mellin transforms of the diagrams. We will not consider the
matter further in this paper however.

It should be clear that our renormalization is completely crossing symmetric.
If we consider the limit $s\ra\i$ at fixed $t$ one identifies the function
$B_s(s,t)$
such that it contains large logarithms of the type $\ln(-s)$ which will
subsequently require a multiplicative renormalization. Similarly, the function
$B_u(t,u)$, this time in distinction to the function $B_t(s,u)$,
will contain large logarithms of the type $\ln s$. A renormalization
procedure totally analogous to the above will yield
\be
\tilde\Gamma^{\ss\f\f\f\f}_B(s,t)={g^2(s)\o -\k} \, \frac{G(t)
(1+e^{i\pi\alpha(t)})}{Z(t,g(\k),m(\k),\k)}
\left({s\o\k}\right)^{\alpha(t)}+ \: B_t(s,u) \:.
\ee
The procedure is identical in the large-$u$ limit. Naturally in the different
asymptotic limits the diagrams that contribute to the renormalization are
different. For instance, a three-rung ladder diagram in the $s$-direction 
in the large-$s$ limit $\sim g_B^6s^{-2}K'(t)\ln s$, 
whereas in the large-$t$ limit it
is $\sim g_B^6t^{-1}K^2(s)(\ln t)^2$. On the contrary a three-rung ladder in
the $t$-direction varies asymptotically as $\sim g_B^6s^{-1}K^2(t)(\ln s)^2$
for large $s$ and as $\sim g_B^6t^{-2}K'(s)K'(s)\ln t$ for large $t$. Thus a
three-rung $s$-ladder contributes to the leading Regge trajectory in the
large-$t$ limit but to a subleading trajectory in the large-$s$ limit.
Renormalization then is capable of accessing the dual asymptotic limits,
$t\ra\i$ and $s\ra\i$, characteristic of the Veneziano formula and string-like
behaviour.

To calculate explicitly to one loop the scattering amplitude one needs to fix
the renormalization constant $Z_t$ by choosing a normalization
condition. We choose
\be
B_t(s,t=\k,g_B,m,\k)={g^2_B\o -\k} \:. \label{ncone}
\ee
The motivation for a normalization condition of this type is that in the
large-$t$ limit the right hand side of (\ref{ncone}) is what one would
expect when there are no quantum fluctuations. Hence, as we will
see, by this choice all fluctuations in the large-$t$ limit will be absorbed
into the renormalization factor $Z_t$ which the RG will then exponentiate.
It is thus an optimum condition in terms of obtaining the maximum amount
of non-perturbative information in the large-$t$ limit.
We renormalize the mass as in (\ref{massren}), now for $\tg^{\ss\f\f}_B$.
Note that we have chosen not to implement a coupling constant
renormalization.

At one loop, the large-$t$ limit of $B_{B,t}$ is as given in
(\ref{phi3onellarget}), where we have to interpret the $-t$ as $e^{-i\pi}t$,
because the continuation from $t$ on the upper edge of the threshold cut
to $e^{i\pi}t$ in the Euclidean region has to yield a real function there
(for $s < 0$).
With the normalization condition (\ref{ncone}) one finds for large $t$ and
$\k$
\be
Z_t(s,\k)=1-g_B^2 K(s)\ln(e^{-i\pi}\k) \:,
\ee
hence
\be
\gamma_{t}(s)=-g_B^2K(s) \:.
\ee
Using the normalization condition again, the result for $B_t$ is
\be
B_t(s,t,g_B,m,\k)={g^2_B \o -\k}\left({t\o\k}\right)^{\alpha(s)} \:,
\ee
where the Regge trajectory, $\alpha(s)$, is given by
\be
\alpha(s)=g_B^2K(s)-1\label{traj} \:.
\ee

Let us comment here briefly on terms like $\ln t$, where the argument of the
logarithm is dimensionful: In principle such terms should be replaced by
$\ln(t/d)$ and the like, where $d$ will in general be a function of $s$.
For example, in (\ref{phi3onellarget}) the function $d$ could be determined
systematically by extracting the term $\sim t^{-1}$ in the large-$t$ expansion
of the box diagram. However, for asymptotic values of $t$ this function
plays no role, and in fact it drops out when taking the derivative of $Z_t$
with respect to $\ln \k$ and hence does not appear in the renormalization
group equation (\ref{rge}) for $B_t$ at all. It is still present in the
formal expression for $Z_t$.

As for the $u$-contributions, we can take the perturbative expression
(\ref{phi3onellargeu}) with $t = e^{-i\pi} u$ and proceed as above replacing
$t$ by $u$ everywhere. The corresponding normalization condition reads
\be
B_u(s,u=\k,g_B,m,\k)={g^2_B\o -\k} \:.
\ee
Then clearly $Z_u = Z_t$, and in particular we find
the same expression for the Regge trajectory. We then have to continue
from $u$ on the upper edge of the $u$-threshold cut to $u = e^{i\pi} t$,
giving the correct phase factor for the $u$-contributions, cf.\
(\ref{ucontribut}). Note that a different prescription to determine the
$u$-contributions would have led to $Z_u \neq Z_t$, so that the phase
factor would have been hidden in the (undetermined) ratio $Z_u/Z_t$.
The final result for $\f^3$ theory in the large-$t$ limit is
\be
\tilde\Gamma^{\ss\f\f\f\f}_B(s,t)={g^2_B\o -\k} \,
\frac{1+e^{i\pi\alpha(s)}}{Z(s,\k)}
\left({t\o\k}\right)^{\alpha(s)}+ \: \frac{g_B^2}{m^2 - s} \:.
\ee
Only the function $Z(s,\k)$ in the above is not perturbatively well defined.
Within the normal ``philosophy'' of renormalization this means that one has
to measure the two-point scattering amplitude at a fiducial scale for
$t$ and at a fixed value of $s$. This will fix the value of $Z$ for that
value of $s$. The two-point scattering amplitude is then explicitly calculable
for all other values of $t$. If one wishes to know what happens at
another value of $s$ then an appropriate experiment must be carried 
out at that value of $s$ for a fiducial value of
$t$, whereupon once again one can calculate the amplitude at all other
asymptotic values of $t$. Note that here a one-parameter family of
``initial conditions'', as functions of $s$, must be supplied for the RG.
This could be circumvented by implementing two RGs: one associated with
$t$ and another with $s$. In this way only one initial condition at
fixed, fiducial values of both $s$ and $t$ would be needed. Such
multiple RGs have been advocated for similar reasons in other circumstances
\cite{twoRG}.
Throughout the rest of the paper we will leave the $Z$ factors explicit
in our final formulas with the proviso that a set of experiments can
be used to fix them thus leaving expressions for the scattering amplitudes
which consist of explicitly calculated quantities.

For $s=0$ one finds the Regge intercept $\alpha(0)=-1+g_B^2/(16\pi^2m^2)$.
For $g_B^2>32\pi^2m^2$ the resulting cross-section will violate the
Froissart bound. This has been taken to be a serious defect of
``$s$-channel'' approaches \cite{abarbanel}. However, at two loops just
by dimensional analysis one can see that a contribution to the Regge
trajectory of the form $g_B^4/m^4$ will arise. Clearly the unitarity
violating region is beyond the reach of perturbation theory. An analysis
of this region requires a further renormalization --- of the Regge trajectory
itself. After this second renormalization $\alpha(0)$ may continue to
violate the unitarity bound. However, $\alpha(0) > 0$ implies that
$\alpha(s) = 0$ at a negative value of $s$ corresponding to a tachyonic
``bound state''. This indicates the presence of a vacuum condensate
necessitating a shift of the perturbative to the true vacuum of the theory.
We will not pursue this interesting matter further here.

\section{Matrix Renormalization}

In more complicated cases than a pure $\phi^3$ theory
a pure multiplicative renormalization is not sufficient due to the
mixing in the large-$t$ limit of the effective quartic interactions with
different $ijkl$. For instance, for a theory with interactions of the
form $\fd\phi\psi$, where $\phi$ is a charged scalar field and $\psi$ a
neutral one,
this mixing occurs among all effective interactions in the same charge
sector, i.e. there is a superselection rule preventing mixing between different
sectors. Therefore we introduce a matrix multiplicative renormalization
within each sector of definite charge. Considering again the limit $t \to
\infty$, the sectors are to be taken
with respect to the $s$-channel process, i.e.\
mixing occurs between pairs of particles $(ij)$ and $(kl)$ with the same total
charge. The multiplicative renormalization of $B_{B,t}^{ijkl}$ then takes
the form
\begin{equation}\label{BeqZB}
B^{ijkl}_{t}(s,t,g(\kappa),m(\kappa),\kappa)
= \sum_{m,n}Z^{ijmn}_{t}
\, B^{mnkl}_{B,t}(s,t,g_B,m_B,\Lambda) \:.
\end{equation}

We will write this and similar equations in the following in matrix form,
considering $B^{ijkl}_t$ as the entry in row $(ij)$ and column $(kl)$ of
a matrix ${\mathbf B}_t$. For example, in the charge zero sector of the
above bosonic theory the indices are $(\fd\f)$, $(\f\fd)$ and $(\p\p)$.
Eq.\ (\ref{BeqZB}) becomes
\be
{\mathbf B}_{t}(s,t,g(\kappa),m(\kappa),\kappa)
= {\mathbf Z}_{t} \, {\mathbf B}_{B,t}(s,t,g_B,m_B,M_B,\Lambda) \:.
\ee
As in the $\f^3$ case, the $\k$-independence of ${\mathbf B}_{B,t}$ leads
to the RG equation
\begin{equation}\label{RGeq1}
\kappa\frac{d}{d\kappa} {\mathbf B}_{t}(s,t,g(\kappa),m(\kappa),\kappa) =
\mbox{\boldmath $\gamma$}_{t} \,
{\mathbf B}_{t}(s,t,g(\kappa),m(\kappa),\kappa) \:,
\end{equation}
where
\begin{equation}\label{gamma}
\mbox{\boldmath $\gamma$}_{t}=
\frac{d {\mathbf Z}_{t}}{d \ln\kappa} \,
{\mathbf Z}_{t}^{-1} \:.
\end{equation}
The normalization condition should be so chosen that ${\mathbf Z}_{t}$
has an inverse.

Eq.\ (\ref{RGeq1}) can be considered as a set of independent
equations for the columns ${\mathbf b}_t^{(i)}$ of ${\mathbf B}_t$. 
The standard way of solution
then proceeds by decomposing each column into eigenvectors of
$\mbox{\boldmath $\gamma$}_{t}$ (we will not indicate explicitly
the dependences on $g(\k)$ or $m(\k)$ in the following),
\be
{\mathbf b}_t^{(i)}(s,t,\k)  = \sum_j \beta^{(i)}_j(s,t,\k) \,
{\mathbf v}_j(s) \:, \label{decompgeneral}
\ee
where ${\mathbf v}_j$ denotes the eigenvectors. The corresponding 
eigenvalues $\gamma_j$
play the role of the ``scalar'' $\gamma_t$ in the $\f^3$ case and
determine the Regge trajectories. There seems to be no simple argument
to guarantee that $\mbox{\boldmath $\gamma$}_{t}$ can always be
diagonalized, although we have found it to be so in all the theories
considered so far. If $\mbox{\boldmath $\gamma$}_{t}$ should fail to be
diagonalizable, then the solution of the RG equation will typically involve
logarithms of $t$, an interesting prospect at least.

In the remainder of this section, we will pursue the formal solution of
(\ref{RGeq1}) somewhat further to establish several general features
of the results to one loop. In the large-$t$ limit, one-loop perturbation
theory gives an expression of the form
\begin{equation}
{\mathbf B}_{B,t}(s,t) = \frac{g_{B}^{2}}{-t} \,{\mathbf b}_0 +
g_{B}^{4} \, \frac{\ln(e^{-i\pi} t)}{-t} \,{\mathbf b}_1(s) \:,
\label{defb}
\end{equation}
where  ${\mathbf b}_0$ and ${\mathbf b}_1(s)$ arise from the tree and
one-loop contributions, respectively. Typical entries are $1$ for
${\mathbf b}_0$ and $K(s)$, as in (\ref{Kfeynman}), for ${\mathbf b}_1(s)$.
As a consequence of the time reflection invariance of the bosonic theory,
both matrices are symmetric.

The normalization condition
\begin{equation}
{\mathbf B}_{t}(s,t=\kappa,\kappa)=  \frac{g_{B}^{2}}{-\kappa}\,
{\mathbf b}_0 \label{rencondmatrix}
\end{equation}
leads to
\be \label{zfactort}
{\mathbf Z}_t(s,\kappa) =  1 - g_{B}^{2} \, {\mathbf b}_1(s)\,
{\mathbf b}_0^{-1} \, \ln(e^{-i\pi}\k) \:,
\ee
so the RG equation reads
\begin{equation}
\kappa\frac{d}{d\kappa} {\mathbf B}_{t}(s,t,\kappa) =
\mbox{\boldmath $\gamma$}_t(s) \, {\mathbf B}_{t}(s,t,\kappa)
\end{equation}
with
\be
\mbox{\boldmath $\gamma$}_t(s) = -g_{B}^{2} \, {\mathbf b}_1(s)\,
{\mathbf b}_0^{-1} \:. \label{gamfactort}
\end{eqnarray}
Of course, for the formalism to work, ${\mathbf b}_0$ has to be invertible.

Solving the RG equation yields, by use of the normalization condition
(\ref{rencondmatrix}),
\begin{eqnarray}
{\mathbf B}_{t}(s,t,\kappa) &=&
\exp \left(-\mbox{\boldmath $\gamma$}_t(s) \ln \frac{t}{\k} \right)
{\mathbf B}_{t}(s,t,t)
\nonumber\\
&=& \frac{g_B^2}{-t}\, \sum_{k=0}^{\infty} \frac{g_B^{2k}}{k!}
\left(\ln\frac{t}{\kappa}\right)^k \left(
{\mathbf b}_1(s)\,  {\mathbf b}_0^{-1} \right)^k  {\mathbf b}_0
\label{formal} \:.
\end{eqnarray}
From the last form it is obvious that ${\mathbf B}_{t}$ is again a symmetric
matrix, as it should be. Invoking the decomposition (\ref{decompgeneral}) 
of the columns of ${\mathbf B}_{t}$ into eigenvectors of
$\mbox{\boldmath $\gamma$}_t$, (\ref{formal}) can be written in the form
($\gamma_j$ are the eigenvalues of $\mbox{\boldmath $\gamma$}_t$)
\be
{\mathbf b}_t^{(i)}(s,t,\k)  = \sum_j \beta^{(i)}_j(s,t,t) 
\left( \frac{t}{\k} \right)^{- \gamma_j(s)} {\mathbf v}_j(s) \:,
\label{solgenv}
\ee
where the integration constants at $\k = t$ are fixed by 
(\ref{rencondmatrix}) in the following way:
\be
\frac{g_{B}^{2}}{-t} \, {\mathbf b}_0^{(i)} = \sum_j 
\beta^{(i)}_j(s,t,t) \, {\mathbf v}_j(s) \:, \label{decompinitial}
\ee
${\mathbf b}_0^{(i)}$ being the $i$-th column of ${\mathbf b}_0$.

We now come to the $u$-contributions. In the large-$u$ limit one finds
\be
{\mathbf B}_{B,u}(s,u) = \frac{g_{B}^{2}}{-u} \,{\mathbf b}_0' +
g_{B}^{4} \, \frac{\ln(e^{-i\pi} u)}{-u} \,{\mathbf b}_1'(s) \:,
\ee
where the matrices ${\mathbf b}_0', {\mathbf b}_1'$ are related to
${\mathbf b}_0, {\mathbf b}_1$ by crossing symmetry. More precisely,
the $u$-contribution diagrams are obtained from the corresponding
$t$-contribution ones by interchanging the incoming particles. For
the matrices this implies interchanging columns, for example the ones
with indices $(\f\fd)$ and $(\fd\f)$. We can write this formally as
\be
{\mathbf b}_0' = {\mathbf b}_0 \, S \:, \quad {\mathbf b}_1'(s) =
{\mathbf b}_1(s) \, S \label{interchange}
\ee
with a permutation matrix $S$. Alternatively, we could interchange the
outgoing particles, corresponding to interchanging rows of the matrices.
This implies the following property of ${\mathbf b}_0$ and ${\mathbf b}_1$:
\be
S \, {\mathbf b}_0 = {\mathbf b}_0 \, S \:, \quad S \, {\mathbf b}_1(s)
= {\mathbf b}_1(s) \, S \:. \label{Scommute}
\ee
We use as normalization condition for the $u$-contributions that
\be
{\mathbf B}_{u}(s,u=\kappa,\kappa)=  \frac{g_{B}^{2}}{-\kappa}\,
{\mathbf b}_0' \:.
\ee
Now (\ref{interchange}) implies
\be
{\mathbf b}_1'(s) \, {\mathbf b}_0'^{-1} = {\mathbf b}_1(s) \,
{\mathbf b}_0^{-1} \:, \label{gamugamt}
\ee
and hence that ${\mathbf Z}_u = {\mathbf Z}_t$
and $\mbox{\boldmath $\gamma$}_u = \mbox{\boldmath $\gamma$}_t$. In
particular, the Regge trajectories are the same as for the $t$-contributions. 

To see the appearance of signatures in the general case property
(\ref{Scommute}), arising from crossing symmetry, plays a crucial role.
We first deduce, using $S = S^{-1}$, that
\be
S \, \mbox{\boldmath $\gamma$}_t(s) = \mbox{\boldmath $\gamma$}_t(s) \, S
\:, 
\ee
i.e.\ $\mbox{\boldmath $\gamma$}_t$ commutes with $S$. Assuming for the
moment that the eigenvalues of $\mbox{\boldmath $\gamma$}_t$ are
non-degenerate, we conclude that the eigenvectors satisfy
\be
S \, {\mathbf v}_j(s) = \sigma_j {\mathbf v}_j (s) \label{vsignature}
\ee
and, because
$S^2 = 1$, we have $\sigma_j = \pm 1$. In the degenerate case we can
always, by application of the projection operators
\be
\Sigma_{\pm} = \frac{1 \pm S}{2} \:,
\ee
select a basis of the eigenspace consisting of eigenvectors with definite
``signature'' as in (\ref{vsignature}). Applying this to the decomposition
of the $u$-contributions at $\k = t$ corresponding to (\ref{decompinitial}),
\be
\frac{g_{B}^{2}}{-u} \, {\mathbf b}_0'^{(i)} 
= \frac{g_{B}^{2}}{-u} \, S \, {\mathbf b}_0^{(i)}= \sum_j \sigma_j
\beta^{(i)}_j(s,u,u) \, {\mathbf v}_j(s) \:, 
\ee
we obtain for the coefficients of the $u$-contributions
\be
\beta_j'^{(i)}(s,u,u) = \sigma_j \beta^{(i)}_j(s,u,u) \:.
\ee
Consequently, the sum of ${\mathbf B}_t$ and ${\mathbf B}_u$ is of the form
(by columns)
\begin{eqnarray}
\lefteqn{{\mathbf b}_t^{(i)}(s,t,\k) + {\mathbf b}_u^{(i)}(s,u,\k)}
\hspace{2cm} \nonumber \\  
&=& \sum_j \left[ \beta^{(i)}_j(s,t,t) 
\left( \frac{t}{\k} \right)^{- \gamma_j(s)} +
\sigma_j \beta^{(i)}_j(s,u,u) 
\Bigg( \frac{u}{\k} \Bigg)^{- \gamma_j(s)} \right] {\mathbf v}_j(s) \:,
\end{eqnarray}
featuring the signatures $\sigma_j$. This demonstration also shows the
link between the signatures of the Regge trajectories and the
symmetry properties of the eigenvectors of $\mbox{\boldmath $\gamma$}_t$.

Let us finally comment on the effect of the factor ${\mathbf Z}_t^{-1}$
necessary to convert ${\mathbf B}_t$ to the bare quantity
${\mathbf B}_{B,t}$. By virtue of (\ref{zfactort}) and (\ref{gamfactort})
we have
\be
{\mathbf Z}_t^{-1}(s,\k) = 1 - \ln(e^{-i\pi} \k) \, 
\mbox{\boldmath $\gamma$}_t \:,
\ee
so that the application to (\ref{solgenv}) gives, for example,
\begin{eqnarray}
{\mathbf b}_{B,t}^{(i)}(s,t) &=& 
{\mathbf Z}_t^{-1}(s,\k) \, {\mathbf b}_{t}^{(i)}(s,t,\k) \nonumber \\
&=& \sum_j \Big( 1 - \ln(e^{-i\pi} \k) \, \gamma_j \Big) 
\, \beta^{(i)}_j(s,t,t) 
\left( \frac{t}{\k} \right)^{- \gamma_j(s)} {\mathbf v}_j(s) \:,
\ee
an analogous formula holding for the $u$-contributions. We conclude that
the amplitude of each Regge trajectory gets modified by a certain function
$Z^{-1}_j(s,\k)$ which, however, is the same for all contributions to one
and the same trajectory. In particular, the $t$- and $u$-contributions
to a single trajectory get modified by the same function, hence
signature is preserved. The same argument applies if ${\mathbf Z}_t$ is
an arbitrary power series in $\mbox{\boldmath $\gamma$}_t$ (as far as the
matrix structure is concerned), so although the expression for ${\mathbf Z}_t$
is formal, it appears reasonable to assume that the argument
continues to hold for the ``true'' ${\mathbf Z}_t$ to the present 
approximation.

To summarize, we have developed a renormalization scheme that reproduces
from perturbation theory the predictions of Regge theory. 
It has proved necessary to introduce several
non-standard ideas in the renormalization program. In order to fully justify
the renormalization scheme presented we have to demonstrate that it
can be extended to higher orders, and that the one-loop renormalization
group improved result
sums the leading logs to all orders in perturbation theory. We will 
demonstrate in appendix A the consistency of the scheme to two loops,
while in appendix B we show that the
one-loop renormalization sums the leading logs completely.
In the next section we will present explicit results for the case
of $\fd\f\psi$ theory.

\def\ttoinfty{\stackrel{\s t \to \infty}
{\hbox{\overrightarrow{\hspace{1.3cm}}}}}
\def\utoinfty{\stackrel{\s u \to \infty}
{\hbox{\overrightarrow{\hspace{1.3cm}}}}}
\def\stoinfty{\stackrel{\s s \to \infty}
{\hbox{\overrightarrow{\hspace{1.3cm}}}}}

\section{Application to {\boldmath $\phi^{\dag}\phi\psi$} Theory}

In this section we apply the above
method to the case of a bosonic theory with a
charged scalar field $\f$ and a neutral scalar field $\p$, interacting
through a cubic coupling $\fd\f\p$. We denote the masses of the $\f$ and 
$\p$ particles as $m$ and $M$, respectively. They will be renormalized on
mass shell as in (\ref{massren}). We do not implement a 
renormalization for the coupling constant $g_B$.

Let us consider the large-$t$ limit and begin with the charge $Q=+ 1$ 
sector. To illustrate the method described in the last two sections we will 
explain this case in full detail. The possible pairs of particles
corresponding to this charge are $(\f\p)$ and $(\p\f)$ (taking the
charge of the $\f$-particle equal to one). Consequently, we have to consider 
$2\times 2$ matrices with indices $(\f\p)$ and $(\p\f)$ to take into account
the mixing of effective couplings. To one loop, perturbation theory
gives for the matrix of $t$-contributions
\begin{equation}\label{BBt}
{\mathbf B}_{B,t}(s,t)=
\left(
\begin{array}{cc}
        \boxACAC   & \treeACCA   \\[5mm]
        \treeCAAC  & \boxCACA
\end{array}
\right)
\quad\ttoinfty\quad
\left(
\begin{array}{cc}
        \cboxACAC   &  \ctreeACCA   \\[5mm]
        \ctreeCAAC  &  \cboxCACA
\end{array}
\right) \:.
\end{equation}
Let us consider the contraction of the ``d-lines'' (the horizontal
propagators in this case) in the large-$t$ limit in more detail: We have
\begin{eqnarray}
{\mathbf B}^{\ss\f\p\p\f}_{B,t}(s,t) &=&
        \LtreeACCA
 \quad\ttoinfty\quad
        \ctreeACCA
 \quad \equiv \quad \frac{g_B^2}{-t}\: ,
\nonumber\\[5mm]
{\mathbf B}^{\ss\p\f\f\p}_{B,t}(s,t) &=&
        \LtreeCAAC
 \quad\ttoinfty\quad
        \ctreeCAAC
 \quad \equiv \quad \frac{g_B^2}{-t}\: ,
\nonumber\\[5mm]
{\mathbf B}^{\ss\phi\psi\phi\psi}_{B,t}(s,t) &=&
        \LboxACAC
 \quad\ttoinfty\quad
        \LcboxACAC
 \quad \equiv \quad g_B^4 \,\frac{\ln(-t)}{-t} \,K_{Mm}(s) \:,
\nonumber\\[5mm]
{\mathbf B}^{\ss\psi\phi\psi\phi}_{B,t}(s,t) &=&
        \LboxCACA
 \quad\ttoinfty\quad
        \LcboxCACA
 \quad \equiv \quad g_B^4 \,\frac{\ln(-t)}{-t} \,K_{Mm}(s) \:,
\label{contracdiagrs}
\end{eqnarray}
where the one-loop diagrams show the characteristic
and important factorization property in the limit $t \gg s,m^2,M^2$.
The function $K_{Mm}(s)$ is a {\em two}\/-dimensional one-loop
diagram, showing that the effective degrees of freedom in this limit
are different from those at $t\sim s$. One can immediately see this by
calculating the box diagram using the standard Feynman
parameterization trick. Consider as an illustration the simpler diagram
\begin{eqnarray}
\LboxAAAA &\equiv&  \frac{g_B^4}{(4\pi)^2}\,
\int_0^1 \frac{\mbox{d}\alpha_1 \mbox{d}\alpha_2 \mbox{d}\beta_1 
\mbox{d}\beta_2 \,\delta(1- \alpha_1 - \alpha_2 - \beta_1 - \beta_2)}
{\left[ -\alpha_1\alpha_2 \,t + (\alpha_1+\alpha_2)\,M^2 +
(\beta_1+\beta_2)^2\,m^2 - \beta_1\beta_2\,s \right]^2}
\nonumber\\[3mm]
\ttoinfty && 
\frac{g_B^4}{(4\pi)^2}\,
\int_0^1  \mbox{d}\beta_1 \mbox{d}\beta_2 \, \delta(1-\beta_1-\beta_2)
\int_0^\epsilon \frac{\mbox{d}\alpha_1 \mbox{d}\alpha_2}
{\left[-\alpha_1\alpha_2 \,t + (\beta_1+\beta_2)^2\,m^2 - 
\beta_1\beta_2\,s \right]^2}
\nonumber\\[3mm]
&=& \left\{  \frac{g_B^4}{(4\pi)^2}\,\int_0^1
\frac{\mbox{d}\beta_1 \mbox{d}\beta_2 \, \delta(1-\beta_1-\beta_2)}{
(\beta1+\beta_2)^2\,m^2 - \beta_1\beta_2\,s} \right\} \, \frac{\ln(-t)}{-t}
\nonumber \\[3mm]
&=& g_B^4 \, \frac{\ln(-t)}{-t} \, K_m(s) \quad
\equiv \quad \LcboxAAAA \label{factorization}
\end{eqnarray}
where in taking the limit $t\to \infty$ only the ``end point region''
$\alpha_1 \simeq \alpha_2\simeq 0$ has been considered, and from the
second to the third line contributions of $O(t^{-1})$ have been
neglected. The expression
in curly brackets is the integral corresponding to the two-dimensional
one-loop contracted diagram $K_m(s)$ as shown above. Note that in
our notation the graphs include the factor $\ln(-t)/(-t)$ as well as the 
couplings $g_B$ appearing at each vertex. Observe in particular that
by taking the limit the dependence on the mass $M$ of the exchanged
particle has disappeared.

The function $K_m(s)$ coincides with the $K(s)$ defined in
(\ref{Kfeynman}). In the slightly more complicated case where the
two-dimensional propagators have different masses, we get
\begin{eqnarray}
K_{Mm}(s) &=& \frac{1}{16 \pi^2} \int_0^1 \frac{\mbox{d} \beta}
{\beta M^2 + (1 - \beta) m^2 - \beta(1 - \beta) s} \nonumber \\
&=& \frac{1}{4\pi^2} \, \frac{\arctan \left( 
{\ds \frac{s - (M - m)^2}{(M + m)^2 - s}}
\right)^{\frac{1}{2}}}{\left[ \left( s - (M - m)^2 \right) 
\left( (M + m)^2 - s) \right) \right]^{\frac{1}{2}}} \label{Kexplicit}
\end{eqnarray}
for $s < (M + m)^2$. For $M=m$ this formula reduces to the one for $K_m(s)$.

From (\ref{contracdiagrs}) the matrix ${\mathbf B}_{B,t}$ is given by
\begin{equation}
{\mathbf B}_{B,t}(s,t)= \frac{g_B^2}{-t}
\left(
\begin{array}{cc}
g_B^2\, K_{Mm}(s) \ln(e^{-i\pi}t) & 1 \\[3mm]
1 & g_B^2\, K_{Mm}(s) \ln(e^{-i\pi}t)
\end{array}
\right) \:,
\end{equation}
where $-t$ has been taken as $e^{-i\pi}t$ corresponding to $t$ lying on the
upper (``physical'') edge of the threshold cut. We use the normalization 
condition
\begin{equation}\label{RCb}
{\mathbf B}_{t}(s,t=\kappa,\kappa)=
\frac{g_B^2}{-\kappa}
\left( \begin{array}{cc}
0 & 1 \\
1 & 0
\end{array} \right)\:.
\end{equation}
corresponding to the tree contributions, with the motivation that we wish
to absorb as much as possible of the quantum fluctuations into the
renormalization constant. Then ${\mathbf B}_{t} = {\mathbf Z}_{t} \,
{\mathbf B}_{B,t}$ implies
\be
{\mathbf Z}_{t}(s,\kappa)=
\left(
\begin{array}{cc}
1 & -g_B^2 \, K_{Mm}(s) \ln(e^{-i\pi}\kappa) \\[3mm]
-g_B^2 \, K_{Mm}(s) \ln(e^{-i\pi}\kappa) & 1
\end{array}
\right) \:, \label{zetacharge1}
\ee
neglecting terms of $O(g_B^4)$. Hence
\begin{equation}
\mbox{\boldmath $\gamma$}_t(s) =
\frac{d {\mathbf Z}_{t}}{d \ln\kappa} \, {\mathbf Z}_{t}^{-1} 
= \left( \begin{array}{cc}
0 & -g_B^2\, K_{Mm}(s)\\[2mm]
-g_B^2\, K_{Mm}(s) & 0
\end{array} \right) \:.
\end{equation}

To solve the RG equation
\be
\kappa\frac{d}{d\kappa} {\mathbf B}_{t}(s,t,\kappa) =
\mbox{\boldmath $\gamma$}_t(s) \, {\mathbf B}_{t}(s,t,\kappa) \:,
\label{RGeq2}
\ee
we consider it as a set of two equations
for the columns ${\mathbf b}_t^{(i)}$ of the matrix ${\mathbf B}_{t}$.
Each column will be expressed as a linear combination of
the (two) independent eigenvectors of $\mbox{\boldmath $\gamma$}_t$,
\begin{equation}
{\mathbf b}_t^{(i)}(s,t,\kappa) =
\beta^{(i)}_{+}(s,t,\kappa) \, {\mathbf v}_{+}(s) +
\beta^{(i)}_{-}(s,t,\kappa) \, {\mathbf v}_{-}(s) \:.
\end{equation}
The eigenvectors and
corresponding eigenvalues of $\mbox{\boldmath $\gamma$}_t$ are given by
\begin{eqnarray}
{\mathbf v}_{+} =
\left(\begin{array}{r} 1 \\[1mm] 1 \end{array}\right) \:, & &
\gamma_{+}(s)=-g_B^2\,K_{Mm}(s)\:, \nonumber \\
{\mathbf v}_{-} =
\left(\begin{array}{r} -1 \\[1mm] 1 \end{array}\right) \:, & &
\gamma_{-}(s)=g_B^2\, K_{Mm}(s) \:.
\end{eqnarray}
Eq. (\ref{RGeq2}) now becomes a set of four independent differential 
equations for the coefficients $\beta^{(i)}_{\pm}$,
\begin{equation}
\frac{d}{d\ln\k}\, \beta^{(i)}_{\pm}(s,t,\kappa) =
\gamma_{\pm}(s)\, \beta^{(i)}_{\pm}(s,t,\kappa).
\end{equation}
The integration of these equations (from $t$ to $\k$) immediately reveals 
the Regge behaviour:
\begin{equation}
\beta^{(i)}_{\pm}(s,t,\kappa) =
\beta^{(i)}_{\pm}(s,t,t) \, \left( \frac{t}{\kappa}
 \right)^{-\gamma_{\pm}(s)}.
\end{equation}
The normalization condition (\ref{RCb}) fixes the integration
constants $\beta^{(i)}_{\pm}(s,t,t)$. The final result for the
$t$-contributions is
\begin{equation}
{\mathbf B}_t(s,t,\kappa) =
\frac{g^2_B}{-2\kappa}
\left(
\begin{array}{c@{\hspace{1cm}}c}
\ds \left(\frac{t}{\kappa}\right)^{\alpha_+(s)} -
                \left(\frac{t}{\kappa}\right)^{\alpha_-(s)} &
\ds \left(\frac{t}{\kappa}\right)^{\alpha_+(s)} +
                \left(\frac{t}{\kappa}\right)^{\alpha_-(s)} \\[5mm]
\ds \left(\frac{t}{\kappa}\right)^{\alpha_+(s)} +
                \left(\frac{t}{\kappa}\right)^{\alpha_-(s)} &
\ds \left(\frac{t}{\kappa}\right)^{\alpha_+(s)} -
                \left(\frac{t}{\kappa}\right)^{\alpha_-(s)}
\end{array}
\right)
\end{equation}
with the Regge trajectories
\begin{equation}
\alpha_\pm(s)= -\gamma_{\pm}(s) - 1 = \pm g_B^2\,K_{Mm}(s)-1 \:.
\end{equation}

We can repeat the same procedure for the $u$-contributions, considering
this time the large-$u$ limit. The corresponding matrix ${\mathbf B}_{B,u}$ 
from one-loop perturbation theory reads
\begin{equation}\label{BBu}
{\mathbf B}_{B,u}(s,u)=
\left(
\begin{array}{cc}
        \treecrACAC     &    \boxcrACCA         \\[5mm]
        \boxcrCAAC      &    \treecrCACA
\end{array}
\right)
\quad\utoinfty\quad
\left(
\begin{array}{cc}
   \ctreeACAC   &  \cboxACCA     \\[5mm]
   \cboxCAAC    &  \ctreeCACA
\end{array}
\right) \:,
\end{equation}
implying
\begin{equation}
{\mathbf B}_{u}(s,u) = \frac{g_B^2}{-u}
\left(
\begin{array}{cc}
1 & g_B^2\, K_{Mm}(s) \ln(e^{-i\pi}u) \\[3mm]
g_B^2\, K_{Mm}(s) \ln(e^{-i\pi}u) & 1 \end{array}
\right) \:.
\end{equation}
The computation is identical to (\ref{factorization}), just replacing
$t$ by $u$. Imposing the analogous normalization condition
\begin{equation}
{\mathbf B}_{u}(s,u=\kappa,\kappa)=
\frac{g_B^2}{-\kappa}
\left( \begin{array}{cc}
1 & 0 \\
0 & 1
\end{array} \right)
\end{equation}
leads to the expected result that ${\mathbf Z}_u = {\mathbf Z}_t$ and
$\mbox{\boldmath $\gamma$}_u = \mbox{\boldmath $\gamma$}_t$, hence
the RG equation for ${\mathbf B}_{u}$ is the same as that for the
$t$-contributions. However, the columns in the normalization
condition are interchanged, so that finally
\begin{equation}
{\mathbf B}_u(s,u,\kappa) =
\frac{g^2_B}{-2\kappa}
\left(
\begin{array}{c@{\hspace{1cm}}c}
\ds \left(\frac{u}{\kappa}\right)^{\alpha_+(s)} +
                \left(\frac{u}{\kappa}\right)^{\alpha_-(s)} &
\ds \left(\frac{u}{\kappa}\right)^{\alpha_+(s)} -
                \left(\frac{u}{\kappa}\right)^{\alpha_-(s)} \\[5mm]
\ds \left(\frac{u}{\kappa}\right)^{\alpha_+(s)} -
                \left(\frac{u}{\kappa}\right)^{\alpha_-(s)} &
\ds \left(\frac{u}{\kappa}\right)^{\alpha_+(s)} +
                \left(\frac{u}{\kappa}\right)^{\alpha_-(s)}
\end{array}
\right) 
\end{equation}
with the same Regge trajectories as above. The analytic continuation
to $u = -t$, as explained in the last section, amounts to replacing
$u$ by $e^{i\pi}t$.

The $s$-contributions at tree level are independent of $t$ and read
\begin{equation}
{\mathbf B}_s(s) =
  \left(
  \begin{array}{cc}
    \streeACAC & \streeACCA \\[5mm]
    \streeCAAC & \streeCACA
  \end{array}
  \right)
    =
    \frac{g_B^2}{m^2-s}
    \left(
    \begin{array}{cc}
      1 & 1 \\
      1 & 1
    \end{array}
   \right).
\end{equation}
The corresponding two-particle scattering amplitudes 
${\tilde \Gamma}_B^{ijkl}$ are then given by the matrix
\be
{\widetilde{\mbox{\boldmath $\Gamma$}}_B}(s,t) = {\mathbf B}_s(s)
+ {\mathbf Z}^{-1}(s,\kappa)
\left(
{\mathbf B}_t(s,t,\kappa) +\ {\mathbf B}_u(s,t,\kappa)
\right) \:.
\end{eqnarray}
Taking the formal expression for ${\mathbf Z} = {\mathbf Z}_t$ from 
(\ref{zetacharge1}), we have
\be
{\mathbf Z}^{-1}(s,\kappa) \, {\mathbf v}_{\pm} = \Big( {\mathbf 1} -
\ln(e^{-i\pi}\k) \mbox{\boldmath $\gamma$}_t(s) \Big) \, {\mathbf v}_{\pm}
= \Big( 1 \pm g_B^2 \, K_{Mm}(s) \, \ln(e^{-i\pi}\k) \Big) \,
{\mathbf v}_{\pm} \:.
\ee
As discussed in the last section we assume that the ``true''
${\mathbf Z}^{-1}$ acts similarly, leading to the same functions of $s$
and $\k$  for all contributions corresponding to the same eigenvector and
hence to the same Regge trajectory. The final result for the
charge one sector is then
\begin{eqnarray}
\lefteqn{\gt_B^{\ss \f\p\f\p}(s,t) \;\:=\;\: \gt_B^{\ss \p\f\p\f}(s,t)} 
\hspace{1.2cm} \nonumber \\
&=& \frac{g_B^2}{m^2-s} \:-\: \frac{g_B^2}{2\k} \, 
\frac{1 + e^{i\pi\alpha_+(s)}}{Z_+(s,\k)}
\left(\frac{t}{\kappa}\right)^{\alpha_+(s)} +\:
\frac{g_B^2}{2\k} \, \frac{1 - e^{i\pi\alpha_-(s)}}{Z_-(s,\k)}
\left(\frac{t}{\kappa}\right)^{\alpha_-(s)} \:, 
\end{eqnarray}
\begin{eqnarray}
\lefteqn{\gt_B^{\ss \f\p\p\f}(s,t) \;\:=\;\: \gt_B^{\ss \p\f\f\p}(s,t)} 
\hspace{1.2cm} \nonumber \\
&=& \frac{g_B^2}{m^2-s} \:-\: \frac{g_B^2}{2\k} \, 
\frac{1 + e^{i\pi\alpha_+(s)}}{Z_+(s,\k)}
\left(\frac{t}{\kappa}\right)^{\alpha_+(s)} -\:
\frac{g_B^2}{2\k} \, \frac{1 - e^{i\pi\alpha_-(s)}}{Z_-(s,\k)}
\left(\frac{t}{\kappa}\right)^{\alpha_-(s)} \:.
\end{eqnarray}
We should think of the $s$-dependent factors $Z_+$ and $Z_-$ as
fixed by experiment, for example by measuring the functions
$\gt_B^{\ss \f\p\f\p}$ and $\gt_B^{\ss \f\p\p\f}$ at some fiducial
value $t = \k$.

All the general properties demonstrated in the last section can be
verified explicitly for this example, as for instance the symmetry 
properties of the emerging matrices. In particular, we see how the
signature of the Regge trajectories is related to the symmetry of
the corresponding eigenvectors under the exchange of components.
Of course, an identical calculation holds for the charge
$Q = -1$ sector (just reverse the sense of the arrows in the diagrams).


Now we turn to the zero charge sector. In this case, mixing will occur
between the pairs of particles $(\fd\f)$, $(\f\fd)$ and $(\p\p)$, hence
the mixing matrices are $3\times 3$. To one loop the $t$-contribution
matrix has the form
\begin{eqnarray}
\lefteqn{{\mathbf B}_{B,t}(s,t) \;\:=\;\: \frac{g_B^2}{-t}
\left(
\begin{array}{ccc}
1&0&1\\
0&1&1\\
1&1&0
\end{array}
\right)} \hspace{0.8cm} \nonumber \\[3mm]
&& {}+\: g_B^4 \, \frac{\ln(e^{-i\pi}t)}{-t}
\left(
\begin{array}{c@{\hspace{6mm}}c@{\hspace{6mm}}c}
K_{m}(s) + K_{M}(s) & K_{M}(s) & K_{m}(s) \\[3mm]
K_M(s) & K_m(s) + K_M(s) & K_m(s) \\[3mm]
K_m(s) & K_m(s) & 2 K_m(s)
\end{array}
\right) \label{chargezeroonel}
\end{eqnarray}
in the large-$t$ limit. The corresponding diagrams, as well as several details
of the following calculation, are given in appendix C.
Using again a normalization condition corresponding to the tree
contributions we obtain
\begin{equation}
\mbox{\boldmath $\gamma$}_t(s)=
-g_B^2
\left(
\begin{array}{ccc}
        K_m(s)&0&K_M(s)\\[3mm]
        0&K_m(s)&K_M(s) \\[3mm]
        K_m(s)& K_m(s) & 0
\end{array}
\right) \:. \label{gammafdfp}
\end{equation}
Observe that $\mbox{\boldmath $\gamma$}_t$ is not symmetric,
but that nevertheless
\be
\mbox{\boldmath $\gamma$}_t(s) \, S = S \, \mbox{\boldmath $\gamma$}_t(s)
\:,
\ee
where the permutation matrix 
\be
S = \left(
\begin{array}{ccc}
0&1&0\\
1&0&0\\
0&0&1
\end{array}
\right)
\ee
exchanges the first two rows or columns. The matrix 
$\mbox{\boldmath $\gamma$}_t$ possesses three linear independent
eigenvectors with
\be
S \, {\mathbf v}_{\pm}(s) = {\mathbf v}_{\pm}(s) \:, \quad
S \, {\mathbf v}_0(s) = - {\mathbf v}_0(s) \:,
\ee
so that we expect the corresponding trajectories $\alpha_{\pm}$ to
have positive, and $\alpha_0$ to have negative, signature.

The matrix of $u$-contributions can be obtained from
${\mathbf B}_{B,t}$ by interchanging the first two rows (or columns),
and replacing $t$ by $u$. Let us proceed here directly to the final
result for the charge zero sector, refering to appendix C for 
the intermediate steps. The scattering amplitudes turn out to be in the 
present approximation 
\begin{eqnarray}
\lefteqn{\gt_B^{\ss\phi^{\dag}\phi\phi^{\dag}\phi} (s,t) \;\:=\;\:
\gt_B^{\ss\f\fd\f\fd} (s,t) \;\:=\;\: 
\frac{g_B^2}{M^2- s} \:-\: \frac{g_B^2}{\k} \,
 \frac{C_{+}(s) (1 + e^{i\pi\alpha_+(s)})}{Z_+(s,\k)}
\left( \frac{t}{\kappa} \right)^{\alpha_{+}(s)}} \hspace{15mm} \nonumber \\ 
& & {}-\: \frac{g_B^2}{\k} \,
 \frac{C_{-}(s) (1 + e^{i\pi\alpha_-(s)})}{Z_-(s,\k)}
   \left( \frac{t}{\kappa} \right)^{\alpha_{-}(s)}
\:-\: \frac{g_B^2}{2\k} \,
 \frac{(1 - e^{i\pi\alpha_0(s)})}{Z_0(s,\k)}
   \left( \frac{t}{\kappa} \right)^{\alpha_{0}(s)} \:,
\label{scatzerocharge1}
\end{eqnarray}
\begin{eqnarray}
\lefteqn{\gt_B^{\ss\phi^{\dag}\phi\phi\phi^{\dag}} (s,t) \;\:=\;\:
\gt_B^{\ss\f\fd\fd\f} (s,t) \;\:=\;\: 
\frac{g_B^2}{M^2- s} \:-\: \frac{g_B^2}{\k} \,
 \frac{C_{+}(s) (1 + e^{i\pi\alpha_+(s)})}{Z_+(s,\k)}
\left( \frac{t}{\kappa} \right)^{\alpha_{+}(s)}} \hspace{15mm} \nonumber \\ 
& & {}-\: \frac{g_B^2}{\k} \,
 \frac{C_{-}(s) (1 + e^{i\pi\alpha_-(s)})}{Z_-(s,\k)}
   \left( \frac{t}{\kappa} \right)^{\alpha_{-}(s)}
\:+\: \frac{g_B^2}{2\k} \,
 \frac{(1 - e^{i\pi\alpha_0(s)})}{Z_0(s,\k)}
   \left( \frac{t}{\kappa} \right)^{\alpha_{0}(s)} \:,
\end{eqnarray}
\begin{eqnarray}
\lefteqn{\gt_B^{\ss\fd\f\p\p} (s,t) \;\:=\;\:
\gt_B^{\ss\f\fd\p\p} (s,t) \;\:=\;\: \gt_B^{\ss\p\p\fd\f} (s,t) \;\:=\;\:
\gt_B^{\ss\p\p\f\fd} (s,t)} \hspace{1mm} \nonumber \\ 
&=& -\: \frac{g_B^2}{\k} \,
 \frac{D_{+}(s) (1 + e^{i\pi\alpha_+(s)})}{Z_+(s,\k)}
   \left( \frac{t}{\kappa} \right)^{\alpha_{+}(s)}
\:-\: \frac{g_B^2}{\k} \,
 \frac{D_{-}(s) (1 + e^{i\pi\alpha_-(s)})}{Z_-(s,\k)}
   \left( \frac{t}{\kappa} \right)^{\alpha_{-}(s)} \:,
\end{eqnarray}
\begin{eqnarray}
\lefteqn{\gt_B^{\ss\p\p\p\p} (s,t)} \hspace{6mm} \nonumber \\
&=& -\: \frac{g_B^2}{\k} \,
 \frac{E(s) (1 + e^{i\pi\alpha_+(s)})}{Z_+(s,\k)}
   \left( \frac{t}{\kappa} \right)^{\alpha_{+}(s)}
\:+\: \frac{g_B^2}{\k} \,
 \frac{E(s) (1 + e^{i\pi\alpha_-(s)})}{Z_-(s,\k)}
   \left( \frac{t}{\kappa} \right)^{\alpha_{-}(s)} \:, 
\label{scatzerocharge4}
\end{eqnarray}
where we have also included the $s$-contributions. 
The Regge trajectories with the expected signatures are explicitly
\begin{eqnarray}
\alpha_{\pm}(s) &=&
\frac{g_B^2\,K_m(s)}{2}\left(1 \pm \left( 1+8\frac{K_M(s)}{K_m(s)} 
\right)^{\frac{1}{2}} \right) -1 \:, \nonumber \\ 
\alpha_0(s) &=&
g_B^2 \, K_m(s) -1 \:, \label{chargezerotraj}
\end{eqnarray}
and the amplitudes $C_{\pm}$, $D_{\pm}$ and $E$ are given in the
appendix. We have found a surprisingly rich structure here. Observe in
particular that the negative-signature trajectory $\alpha_0$ does not
mix with $(\p\p)$-states.


The last case to be considered is the charge $Q = 2$ sector. In this case 
we do not need to implement a matrix renormalization. 
The $t$-contributions are given by
\begin{eqnarray}
B_{B,t}^{\ss\phi\phi\phi\phi}(s,t) &=& 
\treeAAAA + \boxAAAA \nonumber \\[3mm]
\ttoinfty && \ctreeAAAA + \cboxAAAA 
\quad\equiv\quad \frac{g_B^2}{-t} + 
g_B^4 \, \frac{\ln(e^{-i\pi}t)}{-t} \, K_m(s) \:,
\end{eqnarray}
the same as (\ref{phi3onellarget}) for $\f^3$ theory. Consequently
the whole renormalization procedure is exactly the same as in
section 3. For better comparison with the limit of large $s$ to be discussed
below, we also present the graphs for the $u$-contributions:
\begin{eqnarray}
B_{B,u}^{\ss\phi\phi\phi\phi}(s,u) &=& 
\treecrAAAA + \boxcrAAAA \nonumber \\[3mm]
\utoinfty && \ctreeAAAA + \cboxAAAA 
\quad\equiv\quad \frac{g_B^2}{-u} + 
g_B^4 \, \frac{\ln(e^{-i\pi}u)}{-u} \, K_m(s) \:.
\end{eqnarray}
The result for the scattering amplitude is then
\begin{equation}
\tilde\Gamma^{\ss\f\f\f\f}_B(s,t)={g_B^2 \o -\k} \, \frac{
1+e^{i\pi\alpha(s)}}{Z(s,\k)}
\left({t\o\k}\right)^{\alpha(s)} \:, \quad
\alpha(s) = g_B^2\,K_m(s) -1\:.
\end{equation}
Note that there are no $s$-contributions here, in contradistinction to
$\f^3$ theory. The Regge trajectory is the same as $\alpha_0$ in the
zero charge sector, the reason being the absence of mixing with 
$(\p\p)$-states there. As is apparent from the diagrams, we obtain 
identical results in the charge $Q = -2$ sector.

We remark that our results for $\fd\f\p$ theory also apply to several
similar bosonic theories. For example, the corresponding results for the
Wick-Cutkosky model (for the process $\f_1\f_2 \to \f_1\f_2$)
can be obtained from the charge two sector of
$\fd\f\p$ theory just by ignoring the $u$-contributions because of the
absence of exchange interactions in this case. Slightly different
trajectories are obtained in a $\f^2\p$ theory, where the $\f$-particle
is uncharged. The corresponding results are given in ref.\ \cite{plbus}.

In the above we have exclusively considered the asymptotic large-$t$
behaviour of the theory.
If we are interested in the asymptotic behaviour of the two-particle
on-shell scattering amplitude in the large-$s$ limit we have to
consider that the mixing among the different quartic interactions has
to be seen in sectors of charge according to the processes in the
$t$-channel. For example in the charge $Q_t= 1$ sector the mixing
occurs between the following quartic interactions:
\begin{equation}
\begin{array}{c}
         t \leftarrow \\[3mm]
         s \uparrow
\end{array}
\quad
\left(
\begin{array}{cc}
	 \genABCC & \genACCA \\[5mm]
	 \genCBBC &\genCCBA
\end{array}
\right) \:,
\end{equation}
where now the indices $(\f\p)$ and $(\p\f)$ have been arranged according 
to the outgoing (rows) and incoming (columns) particles in the $t$-channel.
To coincide with the order of the indices for the matrix, the incoming
particles are shown as entering the diagrams from the right.

For illustration, we present the diagrams for the simplest case of charge 
$Q_t= 2$ in the large-$s$ limit. To one loop we have for the 
$s$-contributions
\begin{eqnarray}
{B}^{\ss\phi\phi\phi\phi}_{B,s}(s,t)  &=&
\streeABBA + \boxABBA \nonumber \\[3mm]
\stoinfty && \ctreeABBA + \cboxSABBA 
\quad\equiv\quad \frac{g_B^2}{-s} +
g_B^4 \, \frac{\ln(e^{-i\pi}s)}{-s} \, K_m(t) \:,
\end{eqnarray}
while the $u$-contributions turn out to be
\begin{eqnarray}
{B}^{\ss\phi\phi\phi\phi}_{B,u}(t,u) &=&
\streecrABBA + \sboxcrABBA \nonumber \\[3mm]
\utoinfty && \ctreeABBA + \cboxSABBA 
\quad\equiv\quad \frac{g_B^2}{-u} +
g_B^4 \, \frac{\ln(e^{-i\pi}u)}{-u} \, K_m(t) \:.
\end{eqnarray}
The $t$-contributions are suppressed in this limit (there
is no tree-level diagram here). Observe how the association of each
one-loop diagram with a certain tree diagram is different from the 
corresponding one in the large-$t$ limit.

Imposing the respective normalization conditions
\be
{B}^{\ss\phi\phi\phi\phi}_{s}(s=\k,t,\k) = 
{B}^{\ss\phi\phi\phi\phi}_{u}(t,u=\k,\k) = \frac{g_B^2}{-\k}
\ee
and proceeding as before, we obtain the expected result
\begin{equation}
\tilde\Gamma^{\ss\f\f\f\f}_B(s,t)={g_B^2 \o -\k} \, \frac{
1+e^{i\pi\alpha(t)}}{Z(t,\k)}
\left({s\o\k}\right)^{\alpha(t)} \:, \quad
\alpha(t) = g_B^2\,K_m(t) -1
\end{equation}
in the limit $s \to \infty$.

Let us conclude this section with a brief discussion of the Regge 
trajectories found in the present approximation.  A much more exhaustive 
discussion of their qualitative and quantitative features will be
given elsewhere. Of the trajectories found above, $\alpha_+$ (in the
sectors with charge one and zero) and $\alpha_0 = \alpha$ correspond to
attractive interactions, while the $\alpha_-$-trajectories are repulsive.
The appearance of repulsive trajectories can be attributed to the
presence of exchange interactions. The other salient qualitative feature
is the divergence of the trajectories at $s = 4M^2$, $4m^2$ or $(M + m)^2$
(depending on the charge sector), as can be read off from the explicit
expression (\ref{Kexplicit}). This divergence at threshold is inappropriate
in the case of interactions through exchange of massive particles 
and has to be considered a defect of the lowest order
approximation. It would imply for instance the existence of an
infinite number of bound states as realized in the case of zero mass
exchange particles producing an interaction of infinite range (barring
confinement, of course). Technically, the problem can be traced to the
loss of information about the mass of the exchange particle
in the process of taking the limit $t \to \infty$, as remarked below
eq.\ (\ref{factorization}). The mass of the
exchange particle will enter at two loops, however, whether this
is sufficient to cut off the spectrum at a finite number of bound states
remains to be seen.
\begin{figure}[t]
\begin{center}
\unitlength1cm
\begin{picture}(16,10.5)
\put(14,0.7){$\mbox{Re} \, l$}
\put(5.2,9.7){$\mbox{Im} \, l$}
\put(8,2){$\alpha_+(s)$}
\put(0,-1){\psfig{figure=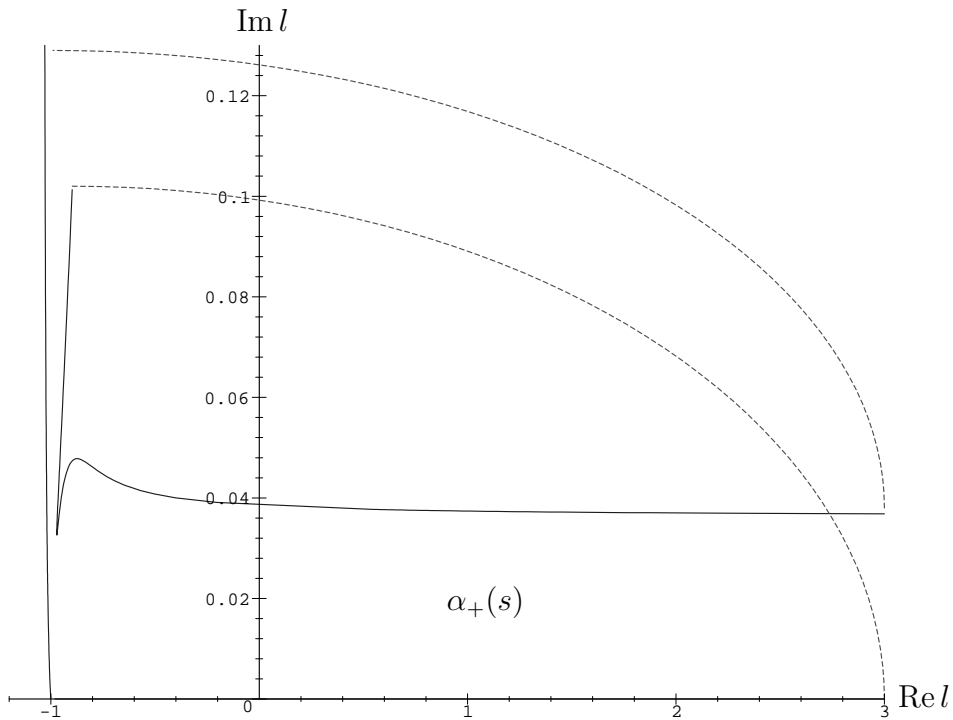,width=16cm,angle=-90}}
\end{picture}
\parbox{14cm}{\caption{The trajectory $\alpha_+$ for the zero charge sector
in the complex $l$-plane for the values $M=1$, $m=2$ and
$g_B^2/(16\pi^2) = 1/25$. The trajectory starts out at $l = -1$ following 
the real axis to the right and goes off to infinity at $s = 4 M^2$ and
$s = 4 m^2$ (represented by the dashed lines).}}
\end{center}
\end{figure}

The trajectories $\alpha_+$ in the charge one sector and $\alpha_0 =
\alpha$ show a qualitative 
behaviour similar to the ``typical'' trajectory depicted
in fig.\ 1, except for the fact that they go off to infinity at
threshold. Interestingly, $\alpha_+$ in the zero charge sector behaves
differently owing to the existence of two thresholds
at $4M^2$ and $4m^2$. Consequently, it produces two series of bound
states and resonances (within ranges of values for the coupling constant)
with overlapping angular momenta. The trajectory is represented in
fig.\ 2 in the complex $l$-plane, while fig.\ 3 shows a plot of its real and
imaginary parts as functions of $s$, so that the masses of bound states
and resonances can in principle be read off from the points where
$\mbox{Re}(\alpha_+(s))$ equals an (even) positive integer. We have used 
different coupling constants in the two cases in order to make the overall
qualitative features clearly visible.
\begin{figure}[t]
\begin{center}
\unitlength1cm
\begin{picture}(16,10)
\put(14.1,3){$s$}
\put(7.1,2){$\mbox{Re}(\alpha_+(s))$}
\put(7.1,4.7){$\mbox{Im}(\alpha_+(s))$}
\put(0,-1){\psfig{figure=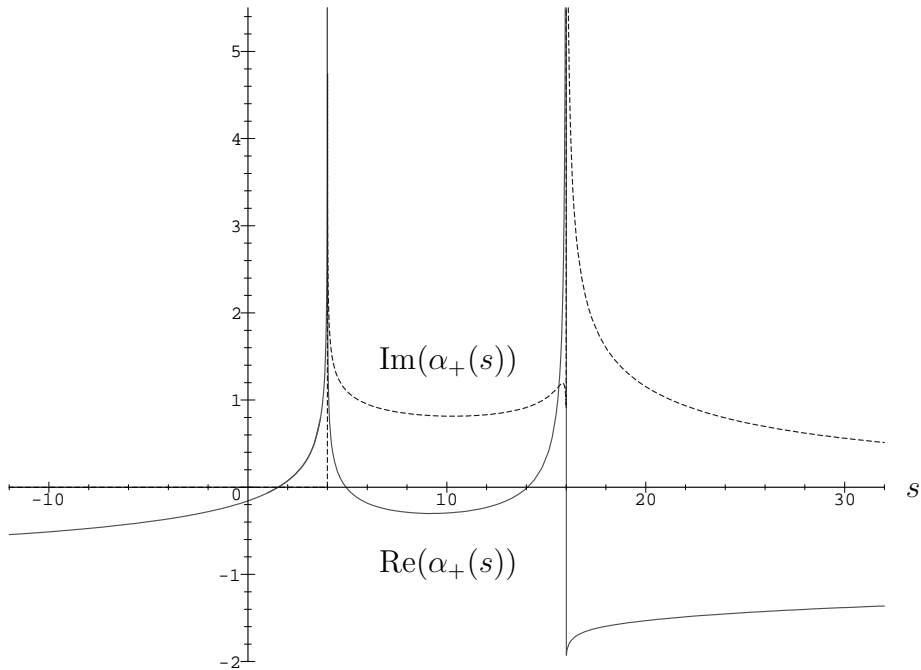,width=16cm,angle=-90}}
\end{picture}
\parbox{14cm}{\caption{Real (solid line) and imaginary (dashed line) 
parts of $\alpha_+$ as functions
of $s$ for the values $M=1$, $m=2$ and $g_B^2/(16\pi^2) = 1$. 
Note that the values of $s$ where $\mbox{Re}(\alpha_+(s))$ crosses
an integer with negative slope do {\em not} correspond to physical
states, but rather to so-called ``ghost'' states.}}
\end{center}
\end{figure}

The trajectory $\alpha$ had been found before from the Bethe-Salpeter
equation \cite{lees} and by direct summation of leading logs \cite{polk}.
We can compare the bound state spectrum with a direct solution of the
Bethe-Salpeter equation in the ladder approximation. For the Wick-Cutkosky
model with a massless exchange particle  
we find good agreement for small values
of the coupling constant. The other Regge trajectories $\alpha_{\pm}$
represent new results and would have been very difficult to obtain from either
the Bethe-Salpeter equation or a summation of leading logs (in the latter
case due to the difficulty in determining the combinatorial factors of the
diagrams in the charge zero sector to all orders).

\section{Discussion and Conclusions}

We have introduced and developed in this paper a methodology
based on environmentally friendly renormalization for
investigating high momentum behaviour in quantum field theory
in the asymmetric case where $t\gg s$ or $s\gg t$, i.e.\ $x \ll 1$. 
We have seen
that renormalization in this limit introduces several rather
novel features. First of all, we saw that it was necessary to
introduce a renormalization to remove a finite term due
to the fact that in the asymmetric limits $t\ra\i$ for fixed $s$,
or $s\ra\i$ for fixed $t$ the term is logarithmically divergent.
We showed that to do this consistently and succesfully in
perturbation theory an overall renormalization of the Greens
function or vertex function was not sufficient, rather one
was obliged to identify appropriate subsets of diagrams contained
in the Greens functions and renormalize them accordingly.
An appropriate normalization condition then assured that as much as
possible of the perturbative series was exponentiated via the RG.
For a $\fd\f\psi$ interaction we saw that due to mixing of the
different four-point effective couplings it was neccessary to
implement a matrix renormalization.

The
renormalizations we implemented naturally preserve crossing symmetry,
an important advantage compared to the Bethe-Salpeter equation in
the ladder approximation. In the case of a gauge theory our
methodology would also naturally preserve gauge invariance, once again
an advantage over the Bethe-Salpeter equation. We saw that an
environmentally friendly RG could be used to derive a scaling form
for the two-particle scattering amplitude that was formally the same
as that derived by Regge theory. We then derived explicit scaling forms
to one loop for a simple cubic theory and for a charged scalar field
interacting with a neutral scalar field. In particular we derived
the Regge trajectories for these theories to one loop, corresponding
to the summation of all leading logs.

We were also able to extract the signatures of the Regge trajectories.
Their physical significance from the viewpoint of the corresponding
bound state and resonance physics, as well as related qualitative
properties of the trajectories, will be discussed fully in a future
publication. As far as quantitative implications are concerned, 
in the cases where a direct comparison can be made, our results are
in good agreement with those of the Bethe-Salpeter equation when the
mass of the exchange particle is small. For large masses there is a
significant deviation, this disadvantage being associated with what
is also the principle advantage
of going to the Regge limit to examine bound state phenomena --- the
factorization of Feynman diagrams. It is this factorization that allowed for 
a simple
multiplicative renormalization once the relevant functions that do
multiplicatively renormalize were identified. Our results would thus seem to 
be more
promising in this regard for theories with massless interchange such as
QED, QCD and gravity.

Some extensions of the present work are immediate and will be treated in
the very near future. Among them is the incorporation of particles
with spin into the formalism, including the phenomenon of Reggeization
for spin one-half fermions. The methodology presented
in this paper appears to be equally applicable to theories with 
fermions and/or gauge bosons, the key being to identify the subsets of
diagrams that need renormalization due to their divergent behaviour in
the high-momentum limit. Another point is the formal divergence of the
trajectories for massless particles propagating in the non-contracted
direction. Here an implicit equation for the Regge trajectory like the
one found by Polkinghorne \cite{logs} through summation of all logs of
ladder diagrams could turn out to be important. Of course, within our 
framework it should be established from RG arguments. Such an equation,
and the extension to higher loop orders, may also be
important in resolving the issue of the independence of the trajectories
to one-loop order with respect to the mass of the exchange particle.

Let us finally comment on some other interesting aspects of the present
formalism, which we plan to explore in future applications.
One noteworthy property is the ``duality'' implicit in our renormalization.
We saw that it was possible to use the RG to resum the ``dual'' limits
$s\ra\i$, $t$ fixed and $t\ra\i$, $s$ fixed yielding in these asymptotic
limits behaviour of the form $s^{\alpha(t)}$ and $t^{\alpha(s)}$ respectively.
In these limits the RG
sums up ``dual'' sets of ladder diagrams, however, each limit requires
a different normalization condition. Naturally, it would be very interesting
to look for a uniform renormalization that accesses both dual limits at one
and the same time.

Another interesting aspect of our methodology is the ability to
access the kinematic dimensional reduction characteristic of the Regge limit.
This limit is not only of interest in a theory such as QCD but also could have
a very important role to play in quantum gravity as has been emphasized
by 't Hooft \cite{gerarddimred}, and especially in the question of black hole
formation
as has been advocated in ref.\ \cite{blackhole} where the use of an RG that is
capable of coarse graining between ``hard'' particles (that contribute to the
background metric of the black hole) and ``soft'' particles (that represent
particle production on the background metric) has been proposed
as a method for understanding the physics of gravitational collapse and
proving that it is governed by a unitary evolution. We hope that the type of
RG exhibited in this paper may prove to be a first step in this direction.

In work related to the above, 't Hooft \cite{gerard} also 
showed that certain Planck
scale processes can be calculated using
known laws of physics. Specifically scattering processes with $s\gg t$ can
be calculated which lead to scattering amplitudes that have certain features
in common with string theory scattering amplitudes. This work was extended
by the Verlindes \cite{verlinde} who showed that in the ``Regge'' regime
quantum gravity separates into strongly coupled (longitudinal) and weakly
coupled
(transverse) sectors. The methodology used was basically an eikonal type
approximation. For a scalar theory such as we have considered
here the eikonal approximation is trivial in the limit $s\ra\i$ for fixed $t$;
the contributions of ladder diagrams that are two-particle irreducible in the
$t$-channel yield a Born type limit, whilst ladder diagrams that are
two-particle irreducible in the $s$-channel lead to non-trivial Regge
behaviour. This is obviously a matter that depends very much on the spin of the
interchange particle and is one we will return to in the future.

\section*{Acknowledgements}

It is a pleasure to thank U. Ellwanger and D. O'Connor for interesting
and helpful discussions. We would also like to thank S. Dilcher for help
with the graphics and reading of the manuscript.

\section*{Appendix A}

In this appendix we will formally give the two-loop results, showing that the
renormalization schemes proposed are consistent to two-loop level. We will
consider explicitly only the simplest case of a $\f^3$ theory. 
For $B_{B,t}$ we have in the large-$t$ limit to two loops
\begin{eqnarray}
B_{B,t}(s,t) &=& {g_B^2\o -t}+{g_B^4\o -t}\,K(s)\ln(-t)+
{g_B^6\o -2t}\,K^2(s)\ln^2(-t) \nonumber \\
&& {}+{g_B^6\o-t}\,K'(s)\ln(-t)+
O \left( \frac{g_B^4}{t}, \frac{g_B^6}{t} \right)
\end{eqnarray}
with $-t = e^{i\pi}t$ and $K(s)$ and $K'(s)$ as in section 3.
Using the normalization condition (\ref{ncone}),
\[
B_t(s,t=\k,\k)={g^2_B\o -\k} \:,
\]
one finds
\be
Z_t(s,\k)=1-g^2_B K(s)\ln(-\k)+\frac{g_B^4}{2}\,K^2(s)\ln^2(-\k) -
g^4_B K'(s)\ln(-\k) \:.
\ee
Hence, the renormalized function $B_t$ (before applying the RG) is given by
\be
B_t(s,t,\k)={g_B^2\o -t}+{g_B^4\o -t}\,K(s)\ln \left({t\o\k}\right)+
{g_B^6\o -2t}\,K^2(s)\ln^2 \left({t\o\k}\right) + 
{g_B^6\o -t}\,K'(s)\ln \left({t\o\k}\right) \:.
\ee
As long as the arbitrary scale, $\k$, is chosen such that 
$g_B^2\ln(t/\k)\ll 1$
then perturbative expressions will be well defined, i.e.\ that our proposed
renormalization also works to two-loop order. Utilizing the RG, to two loops
one finds
\be
\gamma_t(s)=-g^2_BK(s)-g_B^4K'(s) \:,
\ee
so the result for the $t$-contributions is
\be
B_t(s,t,\k)={g_B^2\o -\k} \left(\frac{t}{\k}\right)^{\alpha(s)} \:,
\quad \alpha(s)=g^2_BK(s)+g_B^4K'(s)-1 \:.
\ee
The analogous two-loop calculation for $B_u$ goes through in exactly the same
way, showing that the trajectory has positive signature as expected.


\section*{Appendix B}

In this part we demonstrate that the one-loop renormalization scheme 
presented in sections 3 and 4 sums the leading logs to all orders in
perturbation theory. We consider the large-$t$ limit and give the
proof explicitly for the charge zero sector in $\fd\f\p$ theory.
It will be apparent that the same arguments hold in all other cases, too.

It can be shown in general for scalar theories that at a given loop order
the diagrams giving the highest power of $\ln t$ are the ladder diagrams
in $s$-direction. More specifically, a ladder with $k$ loops leads to a 
contribution of the form $\sim (g_B^{2k + 2}/k!) \, t^{-1} K^k(s) (\ln t)^k$ 
(see e.g.\ ref.\ \cite{eden}). For the $t$-contributions, the ladders in 
question are planar diagrams with $k + 1$ ``d-lines'', whose contraction 
leaves a product of facorized two-dimensional bubbles $K(s)$. Depending on 
the specific theory, not necessarily all the $K$'s
represent the same function.

In general it is difficult to write down a formula for the leading
contributions to all orders, including all contributing  
planar ladder diagrams
and their combinatorial factors, and in particular this is the case for
the $\fd\f\p$ theory under consideration. Nevertheless, it is
relatively easy to give a recursive prescription, generating the
contribution to $k+1$ loops from the $k$-loop term just by adding
one further rung to the ladder in every possible way. In this way
we will show that the solution (\ref{formal}) of the RG equation,
\be
{\mathbf B}_{t}(s,t,\kappa) =
 \frac{g_B^2}{-t}\, \sum_{k=0}^{\infty} \frac{g_B^{2k}}{k!} \,
\ln^k(-t) \left( {\mathbf b}_1(s)\,  {\mathbf b}_0^{-1} \right)^k  
{\mathbf b}_0 \:, \label{formalpert}
\ee
produces the correct contributions to every loop order. We have replaced
$\ln(t/\k)$ by $\ln(-t)$ here since the $\k$-dependence does not arise in
perturbation theory without renormalization, and in any case $\ln(-\k)$ is 
negligible compared to $\ln(-t)$ in the large-$t$ limit. In particular 
${\mathbf Z}_{t}$ does not contribute to the leading logs in $t$, so that
${\mathbf B}_{t} = {\mathbf B}_{B,t}$ in the leading log approximation.

From (\ref{formalpert}) we deduce that our renormalization scheme 
yields the recursive formula
\be
{\mathbf B}^{(k+1)}_{t}(s,t,\kappa) = \frac{g_B^2}{k+1} \, \ln(-t) \,
{\mathbf b}_1(s)\,  {\mathbf b}_0^{-1} \, {\mathbf B}^{(k)}_{t}(s,t,\kappa) 
\:, \label{recursion}
\ee
where ${\mathbf B}^{(k)}_{t}$ stands for the $k$-loop term. Referring
to the definition (\ref{defb}) of ${\mathbf b}_0$ and ${\mathbf b}_1$,
we see that the tree-level term is correctly reproduced by
(\ref{formalpert}), and we have to look at the action of the matrix
\be
g_B^2 \ln(-t) \, {\mathbf b}_1(s)\,  {\mathbf b}_0^{-1} = - \ln(-t) \,
\mbox{\boldmath $\gamma$}_{t}(s) = g_B^2 \ln(-t)
\left(
    \begin{array}{ccc}
        K_m(s)         &    0          & K_M(s) \\[2mm]
          0            &    K_m(s)     & K_M(s) \\[2mm]
        K_m(s)         & K_m(s)        &   0
    \end{array}
  \right)
\ee
in (\ref{recursion}) (cf.\ (\ref{gammafdfp}) for the explicit form
of $\mbox{\boldmath $\gamma$}_{t}$).

We may represent this matrix graphically as follows:
\begin{equation}
g_B^2 \ln(-t) \, {\mathbf b}_1(s)\, {\mathbf b}_0^{-1} =
\left(
  \begin{array}{ccc}
    \BAKm &   0   & \BAKM \\[5mm]
      0   & \ABKm & \ABKM \\[5mm]
    \CCKm & \CCKml &  0
   \end{array}
\right), \label{recursematrix}
\end{equation}
indicating that its action consists in closing the outgoing legs of
the $k$-loop terms into two-dimensional loops and adding the correct
upper legs for the $(k + 1)$-loop terms. Note
that the lower vertices indicated in the diagrams are
already present in the $k$-th terms, while the upper vertices are added to
yield the next terms. As an illustration, consider the action of 
(\ref{recursematrix}) on the tree-level terms producing the 
$\fd\f\fd\f$-entry of the matrix of one-loop contributions,
\be
\BAKm \times \ctreeBABA \:+\: \BAKM \times \ctreeCCBA \;\:=\;\:
\cboxBABAh + \cboxBABAv
\ee
(see the next appendix for a diagrammatic representation of
${\mathbf b}_0$ and ${\mathbf b}_1$).

It is obvious that (\ref{recursematrix}) produces all possible terms at
$(k + 1)$-loop order from the $k$-loop terms, thus establishing the
correctness of the leading logs in (\ref{formalpert}) to all orders by
induction. To see that the same argument holds for every scalar theory,
observe that the matrix corresponding to (\ref{recursematrix}) in
the theory considered is
uniquely fixed by the way it produces the one-loop from the tree
contributions (provided that ${\mathbf b}_0$ is invertible, what we
assume here). It will then yield the $(k + 1)$-th term from the $k$-th term
for every $k$.

Turning finally to the $u$-contributions, we have to establish that
the counterpart to (\ref{formalpert}),
\be
{\mathbf B}_{u}(s,t,\kappa) =
 \frac{g_B^2}{t}\, \sum_{k=0}^{\infty} \frac{g_B^{2k}}{k!} \,
(\ln t)^k \left( {\mathbf b}_1'(s)\,  {\mathbf b}_0'^{-1} \right)^k  
{\mathbf b}_0' \:,
\ee
reproduces correctly the leading logs in $t$ for the $u$-contributions.
The argument is identical to the above for the $t$-contributions, 
taking into account eq.\ (\ref{gamugamt}),
\[
{\mathbf b}_1'(s) \, {\mathbf b}_0'^{-1} = {\mathbf b}_1(s) \,
{\mathbf b}_0^{-1} \:.
\]
The only difference between ${\mathbf B}_{u}$ and ${\mathbf B}_t$ is then
(besides the substitution $-t \to t$) in the change of the tree-level matrix 
${\mathbf b}_0 \to {\mathbf b}_0'$. This is entirely reasonable since
on a diagrammatic level it amounts to crossing the incoming legs for
the full series ${\mathbf B}_t$. This remark concludes the proof.


\section*{Appendix C}

In this appendix we will collect together some miscellaneous formulas
needed in the calculation of the charge zero sector in section 5. 
First of all we give the diagrams for the $t$-contributions
in one-loop perturbation theory: 
\begin{eqnarray}
\lefteqn{{\mathbf B}_{B,t}(s,t) \;\:=\;\: 
\left(
\begin{array}{ccc}
\treeBABA & 0           & \treeBACC \\[5mm]
0         & \treeABAB   & \treeABCC \\[5mm]
\treeCCBA & \treeCCAB   & 0
\end{array}
\right)} \nonumber \\[5mm]
&& \hspace{-1cm} {}+ \left(
\begin{array}{c@{\hspace{6mm}}c@{\hspace{6mm}}c}
\boxBABAh + \boxBABAv & \boxBAAB & \boxBACC \\[5mm]
\boxABBA & \boxABABh + \boxABABv & \boxABCC \\[5mm]
\boxCCBA & \boxCCAB & \boxCCCCl + \boxCCCCr
\end{array}
\right) 
\end{eqnarray}
In the limit $t \to \infty$, these diagrams get contracted to yield
\begin{eqnarray}
\lefteqn{{\mathbf B}_{B,t}(s,t) \;\:=\;\: 
\left(
\begin{array}{ccc}
\ctreeBABA & 0           & \ctreeBACC \\[5mm]
0         & \ctreeABAB   & \ctreeABCC \\[5mm]
\ctreeCCBA & \ctreeCCAB   & 0
\end{array}
\right)} \nonumber \\[5mm]
&& \hspace{-1cm} {}+ \left(
\begin{array}{c@{\hspace{6mm}}c@{\hspace{6mm}}c}
\cboxBABAh + \cboxBABAv & \cboxBAAB & \cboxBACC \\[5mm]
\cboxABBA & \cboxABABh + \cboxABABv & \cboxABCC \\[5mm]
\cboxCCBA & \cboxCCAB & \cboxCCCCl + \cboxCCCCr
\end{array}
\right) \label{contracdiagrzerocharge}
\end{eqnarray}
leading to the analytic expressions in (\ref{chargezeroonel}).

The normalization condition
\be
{\mathbf B}_{t}(s,t=\k,\k) = \frac{g_B^2}{-\k}
\left(
\begin{array}{ccc}
1&0&1\\
0&1&1\\
1&1&0
\end{array}
\right)
\ee
gives
\be
{\mathbf Z}_{t}(s,\k) = \left(
\begin{array}{ccc}
1 & 0 & 0 \\
0 & 1 & 0 \\
0 & 0 & 1
\end{array}
\right) 
- g_B^2 \ln(e^{-i\pi}\kappa) \left(
\begin{array}{ccc}
        K_m(s)&0&K_M(s)\\[3mm]
        0&K_m(s)&K_M(s) \\[3mm]
        K_m(s)& K_m(s) & 0
\end{array}
\right) \:,
\ee
hence we arrive at the $\mbox{\boldmath $\gamma$}_t$ in (\ref{gammafdfp}),
\[
\mbox{\boldmath $\gamma$}_t(s)=
-g_B^2
\left(
\begin{array}{ccc}
        K_m(s)&0&K_M(s)\\[3mm]
        0&K_m(s)&K_M(s) \\[3mm]
        K_m(s)& K_m(s) & 0
\end{array}
\right) \:. 
\]
Its (unnormalized) eigenvectors and corresponding eigenvalues are
\begin{eqnarray}
{\mathbf v}_+(s) =
\left(\begin{array}{c}  
\ds 1 + \left( 1+8 \frac{K_M(s)}{K_m(s)} \right)^{\frac{1}{2}} \\[5mm]
\ds 1 + \left( 1+8 \frac{K_M(s)}{K_m(s)} \right)^{\frac{1}{2}} \\[5mm] 
\ds 4 \end{array}\right) \:, & &
\gamma_+(s)= - \, \frac{g_B^2 K_m(s)}{2} 
\left(1 + \left( 1+8 \frac{K_M(s)}{K_m(s)} \right)^{\frac{1}{2}} \right) \:,
\nonumber \\[5mm]
{\mathbf v}_-(s) =
\left(\begin{array}{c} 
\ds 1 - \left( 1+8 \frac{K_M(s)}{K_m(s)} \right)^{\frac{1}{2}} \\[5mm]
\ds 1 - \left( 1+8 \frac{K_M(s)}{K_m(s)} \right)^{\frac{1}{2}} \\[5mm] 
\ds 4 \end{array}\right) \:, & &
\gamma_-(s)= - \, \frac{g^2_B K_m(s)}{2} 
\left( 1 - \left( 1+8 \frac{K_M(s)}{K_m(s)} \right)^{\frac{1}{2}} \right) \:,
\nonumber \\[5mm]
{\mathbf v}_0(s) =
\left(\begin{array}{r} -1\\1\\0 \end{array}\right) \:, & &
\gamma_0(s)= -g_B^2\,K_m(s) \:. 
\end{eqnarray}
The solution of the RG equation yields for ${\mathbf B}_{t}$
\begin{eqnarray*}
\lefteqn{B_t^{\ss\phi^{\dag}\phi\phi^{\dag}\phi} (s,t,\k) \;\:=\;\:
B_t^{\ss\f\fd\f\fd} (s,t,\k)} \hspace{15mm} \\ 
&=& -\: \frac{g_B^2}{\k} \, C_{+}(s)
\left( \frac{t}{\kappa} \right)^{\alpha_{+}(s)} \:-\: \frac{g_B^2}{\k} \,
C_{-}(s) \left( \frac{t}{\kappa} \right)^{\alpha_{-}(s)}
\:-\: \frac{g_B^2}{2\k}
\left( \frac{t}{\kappa} \right)^{\alpha_{0}(s)} \:,
\end{eqnarray*}
\begin{eqnarray*}
\lefteqn{B_t^{\ss\phi^{\dag}\phi\f\fd} (s,t,\k) \;\:=\;\:
B_t^{\ss\f\fd\fd\f} (s,t,\k)} \hspace{15mm} \\ 
&=& -\: \frac{g_B^2}{\k} \, C_{+}(s)
\left( \frac{t}{\kappa} \right)^{\alpha_{+}(s)} \:-\: \frac{g_B^2}{\k} \,
C_{-}(s) \left( \frac{t}{\kappa} \right)^{\alpha_{-}(s)}
\:+\: \frac{g_B^2}{2\k}
\left( \frac{t}{\kappa} \right)^{\alpha_{0}(s)} \:,
\end{eqnarray*}
\begin{eqnarray*}
\lefteqn{B_t^{\ss\fd\f\p\p} (s,t,\k) \;\:=\;\:
B_t^{\ss\f\fd\p\p} (s,t,\k) \;\:=\;\: B_t^{\ss\p\p\fd\f} (s,t,\k) \;\:=\;\:
B_t^{\ss\p\p\f\fd} (s,t,\k)} \hspace{14mm} \\ 
&=& -\: \frac{g_B^2}{\k} \,
D_{+}(s) \left( \frac{t}{\kappa} \right)^{\alpha_{+}(s)}
\:-\: \frac{g_B^2}{\k} \,
D_{-}(s) \left( \frac{t}{\kappa} \right)^{\alpha_{-}(s)} \:, \hspace{29mm}
\end{eqnarray*}
\be
B_t^{\ss\p\p\p\p} (s,t,\k) \;\:=\;\:  
-\: \frac{g_B^2}{\k} \,
E(s) \left( \frac{t}{\kappa} \right)^{\alpha_{+}(s)}
\:+\: \frac{g_B^2}{\k} \,
E(s) \left( \frac{t}{\kappa} \right)^{\alpha_{-}(s)} \:, \hspace{17mm} 
\ee
with the amplitude functions
\begin{eqnarray}
C_{\pm}(s) &=& \frac{1}{4} \left( 1\pm \frac{1+4\frac{K_{M}(s)}{K_m(s)}}
{\sqrt{1+8\frac{K_{M}(s)}{K_m(s)}}}\right) \:, \nonumber
\\[2mm]
D_{\pm}(s) &=& \frac{1}{2} \left( 1\pm \frac{1}
{\sqrt{1+8\frac{K_{M}(s)}{K_m(s)}}}\right) \:, \nonumber
\\[2mm]
E(s) &=& \frac{2}
{\sqrt{1+8\frac{K_{M}(s)}{K_m(s)}}} \:,
\end{eqnarray}
and the Regge trajectories from (\ref{chargezerotraj}),
\begin{eqnarray*}
\alpha_{\pm}(s) &=&
\frac{g_B^2\,K_m(s)}{2}\left(1 \pm \left( 1+8\frac{K_M(s)}{K_m(s)} 
\right)^{\frac{1}{2}} \right) -1 \:, \\ 
\alpha_0(s) &=&
g_B^2 \, K_m(s) -1 \:.
\end{eqnarray*}

Now for the $u$-contributions the diagrams up to one loop are
\begin{eqnarray}
\lefteqn{{\mathbf B}_{B,u}(s,t) \;\:=\;\: 
\left(
\begin{array}{ccc}
0           & \treecrBAAB & \treecrBACC \\[5mm]
\treecrABBA &       0     & \treecrABCC \\[5mm]
\treecrCCBA & \treecrCCAB &      0
\end{array}
\right)} \nonumber \\[5mm]
&& \hspace{-1cm} {}+ \left(
\begin{array}{c@{\hspace{6mm}}c@{\hspace{6mm}}c}
\boxcrBABA & \boxcrBAABh + \boxcrBAABv & \boxcrBACC \\[5mm]
\boxcrABBAh +\boxcrABBAv & \boxcrABAB & \boxcrABCC \\[5mm]
\boxcrCCBA & \boxcrCCAB & \boxCCCCl + \boxCCCCr
\end{array}
\right) 
\end{eqnarray}
The contracted diagrams in the limit $u \to \infty$ lead to the same 
analytic expressions as in (\ref{chargezeroonel}) for the 
$t$-contributions, except that the first and second column (or row) are 
interchanged (and $t$ is replaced by $u$). With the normalization 
condition
\be
{\mathbf B}_{u}(s,u=\k,\k) = \frac{g_B^2}{-\k}
\left(
\begin{array}{ccc}
0&1&1\\
1&0&1\\
1&1&0
\end{array}
\right)
\ee
we then find ${\mathbf Z}_u = {\mathbf Z}_t$ and 
$\mbox{\boldmath $\gamma$}_u = \mbox{\boldmath $\gamma$}_t$. The results
for ${\mathbf B}_u$ read
\begin{eqnarray*}
\lefteqn{B_u^{\ss\phi^{\dag}\phi\phi^{\dag}\phi} (s,u,\k) \;\:=\;\:
B_u^{\ss\f\fd\f\fd} (s,u,\k)} \hspace{15mm} \\ 
&=& -\: \frac{g_B^2}{\k} \, C_{+}(s)
\left( \frac{u}{\kappa} \right)^{\alpha_{+}(s)} \:-\: \frac{g_B^2}{\k} \,
C_{-}(s) \left( \frac{u}{\kappa} \right)^{\alpha_{-}(s)}
\:+\: \frac{g_B^2}{2\k}
\left( \frac{u}{\kappa} \right)^{\alpha_{0}(s)} \:,
\end{eqnarray*}
\begin{eqnarray*}
\lefteqn{B_u^{\ss\phi^{\dag}\phi\f\fd} (s,u,\k) \;\:=\;\:
B_u^{\ss\f\fd\fd\f} (s,u,\k)} \hspace{15mm} \\ 
&=& -\: \frac{g_B^2}{\k} \, C_{+}(s)
\left( \frac{u}{\kappa} \right)^{\alpha_{+}(s)} \:-\: \frac{g_B^2}{\k} \,
C_{-}(s) \left( \frac{u}{\kappa} \right)^{\alpha_{-}(s)}
\:-\: \frac{g_B^2}{2\k}
\left( \frac{u}{\kappa} \right)^{\alpha_{0}(s)} \:,
\end{eqnarray*}
\begin{eqnarray*}
\lefteqn{B_u^{\ss\fd\f\p\p} (s,u,\k) \;\:=\;\:
B_u^{\ss\f\fd\p\p} (s,u,\k) \;\:=\;\: B_u^{\ss\p\p\fd\f} (s,u,\k) \;\:=\;\:
B_u^{\ss\p\p\f\fd} (s,u,\k)} \hspace{14mm} \\ 
&=& -\: \frac{g_B^2}{\k} \,
D_{+}(s) \left( \frac{u}{\kappa} \right)^{\alpha_{+}(s)}
\:-\: \frac{g_B^2}{\k} \,
D_{-}(s) \left( \frac{u}{\kappa} \right)^{\alpha_{-}(s)} \:, \hspace{29mm}
\end{eqnarray*}
\be
B_u^{\ss\p\p\p\p} (s,u,\k) \;\:=\;\:  
-\: \frac{g_B^2}{\k} \,
E(s) \left( \frac{u}{\kappa} \right)^{\alpha_{+}(s)}
\:+\: \frac{g_B^2}{\k} \,
E(s) \left( \frac{u}{\kappa} \right)^{\alpha_{-}(s)} \:. \hspace{17mm}
\ee

Finally, the (finite) $s$-contributions are
\begin{equation}
{\mathbf B}_s(s) \:=\:
\left(
\begin{array}{ccc}
  \streeBABA  &  \streeBAAB & 0 \\[5mm]
  \streeABBA  &  \streeABAB & 0 \\[5mm]
     0        &     0       & 0
\end{array}
\right) \:=\:
 \frac{g_B^2}{M^2- s} \left(
\begin{array}{ccc} 1&1&0\\1&1&0\\0&0&0 \end{array}
\right).
\end{equation}
Adding all the contributions leads to the scattering amplitudes of
eqs.\ (\ref{scatzerocharge1}--\ref{scatzerocharge4}).

\end{fmffile}
\end{document}